\documentclass[twocolumn]{aastex62}
\usepackage[utf8]{inputenc}

\usepackage{acronym,booktabs,multirow,cellspace,tabularx,xspace,mathtools}
\usepackage{breakurl}

\graphicspath{{./figures/}}

\newcommand{\loglikelihoodminus}[1]{\IfEqCase{#1}{{GW190930A}{8.4}{GW190929A}{10.9}{GW190924A}{8.6}{GW190915A}{8.1}{GW190910A}{4.8}{GW190909A}{4.5}{GW190828B}{5.3}{GW190828A}{5.0}{GW190814A}{4.9}{GW190803A}{4.4}{GW190731A}{4.0}{GW190728A}{48007.6}{GW190727A}{5.1}{GW190720A}{9.4}{GW190719A}{4.2}{GW190708A}{4.8}{GW190707A}{7.1}{GW190706A}{5.2}{GW190701A}{3.9}{GW190630A}{5.3}{GW190620A}{5.0}{GW190602A}{4.3}{GW190527A}{6.7}{GW190521B}{5.9}{GW190521A}{11.1}{GW190519A}{17.8}{GW190517A}{5.9}{GW190514A}{4.6}{GW190513A}{4.7}{GW190512A}{5.5}{GW190503A}{4.4}{GW190426A}{5.6}{GW190425A}{5.7}{GW190424A}{3.9}{GW190421A}{4.0}{GW190413B}{5.7}{GW190413A}{4.8}{GW190412A}{10.1}{GW190408A}{5.0}}}
\newcommand{\loglikelihoodmed}[1]{\IfEqCase{#1}{{GW190930A}{-15934.9}{GW190929A}{-11962.2}{GW190924A}{-97031.8}{GW190915A}{-2809.6}{GW190910A}{93.5}{GW190909A}{25.8}{GW190828B}{43.6}{GW190828A}{121.4}{GW190814A}{298.6}{GW190803A}{28.7}{GW190731A}{29.8}{GW190728A}{64.1}{GW190727A}{58.4}{GW190720A}{-23904.6}{GW190719A}{27.0}{GW190708A}{76.7}{GW190707A}{-15883.0}{GW190706A}{72.6}{GW190701A}{55.9}{GW190630A}{114.8}{GW190620A}{65.2}{GW190602A}{73.6}{GW190527A}{22.9}{GW190521B}{322.2}{GW190521A}{-11913.6}{GW190519A}{111.0}{GW190517A}{48.8}{GW190514A}{25.9}{GW190513A}{74.3}{GW190512A}{67.5}{GW190503A}{68.5}{GW190426A}{-389547.0}{GW190425A}{-500483.9}{GW190424A}{45.7}{GW190421A}{48.8}{GW190413B}{42.7}{GW190413A}{28.4}{GW190412A}{-22827.3}{GW190408A}{108.8}}}
\newcommand{\loglikelihoodplus}[1]{\IfEqCase{#1}{{GW190930A}{15973.8}{GW190929A}{12007.8}{GW190924A}{97091.0}{GW190915A}{2897.5}{GW190910A}{4.0}{GW190909A}{3.6}{GW190828B}{4.0}{GW190828A}{4.2}{GW190814A}{3.0}{GW190803A}{2.7}{GW190731A}{2.5}{GW190728A}{13.4}{GW190727A}{5.2}{GW190720A}{23953.2}{GW190719A}{2.8}{GW190708A}{3.3}{GW190707A}{15964.2}{GW190706A}{4.0}{GW190701A}{2.6}{GW190630A}{3.8}{GW190620A}{4.1}{GW190602A}{3.3}{GW190527A}{3.0}{GW190521B}{4.7}{GW190521A}{12013.1}{GW190519A}{9.5}{GW190517A}{4.4}{GW190514A}{2.7}{GW190513A}{4.2}{GW190512A}{3.8}{GW190503A}{3.4}{GW190426A}{4.6}{GW190425A}{4.5}{GW190424A}{2.8}{GW190421A}{2.6}{GW190413B}{3.6}{GW190413A}{4.0}{GW190412A}{23002.9}{GW190408A}{3.7}}}
\newcommand{\chieffminus}[1]{\IfEqCase{#1}{{GW190930A}{0.15}{GW190929A}{0.33}{GW190924A}{0.09}{GW190915A}{0.25}{GW190910A}{0.18}{GW190909A}{0.36}{GW190828B}{0.16}{GW190828A}{0.16}{GW190814A}{0.06}{GW190803A}{0.27}{GW190731A}{0.24}{GW190728A}{0.07}{GW190727A}{0.25}{GW190720A}{0.12}{GW190719A}{0.31}{GW190708A}{0.08}{GW190707A}{0.08}{GW190706A}{0.29}{GW190701A}{0.29}{GW190630A}{0.13}{GW190620A}{0.25}{GW190602A}{0.24}{GW190527A}{0.28}{GW190521B}{0.13}{GW190521A}{0.39}{GW190519A}{0.22}{GW190517A}{0.19}{GW190514A}{0.32}{GW190513A}{0.17}{GW190512A}{0.13}{GW190503A}{0.26}{GW190426A}{0.30}{GW190425A}{0.05}{GW190424A}{0.22}{GW190421A}{0.27}{GW190413B}{0.29}{GW190413A}{0.34}{GW190412A}{0.11}{GW190408A}{0.19}}}
\newcommand{\chieffmed}[1]{\IfEqCase{#1}{{GW190930A}{0.14}{GW190929A}{0.01}{GW190924A}{0.03}{GW190915A}{0.02}{GW190910A}{0.02}{GW190909A}{-0.06}{GW190828B}{0.08}{GW190828A}{0.19}{GW190814A}{0.00}{GW190803A}{-0.03}{GW190731A}{0.06}{GW190728A}{0.12}{GW190727A}{0.11}{GW190720A}{0.18}{GW190719A}{0.32}{GW190708A}{0.02}{GW190707A}{-0.05}{GW190706A}{0.28}{GW190701A}{-0.07}{GW190630A}{0.10}{GW190620A}{0.33}{GW190602A}{0.07}{GW190527A}{0.11}{GW190521B}{0.09}{GW190521A}{0.03}{GW190519A}{0.31}{GW190517A}{0.52}{GW190514A}{-0.19}{GW190513A}{0.11}{GW190512A}{0.03}{GW190503A}{-0.03}{GW190426A}{-0.03}{GW190425A}{0.06}{GW190424A}{0.13}{GW190421A}{-0.06}{GW190413B}{-0.03}{GW190413A}{-0.01}{GW190412A}{0.25}{GW190408A}{-0.03}}}
\newcommand{\chieffplus}[1]{\IfEqCase{#1}{{GW190930A}{0.31}{GW190929A}{0.34}{GW190924A}{0.30}{GW190915A}{0.20}{GW190910A}{0.18}{GW190909A}{0.37}{GW190828B}{0.16}{GW190828A}{0.15}{GW190814A}{0.06}{GW190803A}{0.24}{GW190731A}{0.24}{GW190728A}{0.20}{GW190727A}{0.26}{GW190720A}{0.14}{GW190719A}{0.29}{GW190708A}{0.10}{GW190707A}{0.10}{GW190706A}{0.26}{GW190701A}{0.23}{GW190630A}{0.12}{GW190620A}{0.22}{GW190602A}{0.25}{GW190527A}{0.28}{GW190521B}{0.10}{GW190521A}{0.32}{GW190519A}{0.20}{GW190517A}{0.19}{GW190514A}{0.29}{GW190513A}{0.28}{GW190512A}{0.12}{GW190503A}{0.20}{GW190426A}{0.32}{GW190425A}{0.11}{GW190424A}{0.22}{GW190421A}{0.22}{GW190413B}{0.25}{GW190413A}{0.29}{GW190412A}{0.08}{GW190408A}{0.14}}}
\newcommand{\totalmasssourceminus}[1]{\IfEqCase{#1}{{GW190930A}{1.5}{GW190929A}{25.2}{GW190924A}{1.0}{GW190915A}{6.4}{GW190910A}{9.1}{GW190909A}{17.6}{GW190828B}{4.4}{GW190828A}{4.8}{GW190814A}{0.9}{GW190803A}{9.0}{GW190731A}{11.3}{GW190728A}{1.3}{GW190727A}{8.0}{GW190720A}{2.3}{GW190719A}{10.7}{GW190708A}{1.8}{GW190707A}{1.3}{GW190706A}{13.9}{GW190701A}{9.5}{GW190630A}{4.8}{GW190620A}{13.1}{GW190602A}{15.6}{GW190527A}{9.8}{GW190521B}{4.8}{GW190521A}{23.5}{GW190519A}{14.8}{GW190517A}{9.6}{GW190514A}{10.8}{GW190513A}{5.9}{GW190512A}{3.5}{GW190503A}{8.3}{GW190426A}{1.5}{GW190425A}{0.1}{GW190424A}{10.7}{GW190421A}{9.2}{GW190413B}{11.9}{GW190413A}{9.7}{GW190412A}{3.7}{GW190408A}{3.0}}}
\newcommand{\totalmasssourcemed}[1]{\IfEqCase{#1}{{GW190930A}{20.3}{GW190929A}{104.3}{GW190924A}{13.9}{GW190915A}{59.9}{GW190910A}{79.6}{GW190909A}{75.0}{GW190828B}{34.4}{GW190828A}{58.0}{GW190814A}{25.8}{GW190803A}{64.5}{GW190731A}{70.1}{GW190728A}{20.6}{GW190727A}{67.1}{GW190720A}{21.5}{GW190719A}{57.8}{GW190708A}{30.9}{GW190707A}{20.1}{GW190706A}{104.1}{GW190701A}{94.3}{GW190630A}{59.1}{GW190620A}{92.1}{GW190602A}{116.3}{GW190527A}{59.1}{GW190521B}{74.7}{GW190521A}{163.9}{GW190519A}{106.6}{GW190517A}{63.5}{GW190514A}{67.2}{GW190513A}{53.9}{GW190512A}{35.9}{GW190503A}{71.7}{GW190426A}{7.2}{GW190425A}{3.4}{GW190424A}{72.6}{GW190421A}{72.9}{GW190413B}{78.8}{GW190413A}{58.6}{GW190412A}{38.4}{GW190408A}{43.0}}}
\newcommand{\totalmasssourceplus}[1]{\IfEqCase{#1}{{GW190930A}{8.9}{GW190929A}{34.9}{GW190924A}{5.1}{GW190915A}{7.5}{GW190910A}{9.3}{GW190909A}{55.9}{GW190828B}{5.4}{GW190828A}{7.7}{GW190814A}{1.0}{GW190803A}{12.6}{GW190731A}{15.8}{GW190728A}{4.5}{GW190727A}{11.7}{GW190720A}{4.3}{GW190719A}{18.3}{GW190708A}{2.5}{GW190707A}{1.9}{GW190706A}{20.2}{GW190701A}{12.1}{GW190630A}{4.6}{GW190620A}{18.5}{GW190602A}{19.0}{GW190527A}{21.3}{GW190521B}{7.0}{GW190521A}{39.2}{GW190519A}{13.5}{GW190517A}{9.6}{GW190514A}{18.7}{GW190513A}{8.6}{GW190512A}{3.8}{GW190503A}{9.4}{GW190426A}{3.5}{GW190425A}{0.3}{GW190424A}{13.3}{GW190421A}{13.4}{GW190413B}{17.4}{GW190413A}{13.3}{GW190412A}{3.8}{GW190408A}{4.2}}}
\newcommand{\chipminus}[1]{\IfEqCase{#1}{{GW190930A}{0.24}{GW190929A}{0.45}{GW190924A}{0.18}{GW190915A}{0.39}{GW190910A}{0.32}{GW190909A}{0.38}{GW190828B}{0.23}{GW190828A}{0.31}{GW190814A}{0.03}{GW190803A}{0.33}{GW190731A}{0.30}{GW190728A}{0.20}{GW190727A}{0.36}{GW190720A}{0.22}{GW190719A}{0.30}{GW190708A}{0.24}{GW190707A}{0.23}{GW190706A}{0.28}{GW190701A}{0.31}{GW190630A}{0.23}{GW190620A}{0.28}{GW190602A}{0.31}{GW190527A}{0.34}{GW190521B}{0.29}{GW190521A}{0.44}{GW190519A}{0.29}{GW190517A}{0.29}{GW190514A}{0.34}{GW190513A}{0.22}{GW190512A}{0.17}{GW190503A}{0.29}{GW190426A}{0.00}{GW190425A}{0.27}{GW190424A}{0.38}{GW190421A}{0.36}{GW190413B}{0.41}{GW190413A}{0.31}{GW190412A}{0.16}{GW190408A}{0.31}}}
\newcommand{\chipmed}[1]{\IfEqCase{#1}{{GW190930A}{0.34}{GW190929A}{0.59}{GW190924A}{0.24}{GW190915A}{0.55}{GW190910A}{0.40}{GW190909A}{0.52}{GW190828B}{0.31}{GW190828A}{0.43}{GW190814A}{0.04}{GW190803A}{0.43}{GW190731A}{0.39}{GW190728A}{0.29}{GW190727A}{0.47}{GW190720A}{0.33}{GW190719A}{0.43}{GW190708A}{0.29}{GW190707A}{0.29}{GW190706A}{0.39}{GW190701A}{0.42}{GW190630A}{0.32}{GW190620A}{0.43}{GW190602A}{0.41}{GW190527A}{0.44}{GW190521B}{0.40}{GW190521A}{0.68}{GW190519A}{0.44}{GW190517A}{0.49}{GW190514A}{0.47}{GW190513A}{0.30}{GW190512A}{0.22}{GW190503A}{0.38}{GW190426A}{0.00}{GW190425A}{0.34}{GW190424A}{0.52}{GW190421A}{0.48}{GW190413B}{0.56}{GW190413A}{0.41}{GW190412A}{0.30}{GW190408A}{0.39}}}
\newcommand{\chipplus}[1]{\IfEqCase{#1}{{GW190930A}{0.40}{GW190929A}{0.32}{GW190924A}{0.40}{GW190915A}{0.36}{GW190910A}{0.39}{GW190909A}{0.39}{GW190828B}{0.38}{GW190828A}{0.36}{GW190814A}{0.04}{GW190803A}{0.42}{GW190731A}{0.46}{GW190728A}{0.37}{GW190727A}{0.40}{GW190720A}{0.43}{GW190719A}{0.37}{GW190708A}{0.43}{GW190707A}{0.39}{GW190706A}{0.39}{GW190701A}{0.41}{GW190630A}{0.32}{GW190620A}{0.37}{GW190602A}{0.42}{GW190527A}{0.43}{GW190521B}{0.32}{GW190521A}{0.26}{GW190519A}{0.34}{GW190517A}{0.30}{GW190514A}{0.39}{GW190513A}{0.39}{GW190512A}{0.36}{GW190503A}{0.41}{GW190426A}{0.00}{GW190425A}{0.43}{GW190424A}{0.38}{GW190421A}{0.39}{GW190413B}{0.37}{GW190413A}{0.41}{GW190412A}{0.19}{GW190408A}{0.38}}}
\newcommand{\spinoneyminus}[1]{\IfEqCase{#1}{{GW190930A}{0.47}{GW190929A}{0.71}{GW190924A}{0.35}{GW190915A}{0.68}{GW190910A}{0.48}{GW190909A}{0.66}{GW190828B}{0.41}{GW190828A}{0.51}{GW190814A}{0.04}{GW190803A}{0.58}{GW190731A}{0.52}{GW190728A}{0.37}{GW190727A}{0.61}{GW190720A}{0.49}{GW190719A}{0.55}{GW190708A}{0.43}{GW190707A}{0.39}{GW190706A}{0.50}{GW190701A}{0.52}{GW190630A}{0.36}{GW190620A}{0.53}{GW190602A}{0.52}{GW190527A}{0.61}{GW190521B}{0.44}{GW190521A}{0.76}{GW190519A}{0.55}{GW190517A}{0.58}{GW190514A}{0.56}{GW190513A}{0.41}{GW190512A}{0.29}{GW190503A}{0.48}{GW190426A}{0.00}{GW190425A}{0.48}{GW190424A}{0.64}{GW190421A}{0.58}{GW190413B}{0.70}{GW190413A}{0.54}{GW190412A}{0.39}{GW190408A}{0.48}}}
\newcommand{\spinoneymed}[1]{\IfEqCase{#1}{{GW190930A}{0.002}{GW190929A}{0.005}{GW190924A}{0.0009}{GW190915A}{0.00}{GW190910A}{0.0008}{GW190909A}{-0.01}{GW190828B}{0.0010}{GW190828A}{0.00}{GW190814A}{0.0008}{GW190803A}{0.0007}{GW190731A}{0.0007}{GW190728A}{0.004}{GW190727A}{0.0002}{GW190720A}{0.0009}{GW190719A}{0.004}{GW190708A}{0.00}{GW190707A}{0.00}{GW190706A}{0.00}{GW190701A}{0.003}{GW190630A}{0.0002}{GW190620A}{-0.01}{GW190602A}{0.00}{GW190527A}{0.00}{GW190521B}{0.00}{GW190521A}{0.0003}{GW190519A}{0.00}{GW190517A}{-0.01}{GW190514A}{0.00}{GW190513A}{0.0005}{GW190512A}{0.00}{GW190503A}{0.00}{GW190426A}{0.00}{GW190425A}{0.003}{GW190424A}{0.00}{GW190421A}{0.0008}{GW190413B}{0.00}{GW190413A}{0.003}{GW190412A}{0.06}{GW190408A}{0.001}}}
\newcommand{\spinoneyplus}[1]{\IfEqCase{#1}{{GW190930A}{0.48}{GW190929A}{0.70}{GW190924A}{0.36}{GW190915A}{0.68}{GW190910A}{0.50}{GW190909A}{0.64}{GW190828B}{0.43}{GW190828A}{0.53}{GW190814A}{0.04}{GW190803A}{0.55}{GW190731A}{0.54}{GW190728A}{0.39}{GW190727A}{0.59}{GW190720A}{0.44}{GW190719A}{0.55}{GW190708A}{0.41}{GW190707A}{0.38}{GW190706A}{0.51}{GW190701A}{0.53}{GW190630A}{0.37}{GW190620A}{0.55}{GW190602A}{0.54}{GW190527A}{0.59}{GW190521B}{0.45}{GW190521A}{0.75}{GW190519A}{0.54}{GW190517A}{0.57}{GW190514A}{0.59}{GW190513A}{0.41}{GW190512A}{0.29}{GW190503A}{0.49}{GW190426A}{0.00}{GW190425A}{0.48}{GW190424A}{0.63}{GW190421A}{0.60}{GW190413B}{0.69}{GW190413A}{0.55}{GW190412A}{0.33}{GW190408A}{0.49}}}
\newcommand{\finalmassdetminus}[1]{\IfEqCase{#1}{{GW190930A}{0.9}{GW190929A}{24.5}{GW190924A}{0.8}{GW190915A}{7.4}{GW190910A}{7.1}{GW190909A}{21.3}{GW190828B}{4.2}{GW190828A}{5.2}{GW190814A}{1.0}{GW190803A}{10.9}{GW190731A}{12.8}{GW190728A}{0.7}{GW190727A}{9.8}{GW190720A}{1.2}{GW190719A}{14.1}{GW190708A}{0.7}{GW190707A}{0.5}{GW190706A}{23.7}{GW190701A}{13.4}{GW190630A}{3.3}{GW190620A}{16.2}{GW190602A}{18.3}{GW190527A}{9.5}{GW190521B}{4.8}{GW190521A}{30.4}{GW190519A}{15.4}{GW190517A}{6.4}{GW190514A}{13.9}{GW190513A}{6.7}{GW190512A}{2.8}{GW190503A}{10.8}{GW190424A}{10.0}{GW190421A}{11.3}{GW190413B}{16.7}{GW190413A}{14.0}{GW190412A}{4.7}{GW190408A}{3.4}}}
\newcommand{\finalmassdetmed}[1]{\IfEqCase{#1}{{GW190930A}{22.1}{GW190929A}{144.3}{GW190924A}{14.8}{GW190915A}{74.8}{GW190910A}{97.0}{GW190909A}{114.5}{GW190828B}{42.7}{GW190828A}{75.7}{GW190814A}{26.9}{GW190803A}{95.8}{GW190731A}{104.6}{GW190728A}{22.7}{GW190727A}{99.2}{GW190720A}{23.7}{GW190719A}{90.0}{GW190708A}{34.4}{GW190707A}{22.1}{GW190706A}{171.1}{GW190701A}{124.0}{GW190630A}{66.3}{GW190620A}{130.3}{GW190602A}{163.8}{GW190527A}{80.3}{GW190521B}{88.0}{GW190521A}{256.6}{GW190519A}{146.8}{GW190517A}{79.8}{GW190514A}{108.3}{GW190513A}{70.6}{GW190512A}{43.5}{GW190503A}{87.6}{GW190424A}{96.0}{GW190421A}{103.9}{GW190413B}{129.8}{GW190413A}{89.6}{GW190412A}{42.9}{GW190408A}{53.0}}}
\newcommand{\finalmassdetplus}[1]{\IfEqCase{#1}{{GW190930A}{10.8}{GW190929A}{36.4}{GW190924A}{5.9}{GW190915A}{7.9}{GW190910A}{9.3}{GW190909A}{92.0}{GW190828B}{6.6}{GW190828A}{6.0}{GW190814A}{1.1}{GW190803A}{13.1}{GW190731A}{12.8}{GW190728A}{5.5}{GW190727A}{10.7}{GW190720A}{5.2}{GW190719A}{22.5}{GW190708A}{2.7}{GW190707A}{1.9}{GW190706A}{20.0}{GW190701A}{15.1}{GW190630A}{4.2}{GW190620A}{17.7}{GW190602A}{20.7}{GW190527A}{51.0}{GW190521B}{4.3}{GW190521A}{36.6}{GW190519A}{14.7}{GW190517A}{8.8}{GW190514A}{16.6}{GW190513A}{11.5}{GW190512A}{4.0}{GW190503A}{10.2}{GW190424A}{13.0}{GW190421A}{14.1}{GW190413B}{16.4}{GW190413A}{16.3}{GW190412A}{4.6}{GW190408A}{3.2}}}
\newcommand{\phioneminus}[1]{\IfEqCase{#1}{{GW190930A}{2.79}{GW190929A}{2.78}{GW190924A}{2.82}{GW190915A}{2.86}{GW190910A}{2.79}{GW190909A}{2.98}{GW190828B}{2.78}{GW190828A}{2.84}{GW190814A}{2.65}{GW190803A}{2.80}{GW190731A}{2.83}{GW190728A}{2.76}{GW190727A}{2.83}{GW190720A}{2.81}{GW190719A}{2.77}{GW190708A}{2.90}{GW190707A}{2.85}{GW190706A}{2.84}{GW190701A}{2.77}{GW190630A}{2.80}{GW190620A}{2.89}{GW190602A}{2.85}{GW190527A}{2.88}{GW190521B}{2.87}{GW190521A}{2.79}{GW190519A}{2.84}{GW190517A}{2.90}{GW190514A}{2.82}{GW190513A}{2.82}{GW190512A}{2.84}{GW190503A}{2.83}{GW190426A}{0.00}{GW190425A}{2.73}{GW190424A}{2.84}{GW190421A}{2.82}{GW190413B}{2.83}{GW190413A}{2.76}{GW190412A}{2.36}{GW190408A}{2.77}}}
\newcommand{\phionemed}[1]{\IfEqCase{#1}{{GW190930A}{3.10}{GW190929A}{3.09}{GW190924A}{3.09}{GW190915A}{3.17}{GW190910A}{3.12}{GW190909A}{3.26}{GW190828B}{3.09}{GW190828A}{3.16}{GW190814A}{2.97}{GW190803A}{3.12}{GW190731A}{3.12}{GW190728A}{3.05}{GW190727A}{3.13}{GW190720A}{3.12}{GW190719A}{3.09}{GW190708A}{3.21}{GW190707A}{3.16}{GW190706A}{3.15}{GW190701A}{3.07}{GW190630A}{3.13}{GW190620A}{3.23}{GW190602A}{3.18}{GW190527A}{3.16}{GW190521B}{3.17}{GW190521A}{3.14}{GW190519A}{3.15}{GW190517A}{3.23}{GW190514A}{3.18}{GW190513A}{3.11}{GW190512A}{3.15}{GW190503A}{3.15}{GW190426A}{0.00}{GW190425A}{3.05}{GW190424A}{3.16}{GW190421A}{3.13}{GW190413B}{3.16}{GW190413A}{3.06}{GW190412A}{2.69}{GW190408A}{3.08}}}
\newcommand{\phioneplus}[1]{\IfEqCase{#1}{{GW190930A}{2.86}{GW190929A}{2.90}{GW190924A}{2.86}{GW190915A}{2.79}{GW190910A}{2.85}{GW190909A}{2.71}{GW190828B}{2.87}{GW190828A}{2.82}{GW190814A}{2.95}{GW190803A}{2.85}{GW190731A}{2.88}{GW190728A}{2.92}{GW190727A}{2.83}{GW190720A}{2.84}{GW190719A}{2.86}{GW190708A}{2.79}{GW190707A}{2.83}{GW190706A}{2.83}{GW190701A}{2.88}{GW190630A}{2.83}{GW190620A}{2.75}{GW190602A}{2.77}{GW190527A}{2.83}{GW190521B}{2.80}{GW190521A}{2.80}{GW190519A}{2.83}{GW190517A}{2.76}{GW190514A}{2.77}{GW190513A}{2.86}{GW190512A}{2.84}{GW190503A}{2.81}{GW190426A}{0.00}{GW190425A}{2.90}{GW190424A}{2.78}{GW190421A}{2.82}{GW190413B}{2.79}{GW190413A}{2.90}{GW190412A}{3.20}{GW190408A}{2.87}}}
\newcommand{\phitwominus}[1]{\IfEqCase{#1}{{GW190930A}{2.89}{GW190929A}{2.80}{GW190924A}{2.83}{GW190915A}{2.82}{GW190910A}{2.83}{GW190909A}{2.79}{GW190828B}{2.84}{GW190828A}{2.83}{GW190814A}{2.74}{GW190803A}{2.82}{GW190731A}{2.88}{GW190728A}{2.85}{GW190727A}{2.83}{GW190720A}{2.82}{GW190719A}{2.74}{GW190708A}{2.76}{GW190707A}{2.79}{GW190706A}{2.81}{GW190701A}{2.97}{GW190630A}{2.80}{GW190620A}{2.84}{GW190602A}{2.84}{GW190527A}{2.79}{GW190521B}{2.82}{GW190521A}{2.86}{GW190519A}{2.79}{GW190517A}{2.82}{GW190514A}{2.89}{GW190513A}{2.80}{GW190512A}{2.75}{GW190503A}{2.80}{GW190426A}{0.00}{GW190425A}{2.84}{GW190424A}{2.84}{GW190421A}{2.81}{GW190413B}{2.97}{GW190413A}{2.83}{GW190412A}{2.75}{GW190408A}{2.80}}}
\newcommand{\phitwomed}[1]{\IfEqCase{#1}{{GW190930A}{3.21}{GW190929A}{3.12}{GW190924A}{3.15}{GW190915A}{3.16}{GW190910A}{3.14}{GW190909A}{3.09}{GW190828B}{3.16}{GW190828A}{3.13}{GW190814A}{3.07}{GW190803A}{3.14}{GW190731A}{3.21}{GW190728A}{3.15}{GW190727A}{3.12}{GW190720A}{3.15}{GW190719A}{3.08}{GW190708A}{3.05}{GW190707A}{3.09}{GW190706A}{3.12}{GW190701A}{3.27}{GW190630A}{3.12}{GW190620A}{3.14}{GW190602A}{3.15}{GW190527A}{3.06}{GW190521B}{3.15}{GW190521A}{3.17}{GW190519A}{3.12}{GW190517A}{3.14}{GW190514A}{3.19}{GW190513A}{3.14}{GW190512A}{3.07}{GW190503A}{3.14}{GW190426A}{0.00}{GW190425A}{3.15}{GW190424A}{3.15}{GW190421A}{3.14}{GW190413B}{3.28}{GW190413A}{3.16}{GW190412A}{3.09}{GW190408A}{3.11}}}
\newcommand{\phitwoplus}[1]{\IfEqCase{#1}{{GW190930A}{2.74}{GW190929A}{2.83}{GW190924A}{2.82}{GW190915A}{2.80}{GW190910A}{2.82}{GW190909A}{2.86}{GW190828B}{2.84}{GW190828A}{2.84}{GW190814A}{2.86}{GW190803A}{2.87}{GW190731A}{2.77}{GW190728A}{2.78}{GW190727A}{2.88}{GW190720A}{2.84}{GW190719A}{2.90}{GW190708A}{2.90}{GW190707A}{2.88}{GW190706A}{2.88}{GW190701A}{2.70}{GW190630A}{2.84}{GW190620A}{2.81}{GW190602A}{2.79}{GW190527A}{2.91}{GW190521B}{2.82}{GW190521A}{2.82}{GW190519A}{2.84}{GW190517A}{2.83}{GW190514A}{2.75}{GW190513A}{2.85}{GW190512A}{2.89}{GW190503A}{2.81}{GW190426A}{0.00}{GW190425A}{2.83}{GW190424A}{2.83}{GW190421A}{2.81}{GW190413B}{2.72}{GW190413A}{2.82}{GW190412A}{2.85}{GW190408A}{2.88}}}
\newcommand{\phionetwominus}[1]{\IfEqCase{#1}{{GW190930A}{2.84}{GW190929A}{2.79}{GW190924A}{2.83}{GW190915A}{2.89}{GW190910A}{2.78}{GW190909A}{2.85}{GW190828B}{2.83}{GW190828A}{2.63}{GW190814A}{2.71}{GW190803A}{2.82}{GW190731A}{2.73}{GW190728A}{2.79}{GW190727A}{2.83}{GW190720A}{2.74}{GW190719A}{2.75}{GW190708A}{2.66}{GW190707A}{2.79}{GW190706A}{2.80}{GW190701A}{2.82}{GW190630A}{2.74}{GW190620A}{2.78}{GW190602A}{2.78}{GW190527A}{2.77}{GW190521B}{2.78}{GW190521A}{3.09}{GW190519A}{2.92}{GW190517A}{2.91}{GW190514A}{2.83}{GW190513A}{2.84}{GW190512A}{2.80}{GW190503A}{2.70}{GW190426A}{0.00}{GW190425A}{2.87}{GW190424A}{2.85}{GW190421A}{2.83}{GW190413B}{2.67}{GW190413A}{2.84}{GW190412A}{3.10}{GW190408A}{2.77}}}
\newcommand{\phionetwomed}[1]{\IfEqCase{#1}{{GW190930A}{3.17}{GW190929A}{3.11}{GW190924A}{3.17}{GW190915A}{3.18}{GW190910A}{3.15}{GW190909A}{3.17}{GW190828B}{3.16}{GW190828A}{2.94}{GW190814A}{3.01}{GW190803A}{3.12}{GW190731A}{3.03}{GW190728A}{3.13}{GW190727A}{3.13}{GW190720A}{3.09}{GW190719A}{3.13}{GW190708A}{3.08}{GW190707A}{3.13}{GW190706A}{3.11}{GW190701A}{3.15}{GW190630A}{3.06}{GW190620A}{3.16}{GW190602A}{3.09}{GW190527A}{3.08}{GW190521B}{3.16}{GW190521A}{3.35}{GW190519A}{3.23}{GW190517A}{3.22}{GW190514A}{3.15}{GW190513A}{3.15}{GW190512A}{3.14}{GW190503A}{3.02}{GW190426A}{0.00}{GW190425A}{3.18}{GW190424A}{3.15}{GW190421A}{3.17}{GW190413B}{3.00}{GW190413A}{3.15}{GW190412A}{3.44}{GW190408A}{3.11}}}
\newcommand{\phionetwoplus}[1]{\IfEqCase{#1}{{GW190930A}{2.77}{GW190929A}{2.87}{GW190924A}{2.80}{GW190915A}{2.78}{GW190910A}{2.80}{GW190909A}{2.82}{GW190828B}{2.78}{GW190828A}{3.01}{GW190814A}{2.95}{GW190803A}{2.83}{GW190731A}{2.88}{GW190728A}{2.82}{GW190727A}{2.86}{GW190720A}{2.84}{GW190719A}{2.80}{GW190708A}{2.83}{GW190707A}{2.81}{GW190706A}{2.86}{GW190701A}{2.81}{GW190630A}{2.89}{GW190620A}{2.77}{GW190602A}{2.84}{GW190527A}{2.89}{GW190521B}{2.73}{GW190521A}{2.68}{GW190519A}{2.71}{GW190517A}{2.77}{GW190514A}{2.80}{GW190513A}{2.79}{GW190512A}{2.82}{GW190503A}{2.94}{GW190426A}{0.00}{GW190425A}{2.76}{GW190424A}{2.84}{GW190421A}{2.81}{GW190413B}{2.95}{GW190413A}{2.82}{GW190412A}{2.54}{GW190408A}{2.84}}}
\newcommand{\raminus}[1]{\IfEqCase{#1}{{GW190930A}{5.23224}{GW190929A}{3.05568}{GW190924A}{0.14434}{GW190915A}{0.08865}{GW190910A}{2.79654}{GW190909A}{1.21034}{GW190828B}{0.25804}{GW190828A}{0.19728}{GW190814A}{0.02832}{GW190803A}{0.32018}{GW190731A}{2.05429}{GW190728A}{3.93966}{GW190727A}{0.28275}{GW190720A}{5.03310}{GW190719A}{1.97394}{GW190708A}{2.51716}{GW190707A}{1.47850}{GW190706A}{0.11777}{GW190701A}{0.02938}{GW190630A}{3.00747}{GW190620A}{3.79912}{GW190602A}{0.16873}{GW190527A}{4.81181}{GW190521B}{0.49181}{GW190521A}{3.83925}{GW190519A}{3.53823}{GW190517A}{0.11026}{GW190514A}{2.80859}{GW190513A}{0.16674}{GW190512A}{0.34154}{GW190503A}{0.07308}{GW190426A}{5.11703}{GW190425A}{1.14713}{GW190424A}{2.87584}{GW190421A}{1.90797}{GW190413B}{0.19894}{GW190413A}{2.38305}{GW190412A}{0.06390}{GW190408A}{0.23336}}}
\newcommand{\ramed}[1]{\IfEqCase{#1}{{GW190930A}{5.56800}{GW190929A}{4.57328}{GW190924A}{2.28446}{GW190915A}{3.41338}{GW190910A}{3.71091}{GW190909A}{1.53780}{GW190828B}{2.46668}{GW190828A}{2.55563}{GW190814A}{0.22230}{GW190803A}{1.64028}{GW190731A}{3.03870}{GW190728A}{5.47833}{GW190727A}{1.82934}{GW190720A}{5.19166}{GW190719A}{2.73410}{GW190708A}{2.97596}{GW190707A}{3.54979}{GW190706A}{2.59111}{GW190701A}{0.66145}{GW190630A}{5.86181}{GW190620A}{4.24971}{GW190602A}{1.30570}{GW190527A}{5.13405}{GW190521B}{4.87951}{GW190521A}{3.88408}{GW190519A}{3.58629}{GW190517A}{4.07092}{GW190514A}{3.58258}{GW190513A}{0.89431}{GW190512A}{4.37140}{GW190503A}{1.65900}{GW190426A}{5.27391}{GW190425A}{1.62833}{GW190424A}{3.13318}{GW190421A}{3.50423}{GW190413B}{2.70054}{GW190413A}{2.53690}{GW190412A}{3.81244}{GW190408A}{6.08868}}}
\newcommand{\raplus}[1]{\IfEqCase{#1}{{GW190930A}{0.45017}{GW190929A}{0.94968}{GW190924A}{0.18201}{GW190915A}{0.07354}{GW190910A}{1.06677}{GW190909A}{4.24495}{GW190828B}{3.42033}{GW190828A}{3.27852}{GW190814A}{0.16914}{GW190803A}{1.69538}{GW190731A}{0.37410}{GW190728A}{0.52075}{GW190727A}{4.30388}{GW190720A}{0.96175}{GW190719A}{3.20570}{GW190708A}{2.84449}{GW190707A}{2.17972}{GW190706A}{3.23517}{GW190701A}{0.02994}{GW190630A}{0.12425}{GW190620A}{0.37483}{GW190602A}{0.27462}{GW190527A}{0.80861}{GW190521B}{0.59768}{GW190521A}{2.35925}{GW190519A}{2.66341}{GW190517A}{1.75963}{GW190514A}{1.66161}{GW190513A}{4.13656}{GW190512A}{0.17389}{GW190503A}{0.07931}{GW190426A}{0.85181}{GW190425A}{3.12911}{GW190424A}{2.81897}{GW190421A}{0.15768}{GW190413B}{1.98722}{GW190413A}{0.89215}{GW190412A}{0.03095}{GW190408A}{0.08337}}}
\newcommand{\phijlminus}[1]{\IfEqCase{#1}{{GW190930A}{2.81}{GW190929A}{3.08}{GW190924A}{2.69}{GW190915A}{2.76}{GW190910A}{2.83}{GW190909A}{2.84}{GW190828B}{2.87}{GW190828A}{3.02}{GW190814A}{1.87}{GW190803A}{2.70}{GW190731A}{2.69}{GW190728A}{2.87}{GW190727A}{2.85}{GW190720A}{2.86}{GW190719A}{2.89}{GW190708A}{2.84}{GW190707A}{2.89}{GW190706A}{2.70}{GW190701A}{2.57}{GW190630A}{2.93}{GW190620A}{3.11}{GW190602A}{2.78}{GW190527A}{2.80}{GW190521B}{2.82}{GW190521A}{2.81}{GW190519A}{2.86}{GW190517A}{2.16}{GW190514A}{2.83}{GW190513A}{2.69}{GW190512A}{2.82}{GW190503A}{3.64}{GW190426A}{1.43}{GW190425A}{2.88}{GW190424A}{2.83}{GW190421A}{2.83}{GW190413B}{3.08}{GW190413A}{2.65}{GW190412A}{3.64}{GW190408A}{2.54}}}
\newcommand{\phijlmed}[1]{\IfEqCase{#1}{{GW190930A}{3.14}{GW190929A}{3.37}{GW190924A}{3.02}{GW190915A}{3.39}{GW190910A}{3.15}{GW190909A}{3.16}{GW190828B}{3.19}{GW190828A}{3.35}{GW190814A}{2.28}{GW190803A}{3.05}{GW190731A}{3.03}{GW190728A}{3.19}{GW190727A}{3.21}{GW190720A}{3.15}{GW190719A}{3.21}{GW190708A}{3.17}{GW190707A}{3.21}{GW190706A}{3.03}{GW190701A}{3.05}{GW190630A}{3.23}{GW190620A}{3.51}{GW190602A}{3.12}{GW190527A}{3.13}{GW190521B}{3.17}{GW190521A}{3.14}{GW190519A}{3.17}{GW190517A}{2.38}{GW190514A}{3.15}{GW190513A}{3.00}{GW190512A}{3.16}{GW190503A}{3.90}{GW190426A}{1.70}{GW190425A}{3.23}{GW190424A}{3.16}{GW190421A}{3.14}{GW190413B}{3.36}{GW190413A}{2.98}{GW190412A}{3.82}{GW190408A}{2.90}}}
\newcommand{\phijlplus}[1]{\IfEqCase{#1}{{GW190930A}{2.86}{GW190929A}{2.65}{GW190924A}{2.94}{GW190915A}{2.29}{GW190910A}{2.84}{GW190909A}{2.81}{GW190828B}{2.77}{GW190828A}{2.59}{GW190814A}{3.50}{GW190803A}{2.90}{GW190731A}{2.89}{GW190728A}{2.76}{GW190727A}{2.79}{GW190720A}{2.84}{GW190719A}{2.78}{GW190708A}{2.80}{GW190707A}{2.77}{GW190706A}{2.92}{GW190701A}{2.72}{GW190630A}{2.76}{GW190620A}{2.40}{GW190602A}{2.83}{GW190527A}{2.85}{GW190521B}{2.86}{GW190521A}{2.80}{GW190519A}{2.79}{GW190517A}{3.65}{GW190514A}{2.83}{GW190513A}{2.95}{GW190512A}{2.80}{GW190503A}{2.15}{GW190426A}{1.19}{GW190425A}{2.76}{GW190424A}{2.82}{GW190421A}{2.84}{GW190413B}{2.68}{GW190413A}{2.97}{GW190412A}{2.23}{GW190408A}{3.00}}}
\newcommand{\tilttwominus}[1]{\IfEqCase{#1}{{GW190930A}{0.90}{GW190929A}{1.09}{GW190924A}{1.01}{GW190915A}{1.03}{GW190910A}{1.01}{GW190909A}{1.23}{GW190828B}{0.96}{GW190828A}{0.84}{GW190814A}{1.02}{GW190803A}{1.12}{GW190731A}{1.02}{GW190728A}{0.85}{GW190727A}{0.97}{GW190720A}{0.90}{GW190719A}{0.81}{GW190708A}{1.00}{GW190707A}{1.18}{GW190706A}{0.88}{GW190701A}{1.14}{GW190630A}{0.87}{GW190620A}{0.78}{GW190602A}{0.97}{GW190527A}{1.01}{GW190521B}{0.85}{GW190521A}{1.02}{GW190519A}{0.77}{GW190517A}{0.67}{GW190514A}{1.21}{GW190513A}{0.94}{GW190512A}{0.99}{GW190503A}{1.06}{GW190426A}{0.00}{GW190425A}{0.87}{GW190424A}{0.96}{GW190421A}{1.10}{GW190413B}{1.18}{GW190413A}{1.13}{GW190412A}{0.90}{GW190408A}{1.06}}}
\newcommand{\tilttwomed}[1]{\IfEqCase{#1}{{GW190930A}{1.26}{GW190929A}{1.53}{GW190924A}{1.42}{GW190915A}{1.56}{GW190910A}{1.53}{GW190909A}{1.74}{GW190828B}{1.32}{GW190828A}{1.19}{GW190814A}{1.60}{GW190803A}{1.63}{GW190731A}{1.42}{GW190728A}{1.17}{GW190727A}{1.37}{GW190720A}{1.25}{GW190719A}{1.11}{GW190708A}{1.43}{GW190707A}{1.76}{GW190706A}{1.19}{GW190701A}{1.74}{GW190630A}{1.23}{GW190620A}{1.05}{GW190602A}{1.38}{GW190527A}{1.41}{GW190521B}{1.27}{GW190521A}{1.59}{GW190519A}{1.05}{GW190517A}{0.91}{GW190514A}{1.94}{GW190513A}{1.32}{GW190512A}{1.41}{GW190503A}{1.59}{GW190426A}{0.00}{GW190425A}{1.41}{GW190424A}{1.37}{GW190421A}{1.73}{GW190413B}{1.70}{GW190413A}{1.65}{GW190412A}{1.32}{GW190408A}{1.59}}}
\newcommand{\tilttwoplus}[1]{\IfEqCase{#1}{{GW190930A}{1.21}{GW190929A}{1.13}{GW190924A}{1.16}{GW190915A}{1.09}{GW190910A}{1.04}{GW190909A}{1.01}{GW190828B}{1.20}{GW190828A}{1.25}{GW190814A}{1.00}{GW190803A}{1.05}{GW190731A}{1.16}{GW190728A}{1.29}{GW190727A}{1.20}{GW190720A}{1.25}{GW190719A}{1.31}{GW190708A}{1.13}{GW190707A}{0.92}{GW190706A}{1.29}{GW190701A}{0.97}{GW190630A}{1.18}{GW190620A}{1.32}{GW190602A}{1.15}{GW190527A}{1.20}{GW190521B}{1.13}{GW190521A}{1.05}{GW190519A}{1.26}{GW190517A}{1.30}{GW190514A}{0.88}{GW190513A}{1.19}{GW190512A}{1.11}{GW190503A}{1.07}{GW190426A}{3.14}{GW190425A}{0.94}{GW190424A}{1.16}{GW190421A}{0.99}{GW190413B}{1.00}{GW190413A}{1.04}{GW190412A}{1.12}{GW190408A}{1.03}}}
\newcommand{\costhetajnminus}[1]{\IfEqCase{#1}{{GW190930A}{1.54}{GW190929A}{0.76}{GW190924A}{1.67}{GW190915A}{0.50}{GW190910A}{0.88}{GW190909A}{1.06}{GW190828B}{0.71}{GW190828A}{0.36}{GW190814A}{1.35}{GW190803A}{1.39}{GW190731A}{1.33}{GW190728A}{1.30}{GW190727A}{0.98}{GW190720A}{0.20}{GW190719A}{0.92}{GW190708A}{1.17}{GW190707A}{0.43}{GW190706A}{1.14}{GW190701A}{0.41}{GW190630A}{1.29}{GW190620A}{0.60}{GW190602A}{0.82}{GW190527A}{1.25}{GW190521B}{1.05}{GW190521A}{1.35}{GW190519A}{0.81}{GW190517A}{0.34}{GW190514A}{1.02}{GW190513A}{1.65}{GW190512A}{0.91}{GW190503A}{0.23}{GW190426A}{0.84}{GW190425A}{1.43}{GW190424A}{1.02}{GW190421A}{0.80}{GW190413B}{0.63}{GW190413A}{1.35}{GW190412A}{0.35}{GW190408A}{1.65}}}
\newcommand{\costhetajnmed}[1]{\IfEqCase{#1}{{GW190930A}{0.59}{GW190929A}{-0.13}{GW190924A}{0.74}{GW190915A}{-0.45}{GW190910A}{-0.05}{GW190909A}{0.12}{GW190828B}{-0.25}{GW190828A}{-0.62}{GW190814A}{0.65}{GW190803A}{0.44}{GW190731A}{0.37}{GW190728A}{0.33}{GW190727A}{0.02}{GW190720A}{-0.79}{GW190719A}{-0.04}{GW190708A}{0.20}{GW190707A}{-0.55}{GW190706A}{0.20}{GW190701A}{0.84}{GW190630A}{0.34}{GW190620A}{-0.36}{GW190602A}{-0.14}{GW190527A}{0.30}{GW190521B}{0.09}{GW190521A}{0.41}{GW190519A}{-0.01}{GW190517A}{-0.64}{GW190514A}{0.07}{GW190513A}{0.70}{GW190512A}{-0.04}{GW190503A}{-0.75}{GW190426A}{-0.13}{GW190425A}{0.47}{GW190424A}{0.05}{GW190421A}{-0.17}{GW190413B}{-0.33}{GW190413A}{0.41}{GW190412A}{0.75}{GW190408A}{0.70}}}
\newcommand{\costhetajnplus}[1]{\IfEqCase{#1}{{GW190930A}{0.39}{GW190929A}{1.01}{GW190924A}{0.25}{GW190915A}{1.32}{GW190910A}{0.96}{GW190909A}{0.82}{GW190828B}{1.20}{GW190828A}{1.57}{GW190814A}{0.16}{GW190803A}{0.53}{GW190731A}{0.59}{GW190728A}{0.65}{GW190727A}{0.94}{GW190720A}{1.68}{GW190719A}{1.01}{GW190708A}{0.78}{GW190707A}{1.51}{GW190706A}{0.75}{GW190701A}{0.15}{GW190630A}{0.62}{GW190620A}{1.27}{GW190602A}{1.10}{GW190527A}{0.66}{GW190521B}{0.87}{GW190521A}{0.55}{GW190519A}{0.81}{GW190517A}{1.38}{GW190514A}{0.89}{GW190513A}{0.27}{GW190512A}{0.99}{GW190503A}{0.48}{GW190426A}{1.09}{GW190425A}{0.50}{GW190424A}{0.91}{GW190421A}{1.12}{GW190413B}{1.27}{GW190413A}{0.55}{GW190412A}{0.14}{GW190408A}{0.28}}}
\newcommand{\spintwominus}[1]{\IfEqCase{#1}{{GW190930A}{0.37}{GW190929A}{0.44}{GW190924A}{0.32}{GW190915A}{0.43}{GW190910A}{0.33}{GW190909A}{0.43}{GW190828B}{0.38}{GW190828A}{0.37}{GW190814A}{0.46}{GW190803A}{0.40}{GW190731A}{0.40}{GW190728A}{0.35}{GW190727A}{0.41}{GW190720A}{0.45}{GW190719A}{0.49}{GW190708A}{0.28}{GW190707A}{0.28}{GW190706A}{0.44}{GW190701A}{0.40}{GW190630A}{0.34}{GW190620A}{0.50}{GW190602A}{0.45}{GW190527A}{0.45}{GW190521B}{0.37}{GW190521A}{0.52}{GW190519A}{0.48}{GW190517A}{0.52}{GW190514A}{0.48}{GW190513A}{0.39}{GW190512A}{0.32}{GW190503A}{0.40}{GW190426A}{0.009}{GW190425A}{0.25}{GW190424A}{0.42}{GW190421A}{0.42}{GW190413B}{0.45}{GW190413A}{0.41}{GW190412A}{0.43}{GW190408A}{0.32}}}
\newcommand{\spintwomed}[1]{\IfEqCase{#1}{{GW190930A}{0.42}{GW190929A}{0.49}{GW190924A}{0.35}{GW190915A}{0.48}{GW190910A}{0.37}{GW190909A}{0.49}{GW190828B}{0.42}{GW190828A}{0.41}{GW190814A}{0.52}{GW190803A}{0.45}{GW190731A}{0.45}{GW190728A}{0.39}{GW190727A}{0.45}{GW190720A}{0.51}{GW190719A}{0.55}{GW190708A}{0.30}{GW190707A}{0.31}{GW190706A}{0.49}{GW190701A}{0.44}{GW190630A}{0.38}{GW190620A}{0.56}{GW190602A}{0.50}{GW190527A}{0.50}{GW190521B}{0.42}{GW190521A}{0.58}{GW190519A}{0.54}{GW190517A}{0.58}{GW190514A}{0.54}{GW190513A}{0.43}{GW190512A}{0.36}{GW190503A}{0.44}{GW190426A}{0.009}{GW190425A}{0.28}{GW190424A}{0.47}{GW190421A}{0.46}{GW190413B}{0.50}{GW190413A}{0.45}{GW190412A}{0.49}{GW190408A}{0.36}}}
\newcommand{\spintwoplus}[1]{\IfEqCase{#1}{{GW190930A}{0.49}{GW190929A}{0.45}{GW190924A}{0.51}{GW190915A}{0.46}{GW190910A}{0.51}{GW190909A}{0.45}{GW190828B}{0.49}{GW190828A}{0.46}{GW190814A}{0.41}{GW190803A}{0.49}{GW190731A}{0.48}{GW190728A}{0.50}{GW190727A}{0.46}{GW190720A}{0.43}{GW190719A}{0.40}{GW190708A}{0.50}{GW190707A}{0.52}{GW190706A}{0.45}{GW190701A}{0.48}{GW190630A}{0.46}{GW190620A}{0.40}{GW190602A}{0.44}{GW190527A}{0.45}{GW190521B}{0.39}{GW190521A}{0.38}{GW190519A}{0.41}{GW190517A}{0.38}{GW190514A}{0.42}{GW190513A}{0.48}{GW190512A}{0.51}{GW190503A}{0.48}{GW190426A}{0.03}{GW190425A}{0.51}{GW190424A}{0.47}{GW190421A}{0.47}{GW190413B}{0.44}{GW190413A}{0.48}{GW190412A}{0.44}{GW190408A}{0.53}}}
\newcommand{\massonedetminus}[1]{\IfEqCase{#1}{{GW190930A}{2.6}{GW190929A}{28.9}{GW190924A}{2.2}{GW190915A}{7.7}{GW190910A}{6.2}{GW190909A}{17.5}{GW190828B}{8.7}{GW190828A}{4.7}{GW190814A}{1.0}{GW190803A}{9.0}{GW190731A}{10.5}{GW190728A}{2.5}{GW190727A}{8.2}{GW190720A}{3.5}{GW190719A}{16.4}{GW190708A}{2.4}{GW190707A}{1.8}{GW190706A}{17.6}{GW190701A}{10.9}{GW190630A}{6.4}{GW190620A}{15.4}{GW190602A}{15.7}{GW190527A}{11.0}{GW190521B}{5.4}{GW190521A}{20.8}{GW190519A}{12.9}{GW190517A}{8.4}{GW190514A}{10.8}{GW190513A}{12.3}{GW190512A}{6.8}{GW190503A}{9.9}{GW190426A}{2.5}{GW190425A}{0.4}{GW190424A}{8.0}{GW190421A}{8.8}{GW190413B}{14.6}{GW190413A}{12.0}{GW190412A}{6.2}{GW190408A}{4.0}}}
\newcommand{\massonedetmed}[1]{\IfEqCase{#1}{{GW190930A}{14.2}{GW190929A}{111.3}{GW190924A}{9.9}{GW190915A}{46.0}{GW190910A}{56.3}{GW190909A}{73.0}{GW190828B}{31.1}{GW190828A}{43.9}{GW190814A}{24.4}{GW190803A}{57.6}{GW190731A}{64.4}{GW190728A}{14.4}{GW190727A}{58.8}{GW190720A}{15.7}{GW190719A}{60.3}{GW190708A}{20.6}{GW190707A}{13.4}{GW190706A}{112.9}{GW190701A}{74.1}{GW190630A}{41.4}{GW190620A}{84.6}{GW190602A}{101.7}{GW190527A}{52.0}{GW190521B}{51.9}{GW190521A}{153.4}{GW190519A}{95.4}{GW190517A}{50.5}{GW190514A}{64.9}{GW190513A}{49.0}{GW190512A}{29.4}{GW190503A}{55.2}{GW190426A}{6.2}{GW190425A}{2.1}{GW190424A}{56.1}{GW190421A}{61.1}{GW190413B}{80.7}{GW190413A}{55.3}{GW190412A}{34.6}{GW190408A}{31.5}}}
\newcommand{\massonedetplus}[1]{\IfEqCase{#1}{{GW190930A}{14.3}{GW190929A}{39.3}{GW190924A}{7.8}{GW190915A}{11.6}{GW190910A}{8.8}{GW190909A}{84.2}{GW190828B}{8.8}{GW190828A}{7.5}{GW190814A}{1.2}{GW190803A}{14.2}{GW190731A}{13.4}{GW190728A}{8.4}{GW190727A}{12.8}{GW190720A}{7.7}{GW190719A}{26.5}{GW190708A}{5.7}{GW190707A}{3.7}{GW190706A}{20.5}{GW190701A}{15.1}{GW190630A}{8.3}{GW190620A}{20.0}{GW190602A}{19.9}{GW190527A}{32.9}{GW190521B}{7.1}{GW190521A}{45.9}{GW190519A}{14.9}{GW190517A}{14.7}{GW190514A}{17.6}{GW190513A}{12.5}{GW190512A}{6.5}{GW190503A}{10.6}{GW190426A}{4.2}{GW190425A}{0.6}{GW190424A}{14.9}{GW190421A}{14.7}{GW190413B}{19.0}{GW190413A}{17.1}{GW190412A}{5.5}{GW190408A}{6.3}}}
\newcommand{\massratiominus}[1]{\IfEqCase{#1}{{GW190930A}{0.46}{GW190929A}{0.16}{GW190924A}{0.37}{GW190915A}{0.26}{GW190910A}{0.23}{GW190909A}{0.39}{GW190828B}{0.16}{GW190828A}{0.23}{GW190814A}{0.009}{GW190803A}{0.31}{GW190731A}{0.31}{GW190728A}{0.38}{GW190727A}{0.32}{GW190720A}{0.30}{GW190719A}{0.29}{GW190708A}{0.28}{GW190707A}{0.27}{GW190706A}{0.25}{GW190701A}{0.30}{GW190630A}{0.22}{GW190620A}{0.27}{GW190602A}{0.33}{GW190527A}{0.32}{GW190521B}{0.21}{GW190521A}{0.34}{GW190519A}{0.19}{GW190517A}{0.29}{GW190514A}{0.33}{GW190513A}{0.18}{GW190512A}{0.18}{GW190503A}{0.23}{GW190426A}{0.15}{GW190425A}{0.25}{GW190424A}{0.29}{GW190421A}{0.30}{GW190413B}{0.31}{GW190413A}{0.28}{GW190412A}{0.06}{GW190408A}{0.25}}}
\newcommand{\massratiomed}[1]{\IfEqCase{#1}{{GW190930A}{0.64}{GW190929A}{0.30}{GW190924A}{0.57}{GW190915A}{0.69}{GW190910A}{0.82}{GW190909A}{0.62}{GW190828B}{0.42}{GW190828A}{0.82}{GW190814A}{0.112}{GW190803A}{0.75}{GW190731A}{0.72}{GW190728A}{0.66}{GW190727A}{0.79}{GW190720A}{0.59}{GW190719A}{0.58}{GW190708A}{0.75}{GW190707A}{0.72}{GW190706A}{0.58}{GW190701A}{0.76}{GW190630A}{0.68}{GW190620A}{0.62}{GW190602A}{0.71}{GW190527A}{0.64}{GW190521B}{0.78}{GW190521A}{0.75}{GW190519A}{0.61}{GW190517A}{0.68}{GW190514A}{0.75}{GW190513A}{0.50}{GW190512A}{0.54}{GW190503A}{0.65}{GW190426A}{0.25}{GW190425A}{0.67}{GW190424A}{0.81}{GW190421A}{0.79}{GW190413B}{0.69}{GW190413A}{0.69}{GW190412A}{0.28}{GW190408A}{0.76}}}
\newcommand{\massratioplus}[1]{\IfEqCase{#1}{{GW190930A}{0.30}{GW190929A}{0.52}{GW190924A}{0.36}{GW190915A}{0.27}{GW190910A}{0.15}{GW190909A}{0.33}{GW190828B}{0.38}{GW190828A}{0.15}{GW190814A}{0.008}{GW190803A}{0.22}{GW190731A}{0.25}{GW190728A}{0.29}{GW190727A}{0.18}{GW190720A}{0.36}{GW190719A}{0.37}{GW190708A}{0.21}{GW190707A}{0.24}{GW190706A}{0.34}{GW190701A}{0.21}{GW190630A}{0.27}{GW190620A}{0.33}{GW190602A}{0.25}{GW190527A}{0.32}{GW190521B}{0.19}{GW190521A}{0.23}{GW190519A}{0.28}{GW190517A}{0.27}{GW190514A}{0.21}{GW190513A}{0.42}{GW190512A}{0.37}{GW190503A}{0.29}{GW190426A}{0.41}{GW190425A}{0.29}{GW190424A}{0.17}{GW190421A}{0.18}{GW190413B}{0.28}{GW190413A}{0.28}{GW190412A}{0.12}{GW190408A}{0.21}}}
\newcommand{\spinoneminus}[1]{\IfEqCase{#1}{{GW190930A}{0.35}{GW190929A}{0.54}{GW190924A}{0.21}{GW190915A}{0.49}{GW190910A}{0.30}{GW190909A}{0.52}{GW190828B}{0.26}{GW190828A}{0.40}{GW190814A}{0.03}{GW190803A}{0.37}{GW190731A}{0.34}{GW190728A}{0.28}{GW190727A}{0.42}{GW190720A}{0.35}{GW190719A}{0.53}{GW190708A}{0.20}{GW190707A}{0.21}{GW190706A}{0.48}{GW190701A}{0.36}{GW190630A}{0.23}{GW190620A}{0.50}{GW190602A}{0.34}{GW190527A}{0.43}{GW190521B}{0.28}{GW190521A}{0.63}{GW190519A}{0.50}{GW190517A}{0.35}{GW190514A}{0.46}{GW190513A}{0.28}{GW190512A}{0.16}{GW190503A}{0.31}{GW190426A}{0.14}{GW190425A}{0.25}{GW190424A}{0.47}{GW190421A}{0.41}{GW190413B}{0.52}{GW190413A}{0.36}{GW190412A}{0.22}{GW190408A}{0.31}}}
\newcommand{\spinonemed}[1]{\IfEqCase{#1}{{GW190930A}{0.39}{GW190929A}{0.64}{GW190924A}{0.24}{GW190915A}{0.55}{GW190910A}{0.34}{GW190909A}{0.58}{GW190828B}{0.28}{GW190828A}{0.44}{GW190814A}{0.03}{GW190803A}{0.41}{GW190731A}{0.37}{GW190728A}{0.32}{GW190727A}{0.46}{GW190720A}{0.40}{GW190719A}{0.62}{GW190708A}{0.22}{GW190707A}{0.24}{GW190706A}{0.55}{GW190701A}{0.40}{GW190630A}{0.26}{GW190620A}{0.61}{GW190602A}{0.38}{GW190527A}{0.47}{GW190521B}{0.31}{GW190521A}{0.73}{GW190519A}{0.60}{GW190517A}{0.86}{GW190514A}{0.52}{GW190513A}{0.30}{GW190512A}{0.17}{GW190503A}{0.34}{GW190426A}{0.14}{GW190425A}{0.27}{GW190424A}{0.53}{GW190421A}{0.46}{GW190413B}{0.58}{GW190413A}{0.40}{GW190412A}{0.44}{GW190408A}{0.34}}}
\newcommand{\spinoneplus}[1]{\IfEqCase{#1}{{GW190930A}{0.40}{GW190929A}{0.32}{GW190924A}{0.43}{GW190915A}{0.39}{GW190910A}{0.50}{GW190909A}{0.37}{GW190828B}{0.43}{GW190828A}{0.45}{GW190814A}{0.05}{GW190803A}{0.51}{GW190731A}{0.54}{GW190728A}{0.37}{GW190727A}{0.47}{GW190720A}{0.40}{GW190719A}{0.34}{GW190708A}{0.52}{GW190707A}{0.47}{GW190706A}{0.39}{GW190701A}{0.50}{GW190630A}{0.37}{GW190620A}{0.34}{GW190602A}{0.51}{GW190527A}{0.47}{GW190521B}{0.42}{GW190521A}{0.25}{GW190519A}{0.33}{GW190517A}{0.13}{GW190514A}{0.43}{GW190513A}{0.51}{GW190512A}{0.44}{GW190503A}{0.51}{GW190426A}{0.40}{GW190425A}{0.51}{GW190424A}{0.42}{GW190421A}{0.47}{GW190413B}{0.38}{GW190413A}{0.51}{GW190412A}{0.16}{GW190408A}{0.47}}}
\newcommand{\costiltoneminus}[1]{\IfEqCase{#1}{{GW190930A}{1.08}{GW190929A}{0.83}{GW190924A}{0.97}{GW190915A}{0.82}{GW190910A}{0.93}{GW190909A}{0.78}{GW190828B}{0.99}{GW190828A}{1.09}{GW190814A}{0.90}{GW190803A}{0.80}{GW190731A}{0.99}{GW190728A}{1.13}{GW190727A}{1.02}{GW190720A}{0.97}{GW190719A}{1.02}{GW190708A}{0.88}{GW190707A}{0.68}{GW190706A}{0.99}{GW190701A}{0.70}{GW190630A}{1.02}{GW190620A}{0.84}{GW190602A}{0.99}{GW190527A}{1.02}{GW190521B}{0.97}{GW190521A}{0.89}{GW190519A}{0.74}{GW190517A}{0.37}{GW190514A}{0.50}{GW190513A}{1.10}{GW190512A}{0.93}{GW190503A}{0.75}{GW190426A}{0.00}{GW190425A}{0.65}{GW190424A}{1.00}{GW190421A}{0.75}{GW190413B}{0.76}{GW190413A}{0.90}{GW190412A}{0.47}{GW190408A}{0.73}}}
\newcommand{\costiltonemed}[1]{\IfEqCase{#1}{{GW190930A}{0.47}{GW190929A}{0.02}{GW190924A}{0.19}{GW190915A}{0.06}{GW190910A}{0.08}{GW190909A}{-0.10}{GW190828B}{0.26}{GW190828A}{0.51}{GW190814A}{0.01}{GW190803A}{-0.07}{GW190731A}{0.16}{GW190728A}{0.49}{GW190727A}{0.30}{GW190720A}{0.54}{GW190719A}{0.66}{GW190708A}{0.07}{GW190707A}{-0.19}{GW190706A}{0.66}{GW190701A}{-0.22}{GW190630A}{0.29}{GW190620A}{0.68}{GW190602A}{0.18}{GW190527A}{0.28}{GW190521B}{0.17}{GW190521A}{0.05}{GW190519A}{0.65}{GW190517A}{0.83}{GW190514A}{-0.45}{GW190513A}{0.41}{GW190512A}{0.07}{GW190503A}{-0.16}{GW190426A}{-1.00}{GW190425A}{0.26}{GW190424A}{0.36}{GW190421A}{-0.15}{GW190413B}{-0.06}{GW190413A}{0.01}{GW190412A}{0.70}{GW190408A}{-0.17}}}
\newcommand{\costiltoneplus}[1]{\IfEqCase{#1}{{GW190930A}{0.49}{GW190929A}{0.65}{GW190924A}{0.76}{GW190915A}{0.73}{GW190910A}{0.79}{GW190909A}{0.96}{GW190828B}{0.63}{GW190828A}{0.44}{GW190814A}{0.87}{GW190803A}{0.91}{GW190731A}{0.74}{GW190728A}{0.47}{GW190727A}{0.61}{GW190720A}{0.42}{GW190719A}{0.31}{GW190708A}{0.80}{GW190707A}{0.97}{GW190706A}{0.31}{GW190701A}{1.01}{GW190630A}{0.63}{GW190620A}{0.29}{GW190602A}{0.72}{GW190527A}{0.65}{GW190521B}{0.71}{GW190521A}{0.78}{GW190519A}{0.32}{GW190517A}{0.16}{GW190514A}{1.08}{GW190513A}{0.53}{GW190512A}{0.82}{GW190503A}{0.96}{GW190426A}{2.00}{GW190425A}{0.61}{GW190424A}{0.56}{GW190421A}{0.94}{GW190413B}{0.82}{GW190413A}{0.87}{GW190412A}{0.20}{GW190408A}{0.94}}}
\newcommand{\finalmasssourceminus}[1]{\IfEqCase{#1}{{GW190930A}{1.5}{GW190929A}{25.3}{GW190924A}{1.0}{GW190915A}{6.0}{GW190910A}{8.6}{GW190909A}{16.8}{GW190828B}{4.5}{GW190828A}{4.3}{GW190814A}{0.9}{GW190803A}{8.5}{GW190731A}{10.8}{GW190728A}{1.3}{GW190727A}{7.5}{GW190720A}{2.2}{GW190719A}{10.2}{GW190708A}{1.8}{GW190707A}{1.3}{GW190706A}{13.5}{GW190701A}{8.9}{GW190630A}{4.6}{GW190620A}{12.1}{GW190602A}{14.9}{GW190527A}{9.3}{GW190521B}{4.4}{GW190521A}{22.4}{GW190519A}{13.8}{GW190517A}{8.9}{GW190514A}{10.4}{GW190513A}{5.8}{GW190512A}{3.5}{GW190503A}{7.7}{GW190424A}{10.1}{GW190421A}{8.7}{GW190413B}{11.4}{GW190413A}{9.2}{GW190412A}{3.8}{GW190408A}{2.8}}}
\newcommand{\finalmasssourcemed}[1]{\IfEqCase{#1}{{GW190930A}{19.4}{GW190929A}{101.5}{GW190924A}{13.3}{GW190915A}{57.2}{GW190910A}{75.8}{GW190909A}{72.0}{GW190828B}{33.1}{GW190828A}{54.9}{GW190814A}{25.6}{GW190803A}{61.7}{GW190731A}{67.0}{GW190728A}{19.6}{GW190727A}{63.8}{GW190720A}{20.4}{GW190719A}{54.9}{GW190708A}{29.5}{GW190707A}{19.2}{GW190706A}{99.0}{GW190701A}{90.2}{GW190630A}{56.4}{GW190620A}{87.2}{GW190602A}{110.9}{GW190527A}{56.4}{GW190521B}{71.0}{GW190521A}{156.3}{GW190519A}{101.0}{GW190517A}{59.3}{GW190514A}{64.5}{GW190513A}{51.6}{GW190512A}{34.5}{GW190503A}{68.6}{GW190424A}{68.9}{GW190421A}{69.7}{GW190413B}{75.5}{GW190413A}{56.0}{GW190412A}{37.3}{GW190408A}{41.1}}}
\newcommand{\finalmasssourceplus}[1]{\IfEqCase{#1}{{GW190930A}{9.2}{GW190929A}{33.6}{GW190924A}{5.2}{GW190915A}{7.1}{GW190910A}{8.5}{GW190909A}{54.9}{GW190828B}{5.5}{GW190828A}{7.2}{GW190814A}{1.1}{GW190803A}{11.8}{GW190731A}{14.6}{GW190728A}{4.7}{GW190727A}{10.9}{GW190720A}{4.5}{GW190719A}{17.3}{GW190708A}{2.5}{GW190707A}{1.9}{GW190706A}{18.3}{GW190701A}{11.3}{GW190630A}{4.4}{GW190620A}{16.8}{GW190602A}{17.7}{GW190527A}{20.2}{GW190521B}{6.5}{GW190521A}{36.8}{GW190519A}{12.4}{GW190517A}{9.1}{GW190514A}{17.9}{GW190513A}{8.2}{GW190512A}{3.8}{GW190503A}{8.8}{GW190424A}{12.4}{GW190421A}{12.5}{GW190413B}{16.4}{GW190413A}{12.5}{GW190412A}{3.9}{GW190408A}{3.9}}}
\newcommand{\phaseminus}[1]{\IfEqCase{#1}{{GW190930A}{2.91}{GW190929A}{2.79}{GW190924A}{2.77}{GW190915A}{2.96}{GW190910A}{2.81}{GW190909A}{2.80}{GW190828B}{2.90}{GW190828A}{2.89}{GW190814A}{2.76}{GW190803A}{2.85}{GW190731A}{2.83}{GW190728A}{2.80}{GW190727A}{2.88}{GW190720A}{2.81}{GW190719A}{2.84}{GW190708A}{2.83}{GW190707A}{2.91}{GW190706A}{2.45}{GW190701A}{2.60}{GW190630A}{3.46}{GW190620A}{2.68}{GW190602A}{2.80}{GW190527A}{2.85}{GW190521B}{1.76}{GW190521A}{2.87}{GW190519A}{2.80}{GW190517A}{2.79}{GW190514A}{2.81}{GW190513A}{2.78}{GW190512A}{2.87}{GW190503A}{2.88}{GW190426A}{2.79}{GW190425A}{2.82}{GW190424A}{2.86}{GW190421A}{2.89}{GW190413B}{2.82}{GW190413A}{2.81}{GW190412A}{1.89}{GW190408A}{2.79}}}
\newcommand{\phasemed}[1]{\IfEqCase{#1}{{GW190930A}{3.22}{GW190929A}{3.08}{GW190924A}{3.08}{GW190915A}{3.30}{GW190910A}{3.15}{GW190909A}{3.15}{GW190828B}{3.23}{GW190828A}{3.17}{GW190814A}{3.13}{GW190803A}{3.14}{GW190731A}{3.15}{GW190728A}{3.11}{GW190727A}{3.21}{GW190720A}{3.12}{GW190719A}{3.15}{GW190708A}{3.14}{GW190707A}{3.24}{GW190706A}{3.01}{GW190701A}{2.89}{GW190630A}{3.82}{GW190620A}{2.97}{GW190602A}{3.13}{GW190527A}{3.13}{GW190521B}{2.03}{GW190521A}{3.12}{GW190519A}{3.13}{GW190517A}{3.10}{GW190514A}{3.13}{GW190513A}{3.08}{GW190512A}{3.18}{GW190503A}{3.16}{GW190426A}{3.09}{GW190425A}{3.12}{GW190424A}{3.16}{GW190421A}{3.20}{GW190413B}{3.11}{GW190413A}{3.12}{GW190412A}{2.15}{GW190408A}{3.11}}}
\newcommand{\phaseplus}[1]{\IfEqCase{#1}{{GW190930A}{2.77}{GW190929A}{2.89}{GW190924A}{2.88}{GW190915A}{2.70}{GW190910A}{2.84}{GW190909A}{2.82}{GW190828B}{2.73}{GW190828A}{2.85}{GW190814A}{2.81}{GW190803A}{2.81}{GW190731A}{2.80}{GW190728A}{2.86}{GW190727A}{2.76}{GW190720A}{2.85}{GW190719A}{2.83}{GW190708A}{2.83}{GW190707A}{2.76}{GW190706A}{2.65}{GW190701A}{3.10}{GW190630A}{2.15}{GW190620A}{2.97}{GW190602A}{2.84}{GW190527A}{2.81}{GW190521B}{3.94}{GW190521A}{2.87}{GW190519A}{2.83}{GW190517A}{2.88}{GW190514A}{2.82}{GW190513A}{2.90}{GW190512A}{2.82}{GW190503A}{2.84}{GW190426A}{2.85}{GW190425A}{2.87}{GW190424A}{2.81}{GW190421A}{2.77}{GW190413B}{2.84}{GW190413A}{2.84}{GW190412A}{3.84}{GW190408A}{2.86}}}
\newcommand{\radiatedenergyminus}[1]{\IfEqCase{#1}{{GW190930A}{0.2}{GW190929A}{1.4}{GW190924A}{0.1}{GW190915A}{0.7}{GW190910A}{0.7}{GW190909A}{1.4}{GW190828B}{0.2}{GW190828A}{0.5}{GW190814A}{0.007}{GW190803A}{0.8}{GW190731A}{1.1}{GW190728A}{0.2}{GW190727A}{0.9}{GW190720A}{0.2}{GW190719A}{1.1}{GW190708A}{0.2}{GW190707A}{0.09}{GW190706A}{2.1}{GW190701A}{1.1}{GW190630A}{0.5}{GW190620A}{1.9}{GW190602A}{1.9}{GW190527A}{1.0}{GW190521B}{0.7}{GW190521A}{2.4}{GW190519A}{1.7}{GW190517A}{1.2}{GW190514A}{0.8}{GW190513A}{0.6}{GW190512A}{0.3}{GW190503A}{1.0}{GW190424A}{0.9}{GW190421A}{0.9}{GW190413B}{1.1}{GW190413A}{0.8}{GW190412A}{0.1}{GW190408A}{0.3}}}
\newcommand{\radiatedenergymed}[1]{\IfEqCase{#1}{{GW190930A}{0.9}{GW190929A}{2.7}{GW190924A}{0.6}{GW190915A}{2.7}{GW190910A}{3.8}{GW190909A}{3.0}{GW190828B}{1.2}{GW190828A}{3.1}{GW190814A}{0.2}{GW190803A}{2.9}{GW190731A}{3.2}{GW190728A}{1.0}{GW190727A}{3.3}{GW190720A}{1.0}{GW190719A}{2.9}{GW190708A}{1.4}{GW190707A}{0.9}{GW190706A}{5.3}{GW190701A}{4.1}{GW190630A}{2.8}{GW190620A}{4.9}{GW190602A}{5.4}{GW190527A}{2.7}{GW190521B}{3.7}{GW190521A}{7.6}{GW190519A}{5.6}{GW190517A}{4.1}{GW190514A}{2.7}{GW190513A}{2.2}{GW190512A}{1.5}{GW190503A}{3.1}{GW190424A}{3.6}{GW190421A}{3.3}{GW190413B}{3.4}{GW190413A}{2.6}{GW190412A}{1.1}{GW190408A}{1.9}}}
\newcommand{\radiatedenergyplus}[1]{\IfEqCase{#1}{{GW190930A}{0.1}{GW190929A}{2.8}{GW190924A}{0.06}{GW190915A}{0.7}{GW190910A}{0.9}{GW190909A}{2.2}{GW190828B}{0.3}{GW190828A}{0.7}{GW190814A}{0.006}{GW190803A}{0.9}{GW190731A}{1.2}{GW190728A}{0.09}{GW190727A}{1.1}{GW190720A}{0.1}{GW190719A}{1.7}{GW190708A}{0.1}{GW190707A}{0.08}{GW190706A}{2.3}{GW190701A}{1.1}{GW190630A}{0.5}{GW190620A}{2.0}{GW190602A}{1.8}{GW190527A}{1.5}{GW190521B}{0.6}{GW190521A}{2.9}{GW190519A}{1.7}{GW190517A}{1.3}{GW190514A}{1.1}{GW190513A}{1.1}{GW190512A}{0.3}{GW190503A}{0.9}{GW190424A}{1.2}{GW190421A}{1.0}{GW190413B}{1.1}{GW190413A}{1.0}{GW190412A}{0.2}{GW190408A}{0.3}}}
\newcommand{\masstwodetminus}[1]{\IfEqCase{#1}{{GW190930A}{3.9}{GW190929A}{16.4}{GW190924A}{2.1}{GW190915A}{8.5}{GW190910A}{9.1}{GW190909A}{23.3}{GW190828B}{2.7}{GW190828A}{6.5}{GW190814A}{0.09}{GW190803A}{13.5}{GW190731A}{16.4}{GW190728A}{3.0}{GW190727A}{13.8}{GW190720A}{2.6}{GW190719A}{12.7}{GW190708A}{3.1}{GW190707A}{1.9}{GW190706A}{27.0}{GW190701A}{17.5}{GW190630A}{5.5}{GW190620A}{19.7}{GW190602A}{27.8}{GW190527A}{12.3}{GW190521B}{7.6}{GW190521A}{39.7}{GW190519A}{16.9}{GW190517A}{9.5}{GW190514A}{15.9}{GW190513A}{6.0}{GW190512A}{2.9}{GW190503A}{10.6}{GW190426A}{0.5}{GW190425A}{0.3}{GW190424A}{10.3}{GW190421A}{13.2}{GW190413B}{20.0}{GW190413A}{11.5}{GW190412A}{1.0}{GW190408A}{4.6}}}
\newcommand{\masstwodetmed}[1]{\IfEqCase{#1}{{GW190930A}{9.1}{GW190929A}{33.9}{GW190924A}{5.6}{GW190915A}{31.9}{GW190910A}{45.9}{GW190909A}{47.1}{GW190828B}{13.3}{GW190828A}{36.1}{GW190814A}{2.72}{GW190803A}{42.7}{GW190731A}{45.6}{GW190728A}{9.5}{GW190727A}{46.0}{GW190720A}{9.2}{GW190719A}{34.5}{GW190708A}{15.5}{GW190707A}{9.7}{GW190706A}{66.6}{GW190701A}{56.2}{GW190630A}{28.0}{GW190620A}{53.1}{GW190602A}{71.5}{GW190527A}{32.8}{GW190521B}{40.5}{GW190521A}{114.8}{GW190519A}{59.3}{GW190517A}{34.4}{GW190514A}{48.0}{GW190513A}{24.7}{GW190512A}{15.8}{GW190503A}{36.2}{GW190426A}{1.6}{GW190425A}{1.4}{GW190424A}{44.8}{GW190421A}{47.8}{GW190413B}{55.2}{GW190413A}{38.0}{GW190412A}{9.6}{GW190408A}{23.7}}}
\newcommand{\masstwodetplus}[1]{\IfEqCase{#1}{{GW190930A}{1.8}{GW190929A}{37.2}{GW190924A}{1.5}{GW190915A}{7.0}{GW190910A}{7.0}{GW190909A}{20.9}{GW190828B}{4.9}{GW190828A}{4.6}{GW190814A}{0.08}{GW190803A}{9.8}{GW190731A}{11.8}{GW190728A}{1.8}{GW190727A}{8.4}{GW190720A}{2.4}{GW190719A}{13.3}{GW190708A}{2.0}{GW190707A}{1.4}{GW190706A}{23.9}{GW190701A}{11.7}{GW190630A}{5.5}{GW190620A}{17.1}{GW190602A}{18.6}{GW190527A}{19.4}{GW190521B}{5.7}{GW190521A}{26.0}{GW190519A}{16.1}{GW190517A}{8.2}{GW190514A}{11.1}{GW190513A}{10.6}{GW190512A}{4.7}{GW190503A}{10.4}{GW190426A}{0.9}{GW190425A}{0.3}{GW190424A}{8.3}{GW190421A}{9.4}{GW190413B}{14.9}{GW190413A}{10.7}{GW190412A}{1.7}{GW190408A}{3.6}}}
\newcommand{\masstwosourceminus}[1]{\IfEqCase{#1}{{GW190930A}{3.3}{GW190929A}{10.6}{GW190924A}{1.9}{GW190915A}{6.1}{GW190910A}{7.2}{GW190909A}{12.7}{GW190828B}{2.1}{GW190828A}{4.8}{GW190814A}{0.09}{GW190803A}{8.2}{GW190731A}{9.5}{GW190728A}{2.6}{GW190727A}{8.4}{GW190720A}{2.2}{GW190719A}{7.2}{GW190708A}{2.7}{GW190707A}{1.7}{GW190706A}{13.3}{GW190701A}{12.0}{GW190630A}{5.1}{GW190620A}{12.3}{GW190602A}{17.4}{GW190527A}{8.1}{GW190521B}{6.4}{GW190521A}{23.1}{GW190519A}{11.1}{GW190517A}{7.3}{GW190514A}{8.8}{GW190513A}{4.1}{GW190512A}{2.5}{GW190503A}{8.0}{GW190426A}{0.5}{GW190425A}{0.3}{GW190424A}{7.7}{GW190421A}{8.8}{GW190413B}{10.8}{GW190413A}{6.7}{GW190412A}{0.9}{GW190408A}{3.6}}}
\newcommand{\masstwosourcemed}[1]{\IfEqCase{#1}{{GW190930A}{7.8}{GW190929A}{24.1}{GW190924A}{5.0}{GW190915A}{24.4}{GW190910A}{35.6}{GW190909A}{28.3}{GW190828B}{10.2}{GW190828A}{26.2}{GW190814A}{2.59}{GW190803A}{27.3}{GW190731A}{28.8}{GW190728A}{8.1}{GW190727A}{29.4}{GW190720A}{7.8}{GW190719A}{20.8}{GW190708A}{13.2}{GW190707A}{8.4}{GW190706A}{38.2}{GW190701A}{40.8}{GW190630A}{23.7}{GW190620A}{35.5}{GW190602A}{47.8}{GW190527A}{22.6}{GW190521B}{32.8}{GW190521A}{69.0}{GW190519A}{40.5}{GW190517A}{25.3}{GW190514A}{28.4}{GW190513A}{18.0}{GW190512A}{12.6}{GW190503A}{28.4}{GW190426A}{1.5}{GW190425A}{1.4}{GW190424A}{31.8}{GW190421A}{31.9}{GW190413B}{31.8}{GW190413A}{23.7}{GW190412A}{8.3}{GW190408A}{18.4}}}
\newcommand{\masstwosourceplus}[1]{\IfEqCase{#1}{{GW190930A}{1.7}{GW190929A}{19.3}{GW190924A}{1.4}{GW190915A}{5.6}{GW190910A}{6.3}{GW190909A}{13.4}{GW190828B}{3.6}{GW190828A}{4.6}{GW190814A}{0.08}{GW190803A}{7.8}{GW190731A}{9.7}{GW190728A}{1.7}{GW190727A}{7.1}{GW190720A}{2.3}{GW190719A}{9.0}{GW190708A}{2.0}{GW190707A}{1.4}{GW190706A}{14.6}{GW190701A}{8.7}{GW190630A}{5.2}{GW190620A}{12.2}{GW190602A}{14.3}{GW190527A}{10.5}{GW190521B}{5.4}{GW190521A}{22.7}{GW190519A}{11.0}{GW190517A}{7.0}{GW190514A}{9.3}{GW190513A}{7.7}{GW190512A}{3.6}{GW190503A}{7.7}{GW190426A}{0.8}{GW190425A}{0.3}{GW190424A}{7.6}{GW190421A}{8.0}{GW190413B}{11.7}{GW190413A}{7.3}{GW190412A}{1.6}{GW190408A}{3.3}}}
\newcommand{\decminus}[1]{\IfEqCase{#1}{{GW190930A}{0.66804}{GW190929A}{1.09453}{GW190924A}{0.31095}{GW190915A}{0.43700}{GW190910A}{0.78759}{GW190909A}{1.37456}{GW190828B}{0.42413}{GW190828A}{0.45822}{GW190814A}{0.12765}{GW190803A}{0.76320}{GW190731A}{0.54299}{GW190728A}{1.45903}{GW190727A}{0.53050}{GW190720A}{1.79814}{GW190719A}{1.52271}{GW190708A}{1.12280}{GW190707A}{0.66130}{GW190706A}{1.13851}{GW190701A}{0.08561}{GW190630A}{0.88066}{GW190620A}{1.15741}{GW190602A}{0.22445}{GW190527A}{0.63740}{GW190521B}{0.61963}{GW190521A}{0.40205}{GW190519A}{1.22807}{GW190517A}{0.23138}{GW190514A}{1.30610}{GW190513A}{1.20492}{GW190512A}{0.07267}{GW190503A}{0.08741}{GW190426A}{1.53579}{GW190425A}{0.89977}{GW190424A}{1.08941}{GW190421A}{0.52647}{GW190413B}{0.10088}{GW190413A}{1.21171}{GW190412A}{0.03938}{GW190408A}{0.33259}}}
\newcommand{\decmed}[1]{\IfEqCase{#1}{{GW190930A}{0.64489}{GW190929A}{0.12643}{GW190924A}{0.16308}{GW190915A}{0.64820}{GW190910A}{-0.19299}{GW190909A}{0.45798}{GW190828B}{-0.70225}{GW190828A}{-0.38234}{GW190814A}{-0.43746}{GW190803A}{0.56728}{GW190731A}{-0.84452}{GW190728A}{0.14876}{GW190727A}{-0.69575}{GW190720A}{0.61861}{GW190719A}{0.60387}{GW190708A}{0.31278}{GW190707A}{-0.26771}{GW190706A}{0.49342}{GW190701A}{-0.11450}{GW190630A}{-0.17794}{GW190620A}{0.40157}{GW190602A}{-0.60585}{GW190527A}{-0.67031}{GW190521B}{0.31697}{GW190521A}{-0.79351}{GW190519A}{0.62138}{GW190517A}{-0.77834}{GW190514A}{0.75342}{GW190513A}{0.66794}{GW190512A}{-0.46498}{GW190503A}{-0.88291}{GW190426A}{0.90282}{GW190425A}{-0.13006}{GW190424A}{-0.00051}{GW190421A}{-0.82328}{GW190413B}{-0.53598}{GW190413A}{0.45042}{GW190412A}{0.63309}{GW190408A}{0.92018}}}
\newcommand{\decplus}[1]{\IfEqCase{#1}{{GW190930A}{0.44937}{GW190929A}{0.92641}{GW190924A}{0.26435}{GW190915A}{0.49993}{GW190910A}{0.99066}{GW190909A}{0.80671}{GW190828B}{1.28711}{GW190828A}{1.25309}{GW190814A}{0.03175}{GW190803A}{0.63137}{GW190731A}{1.11036}{GW190728A}{0.43956}{GW190727A}{1.58271}{GW190720A}{0.05969}{GW190719A}{0.55220}{GW190708A}{0.86640}{GW190707A}{1.42731}{GW190706A}{0.43610}{GW190701A}{0.08803}{GW190630A}{0.78734}{GW190620A}{0.77559}{GW190602A}{0.57550}{GW190527A}{0.97030}{GW190521B}{0.25837}{GW190521A}{1.54032}{GW190519A}{0.40124}{GW190517A}{0.96906}{GW190514A}{0.60827}{GW190513A}{0.37574}{GW190512A}{0.32502}{GW190503A}{0.10637}{GW190426A}{0.61722}{GW190425A}{0.96811}{GW190424A}{1.08246}{GW190421A}{0.53382}{GW190413B}{1.13504}{GW190413A}{0.90354}{GW190412A}{0.02643}{GW190408A}{0.08290}}}
\newcommand{\psiminus}[1]{\IfEqCase{#1}{{GW190930A}{1.82}{GW190929A}{1.43}{GW190924A}{1.79}{GW190915A}{1.85}{GW190910A}{2.23}{GW190909A}{1.42}{GW190828B}{1.30}{GW190828A}{2.05}{GW190814A}{0.32}{GW190803A}{2.82}{GW190731A}{2.78}{GW190728A}{1.92}{GW190727A}{2.89}{GW190720A}{1.90}{GW190719A}{1.84}{GW190708A}{2.83}{GW190707A}{1.83}{GW190706A}{2.03}{GW190701A}{1.87}{GW190630A}{1.25}{GW190620A}{1.54}{GW190602A}{2.89}{GW190527A}{2.77}{GW190521B}{1.24}{GW190521A}{1.42}{GW190519A}{2.65}{GW190517A}{2.20}{GW190514A}{2.82}{GW190513A}{2.03}{GW190512A}{2.81}{GW190503A}{2.51}{GW190426A}{1.40}{GW190425A}{1.46}{GW190424A}{2.78}{GW190421A}{2.77}{GW190413B}{2.84}{GW190413A}{2.79}{GW190412A}{2.36}{GW190408A}{2.73}}}
\newcommand{\psimed}[1]{\IfEqCase{#1}{{GW190930A}{2.02}{GW190929A}{1.62}{GW190924A}{2.00}{GW190915A}{2.06}{GW190910A}{3.19}{GW190909A}{1.60}{GW190828B}{1.45}{GW190828A}{2.21}{GW190814A}{0.39}{GW190803A}{3.18}{GW190731A}{3.13}{GW190728A}{2.16}{GW190727A}{3.16}{GW190720A}{2.09}{GW190719A}{2.05}{GW190708A}{3.14}{GW190707A}{2.05}{GW190706A}{2.19}{GW190701A}{2.03}{GW190630A}{1.81}{GW190620A}{1.82}{GW190602A}{3.13}{GW190527A}{3.10}{GW190521B}{1.73}{GW190521A}{1.59}{GW190519A}{3.30}{GW190517A}{2.37}{GW190514A}{3.18}{GW190513A}{2.26}{GW190512A}{3.14}{GW190503A}{3.12}{GW190426A}{1.58}{GW190425A}{1.62}{GW190424A}{3.15}{GW190421A}{3.11}{GW190413B}{3.10}{GW190413A}{3.15}{GW190412A}{2.56}{GW190408A}{3.14}}}
\newcommand{\psiplus}[1]{\IfEqCase{#1}{{GW190930A}{3.53}{GW190929A}{1.36}{GW190924A}{3.51}{GW190915A}{3.52}{GW190910A}{2.39}{GW190909A}{1.40}{GW190828B}{1.52}{GW190828A}{3.56}{GW190814A}{2.62}{GW190803A}{2.75}{GW190731A}{2.84}{GW190728A}{3.57}{GW190727A}{2.83}{GW190720A}{3.47}{GW190719A}{3.54}{GW190708A}{2.87}{GW190707A}{3.56}{GW190706A}{3.34}{GW190701A}{3.64}{GW190630A}{3.38}{GW190620A}{3.56}{GW190602A}{2.94}{GW190527A}{2.81}{GW190521B}{3.41}{GW190521A}{1.38}{GW190519A}{2.15}{GW190517A}{3.46}{GW190514A}{2.79}{GW190513A}{3.41}{GW190512A}{2.71}{GW190503A}{2.59}{GW190426A}{1.36}{GW190425A}{1.38}{GW190424A}{2.77}{GW190421A}{2.85}{GW190413B}{2.87}{GW190413A}{2.74}{GW190412A}{0.44}{GW190408A}{2.70}}}
\newcommand{\totalmassdetminus}[1]{\IfEqCase{#1}{{GW190930A}{1.0}{GW190929A}{26.3}{GW190924A}{0.7}{GW190915A}{8.1}{GW190910A}{7.8}{GW190909A}{22.9}{GW190828B}{4.0}{GW190828A}{5.9}{GW190814A}{1.0}{GW190803A}{11.9}{GW190731A}{14.3}{GW190728A}{0.7}{GW190727A}{10.9}{GW190720A}{1.2}{GW190719A}{15.5}{GW190708A}{0.8}{GW190707A}{0.5}{GW190706A}{27.7}{GW190701A}{14.8}{GW190630A}{3.5}{GW190620A}{18.4}{GW190602A}{20.6}{GW190527A}{10.3}{GW190521B}{5.4}{GW190521A}{34.6}{GW190519A}{17.9}{GW190517A}{7.3}{GW190514A}{15.1}{GW190513A}{6.7}{GW190512A}{2.8}{GW190503A}{11.8}{GW190426A}{1.6}{GW190425A}{0.08}{GW190424A}{10.9}{GW190421A}{12.4}{GW190413B}{18.0}{GW190413A}{15.3}{GW190412A}{4.6}{GW190408A}{3.8}}}
\newcommand{\totalmassdetmed}[1]{\IfEqCase{#1}{{GW190930A}{23.2}{GW190929A}{148.8}{GW190924A}{15.5}{GW190915A}{78.3}{GW190910A}{101.9}{GW190909A}{119.7}{GW190828B}{44.4}{GW190828A}{79.9}{GW190814A}{27.1}{GW190803A}{100.3}{GW190731A}{109.7}{GW190728A}{23.9}{GW190727A}{104.4}{GW190720A}{24.9}{GW190719A}{94.9}{GW190708A}{36.1}{GW190707A}{23.1}{GW190706A}{180.3}{GW190701A}{129.7}{GW190630A}{69.6}{GW190620A}{137.6}{GW190602A}{171.8}{GW190527A}{84.1}{GW190521B}{92.6}{GW190521A}{269.4}{GW190519A}{155.1}{GW190517A}{85.4}{GW190514A}{112.9}{GW190513A}{73.6}{GW190512A}{45.3}{GW190503A}{91.6}{GW190426A}{7.8}{GW190425A}{3.50}{GW190424A}{101.1}{GW190421A}{108.7}{GW190413B}{135.4}{GW190413A}{93.7}{GW190412A}{44.2}{GW190408A}{55.5}}}
\newcommand{\totalmassdetplus}[1]{\IfEqCase{#1}{{GW190930A}{10.5}{GW190929A}{38.6}{GW190924A}{5.7}{GW190915A}{8.4}{GW190910A}{10.4}{GW190909A}{95.3}{GW190828B}{6.4}{GW190828A}{6.9}{GW190814A}{1.1}{GW190803A}{14.1}{GW190731A}{14.3}{GW190728A}{5.3}{GW190727A}{11.9}{GW190720A}{5.0}{GW190719A}{24.4}{GW190708A}{2.5}{GW190707A}{1.8}{GW190706A}{23.3}{GW190701A}{16.4}{GW190630A}{4.2}{GW190620A}{20.1}{GW190602A}{23.2}{GW190527A}{53.7}{GW190521B}{4.8}{GW190521A}{39.8}{GW190519A}{16.7}{GW190517A}{9.6}{GW190514A}{17.8}{GW190513A}{12.7}{GW190512A}{3.9}{GW190503A}{11.2}{GW190426A}{3.7}{GW190425A}{0.3}{GW190424A}{14.4}{GW190421A}{15.3}{GW190413B}{17.9}{GW190413A}{17.8}{GW190412A}{4.5}{GW190408A}{3.5}}}
\newcommand{\thetajnminus}[1]{\IfEqCase{#1}{{GW190930A}{0.74}{GW190929A}{1.20}{GW190924A}{0.57}{GW190915A}{1.52}{GW190910A}{1.19}{GW190909A}{1.12}{GW190828B}{1.51}{GW190828A}{1.92}{GW190814A}{0.24}{GW190803A}{0.87}{GW190731A}{0.93}{GW190728A}{1.02}{GW190727A}{1.28}{GW190720A}{2.01}{GW190719A}{1.34}{GW190708A}{1.15}{GW190707A}{1.87}{GW190706A}{1.06}{GW190701A}{0.42}{GW190630A}{0.97}{GW190620A}{1.52}{GW190602A}{1.43}{GW190527A}{0.99}{GW190521B}{1.20}{GW190521A}{0.87}{GW190519A}{0.94}{GW190517A}{1.52}{GW190514A}{1.23}{GW190513A}{0.58}{GW190512A}{1.30}{GW190503A}{0.57}{GW190426A}{1.43}{GW190425A}{0.85}{GW190424A}{1.26}{GW190421A}{1.44}{GW190413B}{1.55}{GW190413A}{0.87}{GW190412A}{0.25}{GW190408A}{0.59}}}
\newcommand{\thetajnmed}[1]{\IfEqCase{#1}{{GW190930A}{0.94}{GW190929A}{1.70}{GW190924A}{0.74}{GW190915A}{2.03}{GW190910A}{1.62}{GW190909A}{1.45}{GW190828B}{1.83}{GW190828A}{2.24}{GW190814A}{0.86}{GW190803A}{1.12}{GW190731A}{1.19}{GW190728A}{1.23}{GW190727A}{1.55}{GW190720A}{2.48}{GW190719A}{1.61}{GW190708A}{1.37}{GW190707A}{2.15}{GW190706A}{1.37}{GW190701A}{0.58}{GW190630A}{1.22}{GW190620A}{1.94}{GW190602A}{1.71}{GW190527A}{1.26}{GW190521B}{1.48}{GW190521A}{1.15}{GW190519A}{1.58}{GW190517A}{2.26}{GW190514A}{1.50}{GW190513A}{0.79}{GW190512A}{1.61}{GW190503A}{2.41}{GW190426A}{1.70}{GW190425A}{1.08}{GW190424A}{1.52}{GW190421A}{1.74}{GW190413B}{1.90}{GW190413A}{1.15}{GW190412A}{0.72}{GW190408A}{0.79}}}
\newcommand{\thetajnplus}[1]{\IfEqCase{#1}{{GW190930A}{1.90}{GW190929A}{0.97}{GW190924A}{2.04}{GW190915A}{0.78}{GW190910A}{1.13}{GW190909A}{1.34}{GW190828B}{1.02}{GW190828A}{0.70}{GW190814A}{1.48}{GW190803A}{1.73}{GW190731A}{1.66}{GW190728A}{1.66}{GW190727A}{1.31}{GW190720A}{0.50}{GW190719A}{1.25}{GW190708A}{1.54}{GW190707A}{0.77}{GW190706A}{1.43}{GW190701A}{0.55}{GW190630A}{1.60}{GW190620A}{0.90}{GW190602A}{1.17}{GW190527A}{1.55}{GW190521B}{1.37}{GW190521A}{1.65}{GW190519A}{0.95}{GW190517A}{0.64}{GW190514A}{1.33}{GW190513A}{2.02}{GW190512A}{1.22}{GW190503A}{0.52}{GW190426A}{1.19}{GW190425A}{1.77}{GW190424A}{1.36}{GW190421A}{1.13}{GW190413B}{0.95}{GW190413A}{1.64}{GW190412A}{0.44}{GW190408A}{2.03}}}
\newcommand{\redshiftminus}[1]{\IfEqCase{#1}{{GW190930A}{0.06}{GW190929A}{0.17}{GW190924A}{0.04}{GW190915A}{0.10}{GW190910A}{0.10}{GW190909A}{0.33}{GW190828B}{0.10}{GW190828A}{0.15}{GW190814A}{0.010}{GW190803A}{0.24}{GW190731A}{0.26}{GW190728A}{0.07}{GW190727A}{0.22}{GW190720A}{0.06}{GW190719A}{0.29}{GW190708A}{0.07}{GW190707A}{0.07}{GW190706A}{0.27}{GW190701A}{0.12}{GW190630A}{0.07}{GW190620A}{0.20}{GW190602A}{0.17}{GW190527A}{0.20}{GW190521B}{0.10}{GW190521A}{0.28}{GW190519A}{0.14}{GW190517A}{0.14}{GW190514A}{0.31}{GW190513A}{0.13}{GW190512A}{0.10}{GW190503A}{0.11}{GW190426A}{0.03}{GW190425A}{0.02}{GW190424A}{0.19}{GW190421A}{0.21}{GW190413B}{0.30}{GW190413A}{0.24}{GW190412A}{0.03}{GW190408A}{0.10}}}
\newcommand{\redshiftmed}[1]{\IfEqCase{#1}{{GW190930A}{0.15}{GW190929A}{0.38}{GW190924A}{0.12}{GW190915A}{0.30}{GW190910A}{0.28}{GW190909A}{0.62}{GW190828B}{0.30}{GW190828A}{0.38}{GW190814A}{0.05}{GW190803A}{0.55}{GW190731A}{0.55}{GW190728A}{0.18}{GW190727A}{0.55}{GW190720A}{0.16}{GW190719A}{0.64}{GW190708A}{0.18}{GW190707A}{0.16}{GW190706A}{0.71}{GW190701A}{0.37}{GW190630A}{0.18}{GW190620A}{0.49}{GW190602A}{0.47}{GW190527A}{0.44}{GW190521B}{0.24}{GW190521A}{0.64}{GW190519A}{0.44}{GW190517A}{0.34}{GW190514A}{0.67}{GW190513A}{0.37}{GW190512A}{0.27}{GW190503A}{0.27}{GW190426A}{0.08}{GW190425A}{0.03}{GW190424A}{0.39}{GW190421A}{0.49}{GW190413B}{0.71}{GW190413A}{0.59}{GW190412A}{0.15}{GW190408A}{0.29}}}
\newcommand{\redshiftplus}[1]{\IfEqCase{#1}{{GW190930A}{0.06}{GW190929A}{0.49}{GW190924A}{0.04}{GW190915A}{0.11}{GW190910A}{0.16}{GW190909A}{0.41}{GW190828B}{0.10}{GW190828A}{0.10}{GW190814A}{0.009}{GW190803A}{0.26}{GW190731A}{0.31}{GW190728A}{0.05}{GW190727A}{0.21}{GW190720A}{0.12}{GW190719A}{0.33}{GW190708A}{0.06}{GW190707A}{0.07}{GW190706A}{0.32}{GW190701A}{0.11}{GW190630A}{0.10}{GW190620A}{0.23}{GW190602A}{0.25}{GW190527A}{0.34}{GW190521B}{0.07}{GW190521A}{0.28}{GW190519A}{0.25}{GW190517A}{0.24}{GW190514A}{0.33}{GW190513A}{0.13}{GW190512A}{0.09}{GW190503A}{0.11}{GW190426A}{0.04}{GW190425A}{0.01}{GW190424A}{0.23}{GW190421A}{0.19}{GW190413B}{0.31}{GW190413A}{0.29}{GW190412A}{0.03}{GW190408A}{0.06}}}
\newcommand{\iotaminus}[1]{\IfEqCase{#1}{{GW190930A}{0.73}{GW190929A}{1.24}{GW190924A}{0.56}{GW190915A}{1.73}{GW190910A}{1.19}{GW190909A}{1.11}{GW190828B}{1.55}{GW190828A}{1.97}{GW190814A}{0.27}{GW190803A}{0.85}{GW190731A}{0.90}{GW190728A}{1.01}{GW190727A}{1.26}{GW190720A}{2.01}{GW190719A}{1.34}{GW190708A}{1.14}{GW190707A}{1.88}{GW190706A}{1.04}{GW190701A}{0.43}{GW190630A}{0.98}{GW190620A}{1.69}{GW190602A}{1.53}{GW190527A}{0.97}{GW190521B}{1.19}{GW190521A}{0.87}{GW190519A}{0.93}{GW190517A}{1.40}{GW190514A}{1.22}{GW190513A}{0.58}{GW190512A}{1.29}{GW190503A}{0.57}{GW190426A}{1.43}{GW190425A}{0.85}{GW190424A}{1.26}{GW190421A}{1.48}{GW190413B}{1.61}{GW190413A}{0.86}{GW190412A}{0.35}{GW190408A}{0.59}}}
\newcommand{\iotamed}[1]{\IfEqCase{#1}{{GW190930A}{0.94}{GW190929A}{1.70}{GW190924A}{0.74}{GW190915A}{2.13}{GW190910A}{1.62}{GW190909A}{1.47}{GW190828B}{1.84}{GW190828A}{2.27}{GW190814A}{0.85}{GW190803A}{1.09}{GW190731A}{1.15}{GW190728A}{1.23}{GW190727A}{1.53}{GW190720A}{2.47}{GW190719A}{1.61}{GW190708A}{1.37}{GW190707A}{2.15}{GW190706A}{1.33}{GW190701A}{0.59}{GW190630A}{1.24}{GW190620A}{2.03}{GW190602A}{1.79}{GW190527A}{1.24}{GW190521B}{1.46}{GW190521A}{1.15}{GW190519A}{1.59}{GW190517A}{2.21}{GW190514A}{1.48}{GW190513A}{0.79}{GW190512A}{1.61}{GW190503A}{2.43}{GW190426A}{1.70}{GW190425A}{1.09}{GW190424A}{1.52}{GW190421A}{1.76}{GW190413B}{1.97}{GW190413A}{1.13}{GW190412A}{0.84}{GW190408A}{0.78}}}
\newcommand{\iotaplus}[1]{\IfEqCase{#1}{{GW190930A}{1.89}{GW190929A}{1.03}{GW190924A}{2.04}{GW190915A}{0.74}{GW190910A}{1.14}{GW190909A}{1.30}{GW190828B}{1.01}{GW190828A}{0.67}{GW190814A}{1.51}{GW190803A}{1.76}{GW190731A}{1.71}{GW190728A}{1.66}{GW190727A}{1.34}{GW190720A}{0.49}{GW190719A}{1.27}{GW190708A}{1.55}{GW190707A}{0.78}{GW190706A}{1.49}{GW190701A}{0.57}{GW190630A}{1.55}{GW190620A}{0.85}{GW190602A}{1.11}{GW190527A}{1.57}{GW190521B}{1.40}{GW190521A}{1.65}{GW190519A}{0.92}{GW190517A}{0.66}{GW190514A}{1.37}{GW190513A}{2.03}{GW190512A}{1.22}{GW190503A}{0.51}{GW190426A}{1.19}{GW190425A}{1.77}{GW190424A}{1.38}{GW190421A}{1.11}{GW190413B}{0.88}{GW190413A}{1.66}{GW190412A}{0.38}{GW190408A}{2.06}}}
\newcommand{\spinonexminus}[1]{\IfEqCase{#1}{{GW190930A}{0.44}{GW190929A}{0.71}{GW190924A}{0.35}{GW190915A}{0.67}{GW190910A}{0.51}{GW190909A}{0.63}{GW190828B}{0.42}{GW190828A}{0.52}{GW190814A}{0.04}{GW190803A}{0.57}{GW190731A}{0.56}{GW190728A}{0.37}{GW190727A}{0.58}{GW190720A}{0.43}{GW190719A}{0.55}{GW190708A}{0.42}{GW190707A}{0.39}{GW190706A}{0.53}{GW190701A}{0.52}{GW190630A}{0.36}{GW190620A}{0.55}{GW190602A}{0.52}{GW190527A}{0.59}{GW190521B}{0.44}{GW190521A}{0.74}{GW190519A}{0.54}{GW190517A}{0.57}{GW190514A}{0.57}{GW190513A}{0.42}{GW190512A}{0.30}{GW190503A}{0.49}{GW190426A}{0.00}{GW190425A}{0.50}{GW190424A}{0.64}{GW190421A}{0.59}{GW190413B}{0.68}{GW190413A}{0.53}{GW190412A}{0.33}{GW190408A}{0.47}}}
\newcommand{\spinonexmed}[1]{\IfEqCase{#1}{{GW190930A}{0.002}{GW190929A}{0.007}{GW190924A}{0.0001}{GW190915A}{0.00}{GW190910A}{0.00}{GW190909A}{0.002}{GW190828B}{0.00}{GW190828A}{0.00}{GW190814A}{0.00}{GW190803A}{0.00}{GW190731A}{0.0007}{GW190728A}{0.0008}{GW190727A}{0.002}{GW190720A}{0.003}{GW190719A}{0.004}{GW190708A}{0.004}{GW190707A}{0.003}{GW190706A}{0.00}{GW190701A}{0.00}{GW190630A}{0.00}{GW190620A}{0.00}{GW190602A}{0.00}{GW190527A}{0.002}{GW190521B}{0.001}{GW190521A}{-0.02}{GW190519A}{0.004}{GW190517A}{0.0009}{GW190514A}{0.00007}{GW190513A}{0.0006}{GW190512A}{0.0010}{GW190503A}{0.00}{GW190426A}{0.00}{GW190425A}{0.00}{GW190424A}{0.00}{GW190421A}{0.00005}{GW190413B}{0.00}{GW190413A}{0.0005}{GW190412A}{-0.02}{GW190408A}{0.003}}}
\newcommand{\spinonexplus}[1]{\IfEqCase{#1}{{GW190930A}{0.47}{GW190929A}{0.69}{GW190924A}{0.36}{GW190915A}{0.66}{GW190910A}{0.51}{GW190909A}{0.68}{GW190828B}{0.43}{GW190828A}{0.51}{GW190814A}{0.04}{GW190803A}{0.55}{GW190731A}{0.51}{GW190728A}{0.40}{GW190727A}{0.58}{GW190720A}{0.45}{GW190719A}{0.57}{GW190708A}{0.47}{GW190707A}{0.42}{GW190706A}{0.53}{GW190701A}{0.52}{GW190630A}{0.36}{GW190620A}{0.53}{GW190602A}{0.53}{GW190527A}{0.59}{GW190521B}{0.45}{GW190521A}{0.75}{GW190519A}{0.55}{GW190517A}{0.57}{GW190514A}{0.60}{GW190513A}{0.43}{GW190512A}{0.28}{GW190503A}{0.50}{GW190426A}{0.00}{GW190425A}{0.47}{GW190424A}{0.63}{GW190421A}{0.60}{GW190413B}{0.71}{GW190413A}{0.53}{GW190412A}{0.39}{GW190408A}{0.49}}}
\newcommand{\chirpmassdetminus}[1]{\IfEqCase{#1}{{GW190930A}{0.2}{GW190929A}{15.4}{GW190924A}{0.03}{GW190915A}{3.9}{GW190910A}{3.6}{GW190909A}{12.4}{GW190828B}{0.7}{GW190828A}{2.8}{GW190814A}{0.02}{GW190803A}{6.1}{GW190731A}{8.2}{GW190728A}{0.08}{GW190727A}{5.7}{GW190720A}{0.1}{GW190719A}{6.6}{GW190708A}{0.2}{GW190707A}{0.09}{GW190706A}{17.5}{GW190701A}{8.1}{GW190630A}{1.5}{GW190620A}{11.2}{GW190602A}{13.7}{GW190527A}{5.5}{GW190521B}{3.0}{GW190521A}{17.6}{GW190519A}{10.3}{GW190517A}{3.4}{GW190514A}{7.7}{GW190513A}{2.5}{GW190512A}{0.8}{GW190503A}{6.0}{GW190426A}{0.01}{GW190425A}{0.0006}{GW190424A}{4.8}{GW190421A}{6.0}{GW190413B}{9.8}{GW190413A}{6.6}{GW190412A}{0.2}{GW190408A}{1.7}}}
\newcommand{\chirpmassdetmed}[1]{\IfEqCase{#1}{{GW190930A}{9.8}{GW190929A}{52.2}{GW190924A}{6.44}{GW190915A}{33.1}{GW190910A}{43.9}{GW190909A}{49.8}{GW190828B}{17.4}{GW190828A}{34.5}{GW190814A}{6.41}{GW190803A}{42.7}{GW190731A}{46.6}{GW190728A}{10.1}{GW190727A}{44.7}{GW190720A}{10.4}{GW190719A}{38.7}{GW190708A}{15.5}{GW190707A}{9.89}{GW190706A}{75.1}{GW190701A}{55.5}{GW190630A}{29.4}{GW190620A}{57.5}{GW190602A}{72.9}{GW190527A}{34.9}{GW190521B}{39.8}{GW190521A}{114.8}{GW190519A}{65.1}{GW190517A}{35.9}{GW190514A}{48.1}{GW190513A}{29.5}{GW190512A}{18.6}{GW190503A}{38.6}{GW190426A}{2.60}{GW190425A}{1.49}{GW190424A}{43.4}{GW190421A}{46.6}{GW190413B}{57.0}{GW190413A}{39.4}{GW190412A}{15.2}{GW190408A}{23.7}}}
\newcommand{\chirpmassdetplus}[1]{\IfEqCase{#1}{{GW190930A}{0.2}{GW190929A}{19.9}{GW190924A}{0.04}{GW190915A}{3.3}{GW190910A}{4.6}{GW190909A}{32.2}{GW190828B}{0.6}{GW190828A}{2.9}{GW190814A}{0.02}{GW190803A}{6.3}{GW190731A}{6.8}{GW190728A}{0.09}{GW190727A}{5.3}{GW190720A}{0.2}{GW190719A}{9.2}{GW190708A}{0.3}{GW190707A}{0.1}{GW190706A}{11.0}{GW190701A}{7.3}{GW190630A}{1.6}{GW190620A}{9.0}{GW190602A}{10.8}{GW190527A}{21.7}{GW190521B}{2.2}{GW190521A}{15.2}{GW190519A}{7.7}{GW190517A}{4.0}{GW190514A}{7.5}{GW190513A}{5.6}{GW190512A}{0.9}{GW190503A}{5.3}{GW190426A}{0.01}{GW190425A}{0.0008}{GW190424A}{6.0}{GW190421A}{6.6}{GW190413B}{8.6}{GW190413A}{7.7}{GW190412A}{0.2}{GW190408A}{1.4}}}
\newcommand{\cosiotaminus}[1]{\IfEqCase{#1}{{GW190930A}{1.54}{GW190929A}{0.79}{GW190924A}{1.67}{GW190915A}{0.43}{GW190910A}{0.88}{GW190909A}{1.03}{GW190828B}{0.69}{GW190828A}{0.33}{GW190814A}{1.37}{GW190803A}{1.42}{GW190731A}{1.37}{GW190728A}{1.30}{GW190727A}{1.00}{GW190720A}{0.20}{GW190719A}{0.93}{GW190708A}{1.18}{GW190707A}{0.43}{GW190706A}{1.19}{GW190701A}{0.44}{GW190630A}{1.26}{GW190620A}{0.52}{GW190602A}{0.75}{GW190527A}{1.27}{GW190521B}{1.07}{GW190521A}{1.35}{GW190519A}{0.79}{GW190517A}{0.37}{GW190514A}{1.05}{GW190513A}{1.65}{GW190512A}{0.91}{GW190503A}{0.22}{GW190426A}{0.84}{GW190425A}{1.42}{GW190424A}{1.02}{GW190421A}{0.78}{GW190413B}{0.57}{GW190413A}{1.37}{GW190412A}{0.32}{GW190408A}{1.67}}}
\newcommand{\cosiotamed}[1]{\IfEqCase{#1}{{GW190930A}{0.59}{GW190929A}{-0.13}{GW190924A}{0.74}{GW190915A}{-0.53}{GW190910A}{-0.05}{GW190909A}{0.10}{GW190828B}{-0.27}{GW190828A}{-0.65}{GW190814A}{0.66}{GW190803A}{0.47}{GW190731A}{0.41}{GW190728A}{0.34}{GW190727A}{0.04}{GW190720A}{-0.78}{GW190719A}{-0.04}{GW190708A}{0.20}{GW190707A}{-0.55}{GW190706A}{0.24}{GW190701A}{0.83}{GW190630A}{0.32}{GW190620A}{-0.45}{GW190602A}{-0.22}{GW190527A}{0.32}{GW190521B}{0.11}{GW190521A}{0.41}{GW190519A}{-0.02}{GW190517A}{-0.59}{GW190514A}{0.09}{GW190513A}{0.71}{GW190512A}{-0.04}{GW190503A}{-0.76}{GW190426A}{-0.13}{GW190425A}{0.46}{GW190424A}{0.05}{GW190421A}{-0.19}{GW190413B}{-0.39}{GW190413A}{0.42}{GW190412A}{0.67}{GW190408A}{0.71}}}
\newcommand{\cosiotaplus}[1]{\IfEqCase{#1}{{GW190930A}{0.39}{GW190929A}{1.03}{GW190924A}{0.24}{GW190915A}{1.45}{GW190910A}{0.96}{GW190909A}{0.84}{GW190828B}{1.23}{GW190828A}{1.60}{GW190814A}{0.17}{GW190803A}{0.51}{GW190731A}{0.56}{GW190728A}{0.64}{GW190727A}{0.93}{GW190720A}{1.68}{GW190719A}{1.00}{GW190708A}{0.77}{GW190707A}{1.51}{GW190706A}{0.72}{GW190701A}{0.16}{GW190630A}{0.65}{GW190620A}{1.39}{GW190602A}{1.18}{GW190527A}{0.64}{GW190521B}{0.85}{GW190521A}{0.55}{GW190519A}{0.81}{GW190517A}{1.29}{GW190514A}{0.88}{GW190513A}{0.27}{GW190512A}{0.99}{GW190503A}{0.47}{GW190426A}{1.09}{GW190425A}{0.51}{GW190424A}{0.91}{GW190421A}{1.15}{GW190413B}{1.32}{GW190413A}{0.54}{GW190412A}{0.22}{GW190408A}{0.27}}}
\newcommand{\comovingdistminus}[1]{\IfEqCase{#1}{{GW190930A}{257}{GW190929A}{650}{GW190924A}{184}{GW190915A}{401}{GW190910A}{396}{GW190909A}{1128}{GW190828B}{395}{GW190828A}{567}{GW190814A}{41}{GW190803A}{824}{GW190731A}{906}{GW190728A}{285}{GW190727A}{774}{GW190720A}{254}{GW190719A}{965}{GW190708A}{307}{GW190707A}{295}{GW190706A}{859}{GW190701A}{438}{GW190630A}{289}{GW190620A}{723}{GW190602A}{621}{GW190527A}{725}{GW190521B}{409}{GW190521A}{943}{GW190519A}{518}{GW190517A}{538}{GW190514A}{1034}{GW190513A}{484}{GW190512A}{378}{GW190503A}{431}{GW190426A}{143}{GW190425A}{67}{GW190424A}{717}{GW190421A}{759}{GW190413B}{957}{GW190413A}{829}{GW190412A}{134}{GW190408A}{399}}}
\newcommand{\comovingdistmed}[1]{\IfEqCase{#1}{{GW190930A}{658}{GW190929A}{1541}{GW190924A}{507}{GW190915A}{1244}{GW190910A}{1139}{GW190909A}{2329}{GW190828B}{1228}{GW190828A}{1539}{GW190814A}{229}{GW190803A}{2108}{GW190731A}{2120}{GW190728A}{742}{GW190727A}{2119}{GW190720A}{678}{GW190719A}{2399}{GW190708A}{748}{GW190707A}{667}{GW190706A}{2594}{GW190701A}{1498}{GW190630A}{752}{GW190620A}{1893}{GW190602A}{1832}{GW190527A}{1729}{GW190521B}{1003}{GW190521A}{2390}{GW190519A}{1754}{GW190517A}{1389}{GW190514A}{2475}{GW190513A}{1501}{GW190512A}{1120}{GW190503A}{1133}{GW190426A}{344}{GW190425A}{151}{GW190424A}{1578}{GW190421A}{1923}{GW190413B}{2603}{GW190413A}{2232}{GW190412A}{640}{GW190408A}{1198}}}
\newcommand{\comovingdistplus}[1]{\IfEqCase{#1}{{GW190930A}{258}{GW190929A}{1537}{GW190924A}{174}{GW190915A}{403}{GW190910A}{588}{GW190909A}{1139}{GW190828B}{359}{GW190828A}{342}{GW190814A}{37}{GW190803A}{778}{GW190731A}{925}{GW190728A}{182}{GW190727A}{627}{GW190720A}{472}{GW190719A}{915}{GW190708A}{235}{GW190707A}{272}{GW190706A}{867}{GW190701A}{400}{GW190630A}{379}{GW190620A}{725}{GW190602A}{784}{GW190527A}{1068}{GW190521B}{253}{GW190521A}{795}{GW190519A}{816}{GW190517A}{816}{GW190514A}{915}{GW190513A}{453}{GW190512A}{330}{GW190503A}{406}{GW190426A}{154}{GW190425A}{64}{GW190424A}{755}{GW190421A}{598}{GW190413B}{831}{GW190413A}{856}{GW190412A}{105}{GW190408A}{240}}}
\newcommand{\spintwoyminus}[1]{\IfEqCase{#1}{{GW190930A}{0.54}{GW190929A}{0.57}{GW190924A}{0.48}{GW190915A}{0.61}{GW190910A}{0.52}{GW190909A}{0.61}{GW190828B}{0.54}{GW190828A}{0.50}{GW190814A}{0.61}{GW190803A}{0.57}{GW190731A}{0.58}{GW190728A}{0.51}{GW190727A}{0.59}{GW190720A}{0.55}{GW190719A}{0.55}{GW190708A}{0.44}{GW190707A}{0.47}{GW190706A}{0.51}{GW190701A}{0.58}{GW190630A}{0.48}{GW190620A}{0.56}{GW190602A}{0.60}{GW190527A}{0.61}{GW190521B}{0.53}{GW190521A}{0.68}{GW190519A}{0.55}{GW190517A}{0.54}{GW190514A}{0.60}{GW190513A}{0.54}{GW190512A}{0.51}{GW190503A}{0.57}{GW190426A}{0.00}{GW190425A}{0.48}{GW190424A}{0.60}{GW190421A}{0.59}{GW190413B}{0.59}{GW190413A}{0.57}{GW190412A}{0.57}{GW190408A}{0.52}}}
\newcommand{\spintwoymed}[1]{\IfEqCase{#1}{{GW190930A}{0.00}{GW190929A}{0.0008}{GW190924A}{0.00}{GW190915A}{0.00}{GW190910A}{0.00}{GW190909A}{0.003}{GW190828B}{0.00}{GW190828A}{0.0004}{GW190814A}{0.005}{GW190803A}{0.0002}{GW190731A}{0.00}{GW190728A}{0.00}{GW190727A}{0.0009}{GW190720A}{0.00}{GW190719A}{0.003}{GW190708A}{0.003}{GW190707A}{0.002}{GW190706A}{0.001}{GW190701A}{-0.01}{GW190630A}{0.0006}{GW190620A}{0.00010}{GW190602A}{0.00}{GW190527A}{0.004}{GW190521B}{0.00}{GW190521A}{0.00}{GW190519A}{0.001}{GW190517A}{0.00}{GW190514A}{0.00}{GW190513A}{0.00}{GW190512A}{0.002}{GW190503A}{0.00}{GW190426A}{0.00}{GW190425A}{0.00}{GW190424A}{0.00}{GW190421A}{0.00}{GW190413B}{-0.01}{GW190413A}{0.00}{GW190412A}{0.004}{GW190408A}{0.0008}}}
\newcommand{\spintwoyplus}[1]{\IfEqCase{#1}{{GW190930A}{0.51}{GW190929A}{0.59}{GW190924A}{0.49}{GW190915A}{0.60}{GW190910A}{0.52}{GW190909A}{0.57}{GW190828B}{0.54}{GW190828A}{0.50}{GW190814A}{0.61}{GW190803A}{0.58}{GW190731A}{0.55}{GW190728A}{0.50}{GW190727A}{0.56}{GW190720A}{0.56}{GW190719A}{0.55}{GW190708A}{0.46}{GW190707A}{0.48}{GW190706A}{0.53}{GW190701A}{0.55}{GW190630A}{0.49}{GW190620A}{0.54}{GW190602A}{0.58}{GW190527A}{0.59}{GW190521B}{0.52}{GW190521A}{0.69}{GW190519A}{0.55}{GW190517A}{0.55}{GW190514A}{0.60}{GW190513A}{0.55}{GW190512A}{0.55}{GW190503A}{0.58}{GW190426A}{0.00}{GW190425A}{0.48}{GW190424A}{0.60}{GW190421A}{0.59}{GW190413B}{0.62}{GW190413A}{0.56}{GW190412A}{0.58}{GW190408A}{0.53}}}
\newcommand{\tiltoneminus}[1]{\IfEqCase{#1}{{GW190930A}{0.79}{GW190929A}{0.72}{GW190924A}{1.05}{GW190915A}{0.85}{GW190910A}{0.97}{GW190909A}{1.14}{GW190828B}{0.83}{GW190828A}{0.72}{GW190814A}{1.08}{GW190803A}{1.06}{GW190731A}{0.96}{GW190728A}{0.77}{GW190727A}{0.85}{GW190720A}{0.72}{GW190719A}{0.59}{GW190708A}{0.98}{GW190707A}{1.08}{GW190706A}{0.61}{GW190701A}{1.12}{GW190630A}{0.87}{GW190620A}{0.59}{GW190602A}{0.94}{GW190527A}{0.89}{GW190521B}{0.90}{GW190521A}{0.93}{GW190519A}{0.60}{GW190517A}{0.42}{GW190514A}{1.15}{GW190513A}{0.81}{GW190512A}{1.03}{GW190503A}{1.10}{GW190426A}{0.00}{GW190425A}{0.80}{GW190424A}{0.80}{GW190421A}{1.06}{GW190413B}{0.93}{GW190413A}{1.06}{GW190412A}{0.35}{GW190408A}{1.06}}}
\newcommand{\tiltonemed}[1]{\IfEqCase{#1}{{GW190930A}{1.08}{GW190929A}{1.55}{GW190924A}{1.38}{GW190915A}{1.51}{GW190910A}{1.49}{GW190909A}{1.67}{GW190828B}{1.31}{GW190828A}{1.04}{GW190814A}{1.56}{GW190803A}{1.64}{GW190731A}{1.41}{GW190728A}{1.06}{GW190727A}{1.26}{GW190720A}{1.00}{GW190719A}{0.85}{GW190708A}{1.50}{GW190707A}{1.77}{GW190706A}{0.85}{GW190701A}{1.79}{GW190630A}{1.28}{GW190620A}{0.82}{GW190602A}{1.39}{GW190527A}{1.29}{GW190521B}{1.40}{GW190521A}{1.52}{GW190519A}{0.87}{GW190517A}{0.59}{GW190514A}{2.03}{GW190513A}{1.14}{GW190512A}{1.50}{GW190503A}{1.73}{GW190426A}{0.00}{GW190425A}{1.31}{GW190424A}{1.20}{GW190421A}{1.72}{GW190413B}{1.63}{GW190413A}{1.56}{GW190412A}{0.80}{GW190408A}{1.74}}}
\newcommand{\tiltoneplus}[1]{\IfEqCase{#1}{{GW190930A}{1.14}{GW190929A}{0.97}{GW190924A}{1.09}{GW190915A}{0.92}{GW190910A}{1.09}{GW190909A}{0.99}{GW190828B}{1.08}{GW190828A}{1.15}{GW190814A}{1.11}{GW190803A}{0.99}{GW190731A}{1.13}{GW190728A}{1.20}{GW190727A}{1.11}{GW190720A}{1.01}{GW190719A}{1.09}{GW190708A}{1.00}{GW190707A}{0.86}{GW190706A}{1.06}{GW190701A}{0.95}{GW190630A}{1.11}{GW190620A}{0.91}{GW190602A}{1.12}{GW190527A}{1.11}{GW190521B}{1.09}{GW190521A}{1.04}{GW190519A}{0.79}{GW190517A}{0.50}{GW190514A}{0.78}{GW190513A}{1.18}{GW190512A}{1.11}{GW190503A}{0.97}{GW190426A}{3.14}{GW190425A}{0.66}{GW190424A}{1.06}{GW190421A}{0.97}{GW190413B}{0.91}{GW190413A}{1.11}{GW190412A}{0.54}{GW190408A}{0.94}}}
\newcommand{\spintwozminus}[1]{\IfEqCase{#1}{{GW190930A}{0.41}{GW190929A}{0.55}{GW190924A}{0.36}{GW190915A}{0.51}{GW190910A}{0.36}{GW190909A}{0.62}{GW190828B}{0.42}{GW190828A}{0.35}{GW190814A}{0.53}{GW190803A}{0.55}{GW190731A}{0.47}{GW190728A}{0.38}{GW190727A}{0.45}{GW190720A}{0.53}{GW190719A}{0.46}{GW190708A}{0.33}{GW190707A}{0.36}{GW190706A}{0.45}{GW190701A}{0.54}{GW190630A}{0.31}{GW190620A}{0.47}{GW190602A}{0.46}{GW190527A}{0.49}{GW190521B}{0.32}{GW190521A}{0.54}{GW190519A}{0.43}{GW190517A}{0.46}{GW190514A}{0.59}{GW190513A}{0.41}{GW190512A}{0.33}{GW190503A}{0.51}{GW190426A}{0.03}{GW190425A}{0.18}{GW190424A}{0.44}{GW190421A}{0.53}{GW190413B}{0.55}{GW190413A}{0.54}{GW190412A}{0.44}{GW190408A}{0.37}}}
\newcommand{\spintwozmed}[1]{\IfEqCase{#1}{{GW190930A}{0.08}{GW190929A}{0.008}{GW190924A}{0.02}{GW190915A}{0.003}{GW190910A}{0.006}{GW190909A}{-0.04}{GW190828B}{0.06}{GW190828A}{0.11}{GW190814A}{-0.01}{GW190803A}{-0.01}{GW190731A}{0.03}{GW190728A}{0.11}{GW190727A}{0.04}{GW190720A}{0.11}{GW190719A}{0.16}{GW190708A}{0.03}{GW190707A}{-0.03}{GW190706A}{0.11}{GW190701A}{-0.04}{GW190630A}{0.09}{GW190620A}{0.21}{GW190602A}{0.05}{GW190527A}{0.04}{GW190521B}{0.09}{GW190521A}{-0.01}{GW190519A}{0.21}{GW190517A}{0.29}{GW190514A}{-0.13}{GW190513A}{0.06}{GW190512A}{0.03}{GW190503A}{0.00}{GW190426A}{0.00}{GW190425A}{0.03}{GW190424A}{0.06}{GW190421A}{-0.04}{GW190413B}{-0.03}{GW190413A}{-0.02}{GW190412A}{0.07}{GW190408A}{0.00}}}
\newcommand{\spintwozplus}[1]{\IfEqCase{#1}{{GW190930A}{0.50}{GW190929A}{0.57}{GW190924A}{0.48}{GW190915A}{0.47}{GW190910A}{0.39}{GW190909A}{0.51}{GW190828B}{0.54}{GW190828A}{0.48}{GW190814A}{0.52}{GW190803A}{0.47}{GW190731A}{0.54}{GW190728A}{0.48}{GW190727A}{0.53}{GW190720A}{0.54}{GW190719A}{0.61}{GW190708A}{0.39}{GW190707A}{0.34}{GW190706A}{0.62}{GW190701A}{0.42}{GW190630A}{0.44}{GW190620A}{0.57}{GW190602A}{0.56}{GW190527A}{0.60}{GW190521B}{0.36}{GW190521A}{0.53}{GW190519A}{0.56}{GW190517A}{0.53}{GW190514A}{0.44}{GW190513A}{0.54}{GW190512A}{0.45}{GW190503A}{0.46}{GW190426A}{0.03}{GW190425A}{0.30}{GW190424A}{0.52}{GW190421A}{0.41}{GW190413B}{0.49}{GW190413A}{0.48}{GW190412A}{0.57}{GW190408A}{0.38}}}
\newcommand{\massonesourceminus}[1]{\IfEqCase{#1}{{GW190930A}{2.3}{GW190929A}{33.2}{GW190924A}{2.0}{GW190915A}{6.4}{GW190910A}{6.1}{GW190909A}{13.3}{GW190828B}{7.2}{GW190828A}{4.0}{GW190814A}{1.0}{GW190803A}{7.0}{GW190731A}{9.0}{GW190728A}{2.2}{GW190727A}{6.2}{GW190720A}{3.0}{GW190719A}{10.3}{GW190708A}{2.3}{GW190707A}{1.7}{GW190706A}{16.2}{GW190701A}{8.0}{GW190630A}{5.6}{GW190620A}{12.7}{GW190602A}{13.0}{GW190527A}{9.0}{GW190521B}{4.8}{GW190521A}{18.9}{GW190519A}{12.0}{GW190517A}{7.6}{GW190514A}{8.2}{GW190513A}{9.2}{GW190512A}{5.8}{GW190503A}{8.1}{GW190426A}{2.3}{GW190425A}{0.3}{GW190424A}{7.3}{GW190421A}{6.9}{GW190413B}{10.7}{GW190413A}{8.1}{GW190412A}{5.1}{GW190408A}{3.4}}}
\newcommand{\massonesourcemed}[1]{\IfEqCase{#1}{{GW190930A}{12.3}{GW190929A}{80.8}{GW190924A}{8.9}{GW190915A}{35.3}{GW190910A}{43.9}{GW190909A}{45.8}{GW190828B}{24.1}{GW190828A}{32.1}{GW190814A}{23.2}{GW190803A}{37.3}{GW190731A}{41.5}{GW190728A}{12.3}{GW190727A}{38.0}{GW190720A}{13.4}{GW190719A}{36.5}{GW190708A}{17.6}{GW190707A}{11.6}{GW190706A}{67.0}{GW190701A}{53.9}{GW190630A}{35.1}{GW190620A}{57.1}{GW190602A}{69.1}{GW190527A}{36.5}{GW190521B}{42.2}{GW190521A}{95.3}{GW190519A}{66.0}{GW190517A}{37.4}{GW190514A}{39.0}{GW190513A}{35.7}{GW190512A}{23.3}{GW190503A}{43.3}{GW190426A}{5.7}{GW190425A}{2.0}{GW190424A}{40.5}{GW190421A}{41.3}{GW190413B}{47.5}{GW190413A}{34.7}{GW190412A}{30.1}{GW190408A}{24.6}}}
\newcommand{\massonesourceplus}[1]{\IfEqCase{#1}{{GW190930A}{12.4}{GW190929A}{33.0}{GW190924A}{7.0}{GW190915A}{9.5}{GW190910A}{7.6}{GW190909A}{52.7}{GW190828B}{7.0}{GW190828A}{5.8}{GW190814A}{1.1}{GW190803A}{10.6}{GW190731A}{12.2}{GW190728A}{7.2}{GW190727A}{9.5}{GW190720A}{6.7}{GW190719A}{18.0}{GW190708A}{4.7}{GW190707A}{3.3}{GW190706A}{14.6}{GW190701A}{11.8}{GW190630A}{6.9}{GW190620A}{16.0}{GW190602A}{15.7}{GW190527A}{16.4}{GW190521B}{5.9}{GW190521A}{28.7}{GW190519A}{10.7}{GW190517A}{11.7}{GW190514A}{14.7}{GW190513A}{9.5}{GW190512A}{5.3}{GW190503A}{9.2}{GW190426A}{3.9}{GW190425A}{0.6}{GW190424A}{11.1}{GW190421A}{10.4}{GW190413B}{13.5}{GW190413A}{12.6}{GW190412A}{4.7}{GW190408A}{5.1}}}
\newcommand{\geocenttimeminus}[1]{\IfEqCase{#1}{{GW190930A}{0.02}{GW190929A}{0.02}{GW190924A}{0.008}{GW190915A}{0.003}{GW190910A}{0.0}{GW190909A}{0.0}{GW190828B}{0.0}{GW190828A}{0.0}{GW190814A}{0.0009}{GW190803A}{0.0}{GW190731A}{0.0}{GW190728A}{0.03}{GW190727A}{0.0}{GW190720A}{0.01}{GW190719A}{0.0}{GW190708A}{0.0}{GW190707A}{0.03}{GW190706A}{0.0000002}{GW190701A}{0.0}{GW190630A}{0.0000002}{GW190620A}{0.0000002}{GW190602A}{0.0}{GW190527A}{0.0}{GW190521B}{0.0}{GW190521A}{0.04}{GW190519A}{0.0}{GW190517A}{0.0}{GW190514A}{0.0}{GW190513A}{0.010}{GW190512A}{0.0}{GW190503A}{0.0}{GW190426A}{0.03}{GW190425A}{0.009}{GW190424A}{0.0}{GW190421A}{0.0}{GW190413B}{0.0}{GW190413A}{0.0}{GW190412A}{0.001}{GW190408A}{0.0}}}
\newcommand{\geocenttimemed}[1]{\IfEqCase{#1}{{GW190930A}{1253885759.2}{GW190929A}{1253755327.5}{GW190924A}{1253326744.8}{GW190915A}{1252627040.7}{GW190910A}{1252150105.3}{GW190909A}{1252064527.7}{GW190828B}{1251010527.9}{GW190828A}{1251009263.8}{GW190814A}{1249852257.0}{GW190803A}{1248834439.9}{GW190731A}{1248617394.6}{GW190728A}{1248331528.6}{GW190727A}{1248242632.0}{GW190720A}{1247616534.7}{GW190719A}{1247608532.9}{GW190708A}{1246663515.4}{GW190707A}{1246527224.2}{GW190706A}{1246487219.3}{GW190701A}{1246048404.6}{GW190630A}{1245955943.2}{GW190620A}{1245035079.3}{GW190602A}{1243533585.1}{GW190527A}{1242984073.8}{GW190521B}{1242459857.5}{GW190521A}{1242442967.4}{GW190519A}{1242315362.4}{GW190517A}{1242107479.8}{GW190514A}{1241852074.8}{GW190513A}{1241816086.8}{GW190512A}{1241719652.4}{GW190503A}{1240944862.3}{GW190426A}{1240327333.4}{GW190425A}{1240215503.0}{GW190424A}{1240164426.1}{GW190421A}{1239917954.2}{GW190413B}{1239198206.7}{GW190413A}{1239168612.5}{GW190412A}{1239082262.2}{GW190408A}{1238782700.3}}}
\newcommand{\geocenttimeplus}[1]{\IfEqCase{#1}{{GW190930A}{0.002}{GW190929A}{0.03}{GW190924A}{0.008}{GW190915A}{0.002}{GW190910A}{0.0}{GW190909A}{0.0}{GW190828B}{0.0}{GW190828A}{0.0000005}{GW190814A}{0.004}{GW190803A}{0.0}{GW190731A}{0.0}{GW190728A}{0.0010}{GW190727A}{0.0}{GW190720A}{0.01}{GW190719A}{0.05}{GW190708A}{0.0}{GW190707A}{0.009}{GW190706A}{0.0}{GW190701A}{0.0000005}{GW190630A}{0.0}{GW190620A}{0.0}{GW190602A}{0.0}{GW190527A}{0.0}{GW190521B}{0.0}{GW190521A}{0.01}{GW190519A}{0.0}{GW190517A}{0.0}{GW190514A}{0.0}{GW190513A}{0.0}{GW190512A}{0.0}{GW190503A}{0.0}{GW190426A}{0.02}{GW190425A}{0.03}{GW190424A}{0.0}{GW190421A}{0.0}{GW190413B}{0.0}{GW190413A}{0.0}{GW190412A}{0.007}{GW190408A}{0.0}}}
\newcommand{\costilttwominus}[1]{\IfEqCase{#1}{{GW190930A}{1.09}{GW190929A}{0.93}{GW190924A}{1.00}{GW190915A}{0.90}{GW190910A}{0.88}{GW190909A}{0.75}{GW190828B}{1.06}{GW190828A}{1.14}{GW190814A}{0.83}{GW190803A}{0.83}{GW190731A}{0.99}{GW190728A}{1.16}{GW190727A}{1.04}{GW190720A}{1.11}{GW190719A}{1.19}{GW190708A}{0.97}{GW190707A}{0.71}{GW190706A}{1.16}{GW190701A}{0.74}{GW190630A}{1.08}{GW190620A}{1.22}{GW190602A}{1.01}{GW190527A}{1.02}{GW190521B}{1.03}{GW190521A}{0.86}{GW190519A}{1.17}{GW190517A}{1.21}{GW190514A}{0.59}{GW190513A}{1.05}{GW190512A}{0.98}{GW190503A}{0.87}{GW190426A}{0.00}{GW190425A}{0.87}{GW190424A}{1.02}{GW190421A}{0.76}{GW190413B}{0.77}{GW190413A}{0.82}{GW190412A}{1.01}{GW190408A}{0.85}}}
\newcommand{\costilttwomed}[1]{\IfEqCase{#1}{{GW190930A}{0.31}{GW190929A}{0.04}{GW190924A}{0.15}{GW190915A}{0.02}{GW190910A}{0.04}{GW190909A}{-0.17}{GW190828B}{0.24}{GW190828A}{0.37}{GW190814A}{-0.03}{GW190803A}{-0.06}{GW190731A}{0.15}{GW190728A}{0.39}{GW190727A}{0.20}{GW190720A}{0.31}{GW190719A}{0.44}{GW190708A}{0.14}{GW190707A}{-0.19}{GW190706A}{0.37}{GW190701A}{-0.17}{GW190630A}{0.33}{GW190620A}{0.50}{GW190602A}{0.19}{GW190527A}{0.16}{GW190521B}{0.29}{GW190521A}{-0.02}{GW190519A}{0.50}{GW190517A}{0.61}{GW190514A}{-0.36}{GW190513A}{0.25}{GW190512A}{0.16}{GW190503A}{-0.02}{GW190426A}{-1.00}{GW190425A}{0.16}{GW190424A}{0.20}{GW190421A}{-0.16}{GW190413B}{-0.13}{GW190413A}{-0.08}{GW190412A}{0.25}{GW190408A}{-0.02}}}
\newcommand{\costilttwoplus}[1]{\IfEqCase{#1}{{GW190930A}{0.63}{GW190929A}{0.86}{GW190924A}{0.77}{GW190915A}{0.85}{GW190910A}{0.83}{GW190909A}{1.04}{GW190828B}{0.69}{GW190828A}{0.57}{GW190814A}{0.87}{GW190803A}{0.93}{GW190731A}{0.77}{GW190728A}{0.56}{GW190727A}{0.72}{GW190720A}{0.62}{GW190719A}{0.51}{GW190708A}{0.77}{GW190707A}{1.02}{GW190706A}{0.58}{GW190701A}{1.00}{GW190630A}{0.60}{GW190620A}{0.47}{GW190602A}{0.73}{GW190527A}{0.76}{GW190521B}{0.62}{GW190521A}{0.86}{GW190519A}{0.46}{GW190517A}{0.36}{GW190514A}{1.11}{GW190513A}{0.68}{GW190512A}{0.75}{GW190503A}{0.88}{GW190426A}{2.00}{GW190425A}{0.70}{GW190424A}{0.72}{GW190421A}{0.96}{GW190413B}{1.00}{GW190413A}{0.95}{GW190412A}{0.67}{GW190408A}{0.88}}}
\newcommand{\finalspinminus}[1]{\IfEqCase{#1}{{GW190930A}{0.06}{GW190929A}{0.31}{GW190924A}{0.05}{GW190915A}{0.11}{GW190910A}{0.07}{GW190909A}{0.20}{GW190828B}{0.08}{GW190828A}{0.07}{GW190814A}{0.02}{GW190803A}{0.11}{GW190731A}{0.13}{GW190728A}{0.04}{GW190727A}{0.10}{GW190720A}{0.05}{GW190719A}{0.17}{GW190708A}{0.04}{GW190707A}{0.04}{GW190706A}{0.18}{GW190701A}{0.13}{GW190630A}{0.07}{GW190620A}{0.15}{GW190602A}{0.14}{GW190527A}{0.16}{GW190521B}{0.07}{GW190521A}{0.16}{GW190519A}{0.13}{GW190517A}{0.07}{GW190514A}{0.15}{GW190513A}{0.12}{GW190512A}{0.07}{GW190503A}{0.12}{GW190424A}{0.09}{GW190421A}{0.11}{GW190413B}{0.12}{GW190413A}{0.13}{GW190412A}{0.06}{GW190408A}{0.07}}}
\newcommand{\finalspinmed}[1]{\IfEqCase{#1}{{GW190930A}{0.72}{GW190929A}{0.66}{GW190924A}{0.67}{GW190915A}{0.70}{GW190910A}{0.70}{GW190909A}{0.66}{GW190828B}{0.65}{GW190828A}{0.75}{GW190814A}{0.28}{GW190803A}{0.68}{GW190731A}{0.70}{GW190728A}{0.71}{GW190727A}{0.73}{GW190720A}{0.72}{GW190719A}{0.78}{GW190708A}{0.69}{GW190707A}{0.66}{GW190706A}{0.78}{GW190701A}{0.66}{GW190630A}{0.70}{GW190620A}{0.79}{GW190602A}{0.70}{GW190527A}{0.71}{GW190521B}{0.72}{GW190521A}{0.71}{GW190519A}{0.79}{GW190517A}{0.87}{GW190514A}{0.63}{GW190513A}{0.68}{GW190512A}{0.65}{GW190503A}{0.66}{GW190424A}{0.74}{GW190421A}{0.67}{GW190413B}{0.68}{GW190413A}{0.68}{GW190412A}{0.67}{GW190408A}{0.67}}}
\newcommand{\finalspinplus}[1]{\IfEqCase{#1}{{GW190930A}{0.07}{GW190929A}{0.20}{GW190924A}{0.05}{GW190915A}{0.09}{GW190910A}{0.08}{GW190909A}{0.15}{GW190828B}{0.08}{GW190828A}{0.06}{GW190814A}{0.02}{GW190803A}{0.10}{GW190731A}{0.10}{GW190728A}{0.04}{GW190727A}{0.10}{GW190720A}{0.06}{GW190719A}{0.11}{GW190708A}{0.04}{GW190707A}{0.03}{GW190706A}{0.09}{GW190701A}{0.09}{GW190630A}{0.05}{GW190620A}{0.08}{GW190602A}{0.10}{GW190527A}{0.12}{GW190521B}{0.05}{GW190521A}{0.12}{GW190519A}{0.07}{GW190517A}{0.05}{GW190514A}{0.11}{GW190513A}{0.14}{GW190512A}{0.07}{GW190503A}{0.09}{GW190424A}{0.09}{GW190421A}{0.10}{GW190413B}{0.10}{GW190413A}{0.12}{GW190412A}{0.05}{GW190408A}{0.06}}}
\newcommand{\luminositydistanceminus}[1]{\IfEqCase{#1}{{GW190930A}{0.32}{GW190929A}{1.05}{GW190924A}{0.22}{GW190915A}{0.61}{GW190910A}{0.58}{GW190909A}{2.22}{GW190828B}{0.60}{GW190828A}{0.93}{GW190814A}{0.05}{GW190803A}{1.58}{GW190731A}{1.72}{GW190728A}{0.37}{GW190727A}{1.50}{GW190720A}{0.32}{GW190719A}{2.00}{GW190708A}{0.39}{GW190707A}{0.37}{GW190706A}{1.93}{GW190701A}{0.73}{GW190630A}{0.37}{GW190620A}{1.31}{GW190602A}{1.12}{GW190527A}{1.24}{GW190521B}{0.57}{GW190521A}{1.95}{GW190519A}{0.92}{GW190517A}{0.84}{GW190514A}{2.17}{GW190513A}{0.80}{GW190512A}{0.55}{GW190503A}{0.63}{GW190426A}{0.16}{GW190425A}{0.07}{GW190424A}{1.16}{GW190421A}{1.38}{GW190413B}{2.12}{GW190413A}{1.66}{GW190412A}{0.17}{GW190408A}{0.60}}}
\newcommand{\luminositydistancemed}[1]{\IfEqCase{#1}{{GW190930A}{0.76}{GW190929A}{2.13}{GW190924A}{0.57}{GW190915A}{1.62}{GW190910A}{1.46}{GW190909A}{3.77}{GW190828B}{1.60}{GW190828A}{2.13}{GW190814A}{0.24}{GW190803A}{3.27}{GW190731A}{3.30}{GW190728A}{0.87}{GW190727A}{3.30}{GW190720A}{0.79}{GW190719A}{3.94}{GW190708A}{0.88}{GW190707A}{0.77}{GW190706A}{4.42}{GW190701A}{2.06}{GW190630A}{0.89}{GW190620A}{2.81}{GW190602A}{2.69}{GW190527A}{2.49}{GW190521B}{1.24}{GW190521A}{3.92}{GW190519A}{2.53}{GW190517A}{1.86}{GW190514A}{4.13}{GW190513A}{2.06}{GW190512A}{1.43}{GW190503A}{1.45}{GW190426A}{0.37}{GW190425A}{0.16}{GW190424A}{2.20}{GW190421A}{2.88}{GW190413B}{4.45}{GW190413A}{3.55}{GW190412A}{0.74}{GW190408A}{1.55}}}
\newcommand{\luminositydistanceplus}[1]{\IfEqCase{#1}{{GW190930A}{0.36}{GW190929A}{3.65}{GW190924A}{0.22}{GW190915A}{0.71}{GW190910A}{1.03}{GW190909A}{3.27}{GW190828B}{0.62}{GW190828A}{0.66}{GW190814A}{0.04}{GW190803A}{1.95}{GW190731A}{2.39}{GW190728A}{0.26}{GW190727A}{1.54}{GW190720A}{0.69}{GW190719A}{2.59}{GW190708A}{0.33}{GW190707A}{0.38}{GW190706A}{2.59}{GW190701A}{0.76}{GW190630A}{0.56}{GW190620A}{1.68}{GW190602A}{1.79}{GW190527A}{2.48}{GW190521B}{0.40}{GW190521A}{2.19}{GW190519A}{1.83}{GW190517A}{1.62}{GW190514A}{2.65}{GW190513A}{0.88}{GW190512A}{0.55}{GW190503A}{0.69}{GW190426A}{0.18}{GW190425A}{0.07}{GW190424A}{1.58}{GW190421A}{1.37}{GW190413B}{2.48}{GW190413A}{2.27}{GW190412A}{0.14}{GW190408A}{0.40}}}
\newcommand{\spinonezminus}[1]{\IfEqCase{#1}{{GW190930A}{0.29}{GW190929A}{0.43}{GW190924A}{0.24}{GW190915A}{0.41}{GW190910A}{0.34}{GW190909A}{0.56}{GW190828B}{0.23}{GW190828A}{0.31}{GW190814A}{0.05}{GW190803A}{0.46}{GW190731A}{0.33}{GW190728A}{0.27}{GW190727A}{0.33}{GW190720A}{0.29}{GW190719A}{0.44}{GW190708A}{0.25}{GW190707A}{0.29}{GW190706A}{0.38}{GW190701A}{0.51}{GW190630A}{0.21}{GW190620A}{0.39}{GW190602A}{0.34}{GW190527A}{0.36}{GW190521B}{0.23}{GW190521A}{0.58}{GW190519A}{0.37}{GW190517A}{0.36}{GW190514A}{0.54}{GW190513A}{0.24}{GW190512A}{0.25}{GW190503A}{0.44}{GW190426A}{0.51}{GW190425A}{0.12}{GW190424A}{0.35}{GW190421A}{0.49}{GW190413B}{0.48}{GW190413A}{0.52}{GW190412A}{0.23}{GW190408A}{0.42}}}
\newcommand{\spinonezmed}[1]{\IfEqCase{#1}{{GW190930A}{0.15}{GW190929A}{0.008}{GW190924A}{0.02}{GW190915A}{0.02}{GW190910A}{0.01}{GW190909A}{-0.03}{GW190828B}{0.06}{GW190828A}{0.20}{GW190814A}{0.0001}{GW190803A}{-0.01}{GW190731A}{0.03}{GW190728A}{0.13}{GW190727A}{0.10}{GW190720A}{0.20}{GW190719A}{0.36}{GW190708A}{0.009}{GW190707A}{-0.03}{GW190706A}{0.33}{GW190701A}{-0.05}{GW190630A}{0.05}{GW190620A}{0.37}{GW190602A}{0.04}{GW190527A}{0.09}{GW190521B}{0.04}{GW190521A}{0.02}{GW190519A}{0.35}{GW190517A}{0.67}{GW190514A}{-0.18}{GW190513A}{0.09}{GW190512A}{0.005}{GW190503A}{-0.03}{GW190426A}{-0.03}{GW190425A}{0.06}{GW190424A}{0.15}{GW190421A}{-0.04}{GW190413B}{-0.02}{GW190413A}{0.001}{GW190412A}{0.30}{GW190408A}{-0.03}}}
\newcommand{\spinonezplus}[1]{\IfEqCase{#1}{{GW190930A}{0.41}{GW190929A}{0.45}{GW190924A}{0.39}{GW190915A}{0.40}{GW190910A}{0.39}{GW190909A}{0.53}{GW190828B}{0.24}{GW190828A}{0.41}{GW190814A}{0.04}{GW190803A}{0.39}{GW190731A}{0.45}{GW190728A}{0.30}{GW190727A}{0.48}{GW190720A}{0.29}{GW190719A}{0.43}{GW190708A}{0.24}{GW190707A}{0.20}{GW190706A}{0.43}{GW190701A}{0.34}{GW190630A}{0.28}{GW190620A}{0.42}{GW190602A}{0.46}{GW190527A}{0.50}{GW190521B}{0.32}{GW190521A}{0.53}{GW190519A}{0.37}{GW190517A}{0.25}{GW190514A}{0.39}{GW190513A}{0.46}{GW190512A}{0.21}{GW190503A}{0.31}{GW190426A}{0.36}{GW190425A}{0.18}{GW190424A}{0.46}{GW190421A}{0.40}{GW190413B}{0.40}{GW190413A}{0.44}{GW190412A}{0.12}{GW190408A}{0.26}}}
\newcommand{\chirpmasssourceminus}[1]{\IfEqCase{#1}{{GW190930A}{0.5}{GW190929A}{8.2}{GW190924A}{0.2}{GW190915A}{2.7}{GW190910A}{4.1}{GW190909A}{7.5}{GW190828B}{1.0}{GW190828A}{2.1}{GW190814A}{0.06}{GW190803A}{4.1}{GW190731A}{5.2}{GW190728A}{0.3}{GW190727A}{3.7}{GW190720A}{0.8}{GW190719A}{4.0}{GW190708A}{0.6}{GW190707A}{0.5}{GW190706A}{7.0}{GW190701A}{4.9}{GW190630A}{2.1}{GW190620A}{6.5}{GW190602A}{8.5}{GW190527A}{4.2}{GW190521B}{2.5}{GW190521A}{10.6}{GW190519A}{7.1}{GW190517A}{4.0}{GW190514A}{4.8}{GW190513A}{1.9}{GW190512A}{1.0}{GW190503A}{4.2}{GW190426A}{0.08}{GW190425A}{0.02}{GW190424A}{4.6}{GW190421A}{4.2}{GW190413B}{5.4}{GW190413A}{4.1}{GW190412A}{0.3}{GW190408A}{1.2}}}
\newcommand{\chirpmasssourcemed}[1]{\IfEqCase{#1}{{GW190930A}{8.5}{GW190929A}{35.8}{GW190924A}{5.8}{GW190915A}{25.3}{GW190910A}{34.3}{GW190909A}{30.9}{GW190828B}{13.3}{GW190828A}{25.0}{GW190814A}{6.09}{GW190803A}{27.3}{GW190731A}{29.5}{GW190728A}{8.6}{GW190727A}{28.6}{GW190720A}{8.9}{GW190719A}{23.5}{GW190708A}{13.2}{GW190707A}{8.5}{GW190706A}{42.7}{GW190701A}{40.3}{GW190630A}{24.9}{GW190620A}{38.3}{GW190602A}{49.1}{GW190527A}{24.3}{GW190521B}{32.1}{GW190521A}{69.2}{GW190519A}{44.5}{GW190517A}{26.6}{GW190514A}{28.5}{GW190513A}{21.6}{GW190512A}{14.6}{GW190503A}{30.2}{GW190426A}{2.41}{GW190425A}{1.44}{GW190424A}{31.0}{GW190421A}{31.2}{GW190413B}{33.0}{GW190413A}{24.6}{GW190412A}{13.3}{GW190408A}{18.3}}}
\newcommand{\chirpmasssourceplus}[1]{\IfEqCase{#1}{{GW190930A}{0.5}{GW190929A}{14.9}{GW190924A}{0.2}{GW190915A}{3.2}{GW190910A}{4.1}{GW190909A}{17.2}{GW190828B}{1.2}{GW190828A}{3.4}{GW190814A}{0.06}{GW190803A}{5.7}{GW190731A}{7.1}{GW190728A}{0.5}{GW190727A}{5.3}{GW190720A}{0.5}{GW190719A}{6.5}{GW190708A}{0.9}{GW190707A}{0.6}{GW190706A}{10.0}{GW190701A}{5.4}{GW190630A}{2.1}{GW190620A}{8.3}{GW190602A}{9.1}{GW190527A}{9.1}{GW190521B}{3.2}{GW190521A}{17.0}{GW190519A}{6.4}{GW190517A}{4.0}{GW190514A}{7.9}{GW190513A}{3.8}{GW190512A}{1.3}{GW190503A}{4.2}{GW190426A}{0.08}{GW190425A}{0.02}{GW190424A}{5.8}{GW190421A}{5.9}{GW190413B}{8.2}{GW190413A}{5.5}{GW190412A}{0.4}{GW190408A}{1.9}}}
\newcommand{\symmetricmassratiominus}[1]{\IfEqCase{#1}{{GW190930A}{0.11}{GW190929A}{0.07}{GW190924A}{0.09}{GW190915A}{0.03}{GW190910A}{0.01}{GW190909A}{0.09}{GW190828B}{0.04}{GW190828A}{0.01}{GW190814A}{0.006}{GW190803A}{0.03}{GW190731A}{0.04}{GW190728A}{0.07}{GW190727A}{0.03}{GW190720A}{0.06}{GW190719A}{0.06}{GW190708A}{0.03}{GW190707A}{0.03}{GW190706A}{0.05}{GW190701A}{0.03}{GW190630A}{0.03}{GW190620A}{0.04}{GW190602A}{0.04}{GW190527A}{0.06}{GW190521B}{0.01}{GW190521A}{0.04}{GW190519A}{0.03}{GW190517A}{0.04}{GW190514A}{0.04}{GW190513A}{0.04}{GW190512A}{0.03}{GW190503A}{0.03}{GW190426A}{0.08}{GW190425A}{0.03}{GW190424A}{0.02}{GW190421A}{0.03}{GW190413B}{0.04}{GW190413A}{0.04}{GW190412A}{0.02}{GW190408A}{0.02}}}
\newcommand{\symmetricmassratiomed}[1]{\IfEqCase{#1}{{GW190930A}{0.24}{GW190929A}{0.18}{GW190924A}{0.23}{GW190915A}{0.242}{GW190910A}{0.248}{GW190909A}{0.24}{GW190828B}{0.21}{GW190828A}{0.248}{GW190814A}{0.090}{GW190803A}{0.245}{GW190731A}{0.243}{GW190728A}{0.24}{GW190727A}{0.247}{GW190720A}{0.23}{GW190719A}{0.23}{GW190708A}{0.245}{GW190707A}{0.244}{GW190706A}{0.23}{GW190701A}{0.246}{GW190630A}{0.241}{GW190620A}{0.24}{GW190602A}{0.243}{GW190527A}{0.24}{GW190521B}{0.246}{GW190521A}{0.245}{GW190519A}{0.24}{GW190517A}{0.241}{GW190514A}{0.245}{GW190513A}{0.22}{GW190512A}{0.23}{GW190503A}{0.24}{GW190426A}{0.16}{GW190425A}{0.240}{GW190424A}{0.247}{GW190421A}{0.247}{GW190413B}{0.241}{GW190413A}{0.241}{GW190412A}{0.17}{GW190408A}{0.245}}}
\newcommand{\symmetricmassratioplus}[1]{\IfEqCase{#1}{{GW190930A}{0.01}{GW190929A}{0.07}{GW190924A}{0.02}{GW190915A}{0.008}{GW190910A}{0.002}{GW190909A}{0.01}{GW190828B}{0.04}{GW190828A}{0.002}{GW190814A}{0.005}{GW190803A}{0.005}{GW190731A}{0.007}{GW190728A}{0.01}{GW190727A}{0.003}{GW190720A}{0.02}{GW190719A}{0.02}{GW190708A}{0.005}{GW190707A}{0.006}{GW190706A}{0.02}{GW190701A}{0.004}{GW190630A}{0.009}{GW190620A}{0.01}{GW190602A}{0.007}{GW190527A}{0.01}{GW190521B}{0.004}{GW190521A}{0.005}{GW190519A}{0.01}{GW190517A}{0.009}{GW190514A}{0.005}{GW190513A}{0.03}{GW190512A}{0.02}{GW190503A}{0.01}{GW190426A}{0.08}{GW190425A}{0.010}{GW190424A}{0.003}{GW190421A}{0.003}{GW190413B}{0.008}{GW190413A}{0.008}{GW190412A}{0.03}{GW190408A}{0.005}}}
\newcommand{\spintwoxminus}[1]{\IfEqCase{#1}{{GW190930A}{0.52}{GW190929A}{0.60}{GW190924A}{0.49}{GW190915A}{0.59}{GW190910A}{0.52}{GW190909A}{0.58}{GW190828B}{0.55}{GW190828A}{0.50}{GW190814A}{0.62}{GW190803A}{0.58}{GW190731A}{0.57}{GW190728A}{0.46}{GW190727A}{0.55}{GW190720A}{0.58}{GW190719A}{0.56}{GW190708A}{0.46}{GW190707A}{0.46}{GW190706A}{0.53}{GW190701A}{0.55}{GW190630A}{0.49}{GW190620A}{0.54}{GW190602A}{0.60}{GW190527A}{0.61}{GW190521B}{0.51}{GW190521A}{0.66}{GW190519A}{0.57}{GW190517A}{0.53}{GW190514A}{0.60}{GW190513A}{0.53}{GW190512A}{0.50}{GW190503A}{0.55}{GW190426A}{0.00}{GW190425A}{0.47}{GW190424A}{0.59}{GW190421A}{0.59}{GW190413B}{0.59}{GW190413A}{0.58}{GW190412A}{0.57}{GW190408A}{0.53}}}
\newcommand{\spintwoxmed}[1]{\IfEqCase{#1}{{GW190930A}{0.00}{GW190929A}{0.0009}{GW190924A}{0.00}{GW190915A}{0.00}{GW190910A}{0.0003}{GW190909A}{0.0003}{GW190828B}{0.00}{GW190828A}{0.002}{GW190814A}{-0.01}{GW190803A}{0.006}{GW190731A}{0.003}{GW190728A}{0.001}{GW190727A}{0.002}{GW190720A}{0.002}{GW190719A}{0.00}{GW190708A}{0.002}{GW190707A}{0.0004}{GW190706A}{0.003}{GW190701A}{0.003}{GW190630A}{0.00}{GW190620A}{0.00006}{GW190602A}{0.00}{GW190527A}{0.005}{GW190521B}{0.00}{GW190521A}{0.002}{GW190519A}{0.0006}{GW190517A}{0.001}{GW190514A}{0.001}{GW190513A}{0.00}{GW190512A}{0.0009}{GW190503A}{0.001}{GW190426A}{0.00}{GW190425A}{0.0006}{GW190424A}{0.00}{GW190421A}{0.00}{GW190413B}{0.0007}{GW190413A}{0.00}{GW190412A}{-0.01}{GW190408A}{0.002}}}
\newcommand{\spintwoxplus}[1]{\IfEqCase{#1}{{GW190930A}{0.52}{GW190929A}{0.59}{GW190924A}{0.48}{GW190915A}{0.60}{GW190910A}{0.54}{GW190909A}{0.57}{GW190828B}{0.54}{GW190828A}{0.51}{GW190814A}{0.59}{GW190803A}{0.57}{GW190731A}{0.58}{GW190728A}{0.48}{GW190727A}{0.55}{GW190720A}{0.57}{GW190719A}{0.56}{GW190708A}{0.43}{GW190707A}{0.46}{GW190706A}{0.54}{GW190701A}{0.56}{GW190630A}{0.48}{GW190620A}{0.56}{GW190602A}{0.60}{GW190527A}{0.59}{GW190521B}{0.51}{GW190521A}{0.69}{GW190519A}{0.55}{GW190517A}{0.53}{GW190514A}{0.59}{GW190513A}{0.54}{GW190512A}{0.49}{GW190503A}{0.58}{GW190426A}{0.00}{GW190425A}{0.47}{GW190424A}{0.59}{GW190421A}{0.58}{GW190413B}{0.60}{GW190413A}{0.57}{GW190412A}{0.56}{GW190408A}{0.53}}}
\newcommand{\networkoptimalsnrminus}[1]{\IfEqCase{#1}{{GW190814A}{1.7}{GW190426A}{1.8}{GW190425A}{1.7}}}
\newcommand{\networkoptimalsnrmed}[1]{\IfEqCase{#1}{{GW190814A}{24.7}{GW190426A}{8.3}{GW190425A}{12.0}}}
\newcommand{\networkoptimalsnrplus}[1]{\IfEqCase{#1}{{GW190814A}{1.7}{GW190426A}{1.8}{GW190425A}{1.7}}}
\newcommand{\networkmatchedfiltersnrminus}[1]{\IfEqCase{#1}{{GW190814A}{0.2}{GW190426A}{0.6}{GW190425A}{0.4}{GW190412A}{0.4}}}
\newcommand{\networkmatchedfiltersnrmed}[1]{\IfEqCase{#1}{{GW190814A}{24.9}{GW190426A}{8.7}{GW190425A}{12.4}{GW190412A}{19.0}}}
\newcommand{\networkmatchedfiltersnrplus}[1]{\IfEqCase{#1}{{GW190814A}{0.1}{GW190426A}{0.5}{GW190425A}{0.3}{GW190412A}{0.2}}}
\newcommand{\logpriorminus}[1]{\IfEqCase{#1}{{GW190426A}{10.5}{GW190425A}{8.6}}}
\newcommand{\logpriormed}[1]{\IfEqCase{#1}{{GW190426A}{161.3}{GW190425A}{98.4}}}
\newcommand{\logpriorplus}[1]{\IfEqCase{#1}{{GW190426A}{8.6}{GW190425A}{6.7}}}
\newcommand{\PEpercentBNS}[1]{\IfEqCase{#1}{{GW190930A}{0}{GW190929A}{0}{GW190924A}{0}{GW190915A}{0}{GW190910A}{0}{GW190909A}{0}{GW190828B}{0}{GW190828A}{0}{GW190814A}{0}{GW190803A}{0}{GW190731A}{0}{GW190728A}{0}{GW190727A}{0}{GW190720A}{0}{GW190719A}{0}{GW190708A}{0}{GW190707A}{0}{GW190706A}{0}{GW190701A}{0}{GW190630A}{0}{GW190620A}{0}{GW190602A}{0}{GW190527A}{0}{GW190521B}{0}{GW190521A}{0}{GW190519A}{0}{GW190517A}{0}{GW190514A}{0}{GW190513A}{0}{GW190512A}{0}{GW190503A}{0}{GW190426A}{1}{GW190425A}{100}{GW190424A}{0}{GW190421A}{0}{GW190413B}{0}{GW190413A}{0}{GW190412A}{0}{GW190408A}{0}}}
\newcommand{\PEpercentNSBH}[1]{\IfEqCase{#1}{{GW190930A}{0}{GW190929A}{0}{GW190924A}{4}{GW190915A}{0}{GW190910A}{0}{GW190909A}{0}{GW190828B}{0}{GW190828A}{0}{GW190814A}{100}{GW190803A}{0}{GW190731A}{0}{GW190728A}{0}{GW190727A}{0}{GW190720A}{0}{GW190719A}{0}{GW190708A}{0}{GW190707A}{0}{GW190706A}{0}{GW190701A}{0}{GW190630A}{0}{GW190620A}{0}{GW190602A}{0}{GW190527A}{0}{GW190521B}{0}{GW190521A}{0}{GW190519A}{0}{GW190517A}{0}{GW190514A}{0}{GW190513A}{0}{GW190512A}{0}{GW190503A}{0}{GW190426A}{99}{GW190425A}{0}{GW190424A}{0}{GW190421A}{0}{GW190413B}{0}{GW190413A}{0}{GW190412A}{0}{GW190408A}{0}}}
\newcommand{\PEpercentBBH}[1]{\IfEqCase{#1}{{GW190930A}{100}{GW190929A}{100}{GW190924A}{96}{GW190915A}{100}{GW190910A}{100}{GW190909A}{100}{GW190828B}{100}{GW190828A}{100}{GW190814A}{0}{GW190803A}{100}{GW190731A}{100}{GW190728A}{100}{GW190727A}{100}{GW190720A}{100}{GW190719A}{100}{GW190708A}{100}{GW190707A}{100}{GW190706A}{100}{GW190701A}{100}{GW190630A}{100}{GW190620A}{100}{GW190602A}{100}{GW190527A}{100}{GW190521B}{100}{GW190521A}{100}{GW190519A}{100}{GW190517A}{100}{GW190514A}{100}{GW190513A}{100}{GW190512A}{100}{GW190503A}{100}{GW190426A}{0}{GW190425A}{0}{GW190424A}{100}{GW190421A}{100}{GW190413B}{100}{GW190413A}{100}{GW190412A}{100}{GW190408A}{100}}}
\newcommand{\PEpercentMassGap}[1]{\IfEqCase{#1}{{GW190930A}{0}{GW190929A}{0}{GW190924A}{0}{GW190915A}{0}{GW190910A}{0}{GW190909A}{0}{GW190828B}{0}{GW190828A}{0}{GW190814A}{0}{GW190803A}{0}{GW190731A}{0}{GW190728A}{0}{GW190727A}{0}{GW190720A}{0}{GW190719A}{0}{GW190708A}{0}{GW190707A}{0}{GW190706A}{0}{GW190701A}{0}{GW190630A}{0}{GW190620A}{0}{GW190602A}{0}{GW190527A}{0}{GW190521B}{0}{GW190521A}{0}{GW190519A}{0}{GW190517A}{0}{GW190514A}{0}{GW190513A}{0}{GW190512A}{0}{GW190503A}{0}{GW190426A}{0}{GW190425A}{0}{GW190424A}{0}{GW190421A}{0}{GW190413B}{0}{GW190413A}{0}{GW190412A}{0}{GW190408A}{0}}}

\newcolumntype{Y}{>{\centering\arraybackslash}X}

\newacro{ADI}[ADI]{accretion disk instability}
\newacro{BAT}[BAT]{Burst Alert Telescope}
\newacro{BH}[BH]{black hole}
\newacro{BNS}[BNS]{binary neutron star}
\newacro{CSG}[CSG]{circular sine-Gaussian}
\newacro{GBM}[GBM]{Gamma-ray Burst Monitor}
\newacro{GCN}[GCN]{Gamma-ray Coordinates Network}
\newacro{GRB}[GRB]{gamma-ray burst}
\newacro{GW}[GW]{gravitational wave}
\newacro{NS}[NS]{neutron star}
\newacro{NSBH}[NSBH]{neutron star--black hole}
\newacro{O1}[O1]{the first observing run of Advanced LIGO and Advanced Virgo}
\newacro{O2}[O2]{the second observing run of Advanced LIGO and Advanced Virgo}
\newacro{O3a}[O3a]{the first part of the third observing run of Advanced LIGO and Advanced Virgo}
\newacro{SNR}[S/N]{signal-to-noise ratio}
\newacro{VALID}[VALID]{Vetting Automation and Literature Informed Database}

\newcommand{\LALSuite}{\texttt{LALSuite}\xspace}
\newcommand{\PYGRB}{\texttt{PyGRB}\xspace}
\newcommand{\PYCBC}{\texttt{PyCBC}\xspace}
\newcommand{\Xpipeline}{\texttt{X-Pipeline}\xspace}
\newcommand{\Msun}{\ensuremath{M_{\sun}}\xspace}

\newcommand{\OThreeAStart}{{2019 April 1 15:00 UTC}\xspace}
\newcommand{\OThreeAEnd}{{2019 October 1 15:00 UTC}\xspace}
\newcommand{\CBCsInOThreeA}{{39}\xspace}
\newcommand{\PossibleNSBsInOThreeA}{{three}\xspace}
\newcommand{\NSBHAprOThreeA}{{GW190426}\xspace}
\newcommand{\NSBHAugOThreeA}{{GW190814}\xspace}
\newcommand{\BNSOThreeA}{{GW190425}\xspace}

\newcommand{\NSBHAprOThreeAMOne}{$\massonesourcemed{GW190426A}_{ -\massonesourceminus{GW190426A} }^{ +\massonesourceplus{GW190426A} }$\xspace}
\newcommand{\NSBHAprOThreeAMTwo}{$\masstwosourcemed{GW190426A}_{ -\masstwosourceminus{GW190426A} }^{ +\masstwosourceplus{GW190426A} }$\xspace}
\newcommand{\NSBHAugOThreeAMOne}{$\massonesourcemed{GW190814A}_{ -\massonesourceminus{GW190814A} }^{ +\massonesourceplus{GW190814A} }$\xspace}
\newcommand{\NSBHAugOThreeAMTwo}{$\masstwosourcemed{GW190814A}_{ -\masstwosourceminus{GW190814A} }^{ +\masstwosourceplus{GW190814A} }$\xspace}
\newcommand{\BNSOThreeAMOne}{$\massonesourcemed{GW190425A}_{ -\massonesourceminus{GW190425A} }^{ +\massonesourceplus{GW190425A} }$\xspace}
\newcommand{\BNSOThreeAMTwo}{$\masstwosourcemed{GW190425A}_{ -\masstwosourceminus{GW190425A} }^{ +\masstwosourceplus{GW190425A} }$\xspace}
\newcommand{\BNSOThreeAArea}{{\ensuremath{>8000\,\mathrm{deg}^2}}\xspace}
\newcommand{\BNSOThreeALumDist}{{\ensuremath{\sim 160\,\mathrm{Mpc}}}\xspace}
\newcommand{\HBNSRange}{{\ensuremath{108\,\mathrm{Mpc}}\xspace}}
\newcommand{\LBNSRange}{{\ensuremath{135\,\mathrm{Mpc}}\xspace}}
\newcommand{\VBNSRange}{{\ensuremath{45\,\mathrm{Mpc}}\xspace}}
\newcommand{\DoubleDF}{{\ensuremath{81.9\,\%}\xspace}}
\newcommand{\SingleDF}{{\ensuremath{96.9\,\%}\xspace}}

\newcommand{\nAllGRB}{{141}\xspace}
\newcommand{\nLongRun}{{108}\xspace}
\newcommand{\nAmbiguousRun}{{13}\xspace}
\newcommand{\nShortRun}{{20}\xspace}
\newcommand{\nShortAndAmbiguousRun}{{33}\xspace}
\newcommand{\NoDataPyGRB}{{GRB 190605974}}
\newcommand{\nNearbyGalaxy}{{four}\xspace}
\newcommand{\NearbyGRBPosErr}{{\ensuremath{1^\prime\hspace{-0.8ex}.\hspace{0.1ex}9}}\xspace}
\newcommand{\NearbyGRBGalaxyLumDist}{{\ensuremath{165\,\mathrm{Mpc}}}\xspace}
\newcommand{\NearbyGRBGalaxyRedshift}{{\ensuremath{z=0.037}}\xspace}
\newcommand{\NearbyGRBSigma}{{\ensuremath{2.21\sigma}}\xspace}
\newcommand{\nCBC}{{32}\xspace}
\newcommand{\FracCBC}{{97.0\%}\xspace}
\newcommand{\nBurst}{{105}\xspace}
\newcommand{\FracBurst}{{74.5\%}\xspace}

\newcommand{\pvalBurstLowest}{{\ensuremath{5.5\times 10^{-3}}}\xspace}
\newcommand{\nameBurstLowest}{{GRB 190804058}\xspace}
\newcommand{\pvalBurst}{{0.31}\xspace}
\newcommand{\pvalBurstOOne}{{0.75}\xspace}
\newcommand{\pvalBurstOTwo}{{0.75}\xspace}
\newcommand{\DCSGSeventy}{{146}\xspace}
\newcommand{\DCSGHundred}{{104}\xspace}
\newcommand{\DCSGOneFifty}{{73}\xspace}
\newcommand{\DCSGThreeHundred}{{28}\xspace}
\newcommand{\DADIA}{{23}\xspace}
\newcommand{\DADIB}{{123}\xspace}
\newcommand{\DADIC}{{28}\xspace}
\newcommand{\DADID}{{11}\xspace}
\newcommand{\DADIE}{{33}\xspace}

\newcommand{\pvalCBC}{{0.43}\xspace}
\newcommand{\pvalCBCOOne}{{0.57}\xspace}
\newcommand{\pvalCBCOTwo}{{0.30}\xspace}
\newcommand{\DBNS}{{119}\xspace}
\newcommand{\DNSBHGen}{{160}\xspace}
\newcommand{\DNSBHAli}{{231}\xspace}
\newcommand{\pvalCBCLowestFAP}{{\ensuremath{2.7\times 10^{-2}}}\xspace}
\newcommand{\nameCBCLowestFAP}{{GRB 190601325}\xspace}
\newcommand{\BNSExclJuneTen}{{\ensuremath{63\,\mathrm{Mpc}}}\xspace}
\newcommand{\NSBHExclJuneTen}{{\ensuremath{82\,\mathrm{Mpc}}}\xspace}
\newcommand{\NSBHAlignExclJuneTen}{{\ensuremath{114\,\mathrm{Mpc}}}\xspace}

\newcommand\blfootnote[2]{
  \begingroup
  \renewcommand\thefootnote{}\footnote{\noindent{}$^{#1}$~#2}
  \addtocounter{footnote}{#1}
  \addtocounter{footnote}{-1}
  \endgroup
}

\received{2020 October 13}
\revised{2021 March 10}
\accepted{2021 March 10}
\published{2021 July 9}

\begin{document}

\title{Search for Gravitational Waves Associated with Gamma-Ray Bursts
  Detected by Fermi and Swift during the LIGO--Virgo Run
  O3a}

\AuthorCollaborationLimit=3000

\author{R.~Abbott}
\affiliation{LIGO, California Institute of Technology, Pasadena, CA 91125, USA}
\author{T.~D.~Abbott}
\affiliation{Louisiana State University, Baton Rouge, LA 70803, USA}
\author{S.~Abraham}
\affiliation{Inter-University Centre for Astronomy and Astrophysics, Pune 411007, India}
\author{F.~Acernese}
\affiliation{Dipartimento di Farmacia, Universit\`a di Salerno, I-84084 Fisciano, Salerno, Italy}
\affiliation{INFN, Sezione di Napoli, Complesso Universitario di Monte S.Angelo, I-80126 Napoli, Italy}
\author{K.~Ackley}
\affiliation{OzGrav, School of Physics \& Astronomy, Monash University, Clayton 3800, Victoria, Australia}
\author{C.~Adams}
\affiliation{LIGO Livingston Observatory, Livingston, LA 70754, USA}
\author{R.~X.~Adhikari}
\affiliation{LIGO, California Institute of Technology, Pasadena, CA 91125, USA}
\author{V.~B.~Adya}
\affiliation{OzGrav, Australian National University, Canberra, Australian Capital Territory 0200, Australia}
\author{C.~Affeldt}
\affiliation{Max Planck Institute for Gravitational Physics (Albert Einstein Institute), D-30167 Hannover, Germany}
\affiliation{Leibniz Universit\"at Hannover, D-30167 Hannover, Germany}
\author{M.~Agathos}
\affiliation{Theoretisch-Physikalisches Institut, Friedrich-Schiller-Universit\"at Jena, D-07743 Jena, Germany}
\affiliation{University of Cambridge, Cambridge CB2 1TN, UK}
% ...

\collaboration{The LIGO Scientific Collaboration and the Virgo Collaboration$^{203}$}

\date{\today}

\begin{abstract}
  We search for gravitational-wave transients associated with
  gamma-ray bursts (GRBs) detected by the Fermi and Swift
  satellites during the first part of the third observing run of
  Advanced LIGO and Advanced Virgo (\OThreeAStart~-- \OThreeAEnd).
  A total of \nBurst GRBs were analyzed using a search for generic
  gravitational-wave transients; \nCBC GRBs were analyzed
  with a search that specifically targets neutron star binary mergers
  as short GRB progenitors. 
  We find no significant evidence for gravitational-wave signals
  associated with the GRBs that we followed up, nor for a
  population of unidentified subthreshold signals.  We consider
  several source types and signal morphologies, and report for these
  lower bounds on the distance to each GRB.
\end{abstract}
\section{Introduction}
\label{introducion}

\blfootnote{203}{Please direct all correspondence to LSC Spokesperson at \href{mailto:lsc-spokesperson@ligo.org}{lsc-spokesperson@ligo.org}, or Virgo Spokesperson at \href{mailto:virgo-spokesperson@ego-gw.it}{virgo-spokesperson@ego-gw.it}.}

Gamma-ray bursts (GRBs)\acused{GRB} are transient flashes of
gamma radiation of cosmological origin observed at a
rate of ${\gtrsim}1$ per day~\citep{Nakar:2007yr}.  The interaction of
matter with a compact central object, e.g., an accreting
\acused{BH}black hole~\citep[\ac{BH};][]{Woosley:1993wj,
  Popham:1998ab} or a magnetar~\citep{Usov:1992zd, Zhang:2000wx}, is
believed to drive highly relativistic jets which power the prompt
emission of these astrophysical events.  \acp{GRB} are broadly grouped
into two classes --- long and short \acp{GRB} --- depending on the
duration and spectral hardness of their prompt
emission~\citep{Mazets:1981, Norris:1983, Kouveliotou:1993yx}.

Long, soft \acp{GRB} have durations ${\gtrsim}2$\,s and are firmly
associated by optical observations to the collapse of massive
stars~\citep{Galama:1998ea, Hjorth:2003jt, Stanek:2003tw,
  Hjorth:2011zx}.  Gravitational waves \acused{GW} (\acp{GW}) will be
radiated by the core-collapse process, \citep[e.g.,][]{Fryer:2011zz}.
Several models of this process do not yield radiation that is
detectable by the current generation of \ac{GW} interferometers beyond
Galactic distances~\citep{Abbott:2019pxc}.  However,
rotational instabilities and instabilities induced by the additional
presence of an accretion disk as part of the \ac{GRB} engine may
enhance the \ac{GW} emission, making it detectable even for
extragalactic sources~\citep{vanPutten:2001gi,
  Davies:2002hj, Fryer:2001zw, Kobayashi:2002by, Shibata:2003yj,
  Piro:2006ja, Corsi:2009jt, Romero:2010ze, Gossan:2015xda,
  Abbott:2019pxc}.

The unambiguous association~\citep{Monitor:2017mdv} of \ac{NS} binary
merger GW170817~\citep{TheLIGOScientific:2017qsa, Abbott:2018wiz} and
short GRB 170817A~\citep{Goldstein:2017mmi, Savchenko:2017ffs} has
confirmed that compact binary mergers of this kind can produce short
\acp{GRB}.  This milestone in multimessenger astronomy corroborated
the idea first proposed in the 1980s~\citep{Blinnikov:2018boq,
  Paczynski:1986px, Eichler:1989ve, Paczynski:1991aq, Narayan:1992iy}
that the progenitors of short \acp{GRB} are compact binaries
containing \acp{NS}~\citep[for a review of proposed progenitors,
see][]{Lee:2007js, Nakar:2007yr}.  Indirect evidence that had
previously reinforced this idea was due to the observation of a
possible kilonova associated with GRB~130603B~\citep{Berger:2013wna,
  Tanvir:2013pia}, and to numerous studies of the environments of
short \acp{GRB}~\citep[for reviews see][]{Berger:2010qx,
  Berger:2013jza}, starting with the afterglow observation and
host-galaxy association of GRB 050509B~\citep{CastroTirado:2005yk, Gehrels:2005qk, Bloom:2005qx}.

In addition to confirming the origin of \emph{some} short \acp{GRB},
combining data from observations of GW170817 and \ac{GRB} 170817A
allowed for the inference of \emph{basic} properties of short \ac{GRB}
jets.  These include the isotropic equivalent luminosity of the jet,
determined through a redshift measurement made possible by the optical
follow-up of the \ac{GW}
localization~\citep{Monitor:2017mdv,
  Goldstein:2017mmi}, and the geometry of the \ac{GRB}
jets~\citep{Williams_2018, Farah:2019tue, Mogushi:2018ufy}.  The
precise mechanism by which the jet is launched is still unknown,
although it is typically believed to be either neutrino-driven or
magnetically driven (\citeauthor{Nakar:2007yr}~\citeyear[][but see
also]{Nakar:2007yr}~\citeauthor{Liu:2015lfa}~\citeyear[][and
references therein]{Liu:2015lfa}).  Indeed, the scientific debate
about the emission profile of the jet and the subsequent gamma-ray
production mechanism of \ac{GRB} 170817A is still
ongoing~\citep{Hallinan:2017woc, Kasliwal:2017ngb, Lamb:2017ych,
  Troja:2017nqp, Gottlieb:2017pju, Lazzati:2017zsj, Gill:2018kcw, Mooley:2018dlz,
  Zhang:2017lpb, Ghirlanda:2018uyx}.  It is generally
believed that there are symmetric polar outflows of highly
relativistic material that travel parallel to the total angular
momentum of the binary system~\citep{Aloy:2004nh, Kumar:2014upa,
  Murguia-Berthier:2016fys}.  These jets are thought to be collimated
and roughly axisymmetric, emitting preferentially in a narrow opening
angle due to a combination of outflow geometry and relativistic
beaming.
The data from extensive multi-wavelength observation campaigns that
ran for nearly 20 months following the merger~\citep{Fong:2019vgn,
  2020arXiv200602382M, Troja:2020pzf} are in agreement with a
structured jet model, in which the energy and bulk Lorentz factor
gradually decrease with angular distance from the jet symmetry
axis~\citep[e.g.,][]{2001ApJ...552...72D, 2001ARep...45..236L, 
  2002MNRAS.332..945R, 2002ApJ...571..876Z, Ghirlanda:2018uyx,
  Beniamini:2020eza}.
Further, according to one of the models proposed, as the jet drills
through the surrounding merger ejecta it inflates a mildly
relativistic cocoon due to interactions between the material at the
edge of the jet and the ejecta~\citep{Lazzati:2016yxl,
  Gottlieb:2017mqv}.
In this case, it is possible that the cocoon alone could produce the
gamma-rays observed from GRB 170817A~\citep{Gottlieb:2017pju}.
Additional joint detections of \acp{GRB} and \acp{GW} will
significantly aid in our understanding of the underlying
energetics~\citep{Lamb:2017ych, Wu:2018bxg, burns2019summary}, jet
geometry~\citep{Farah:2019tue, Mogushi:2018ufy, Biscoveanu:2019bpy,
  Hayes:2019hso}, and jet ignition
mechanisms~\citep{2018arXiv180207328V,2019PhRvD.100b3005C,
  2019FrPhy..1464402Z} of \ac{BNS} coalescences.

A targeted search for \acp{GW} in sky and time coincidence with
\acp{GRB} enhances our potential of achieving such joint detections.
In this paper we present our results for the targeted \ac{GW}
follow-up of \acp{GRB} reported during \ac{O3a} by the
Fermi~\citep{2009ApJ...702..791M} and
Swift~\citep{2004ApJ...611.1005G, 2005SSRv..120..143B} satellites.
As in the first~\citep{Abbott:2016cjt} and
second~\citep{Monitor:2017mdv, Authors:2019fue} observing runs, two
searches with different assumptions about signal morphology are
applied to the \ac{GW} data: we process all \acp{GRB} with a search
for generic \ac{GW} transients~\citep[\texttt{X-Pipeline};][see
Sec.\,\ref{sec:burst-search} for details]{Sutton:2009gi, Was:2012zq}
and we follow up short \acp{GRB} with an additional, modeled search
for \ac{BNS} and \ac{NSBH} \ac{GW} inspiral
signals~\citep[\texttt{PyGRB};][see Sec.\,\ref{sec:cbc-search} for
details]{Harry:2010fr, Williamson:2014wma}.  These searches were able
to process \nBurst and \nCBC \acp{GRB} in \ac{O3a}, respectively.

The scope of these targeted searches is to enhance our ability to
detect \ac{GW} signals in coincidence with \acp{GRB} with respect to
all-sky searches for transient \ac{GW} signals carried out by the
LIGO Scientific \& Virgo
  Collaboration~\citep{LIGOScientific:2018mvr, gwtc2}.  These may
lead to joint \ac{GW}--\ac{GRB} detections in the case of loud \ac{GW}
events, as for GW170817 and GRB 170817A, but the targeted searches we
report on here aim at uncovering sub-threshold \ac{GW} signals by
exploiting the time and localization information of the \acp{GRB}
themselves.  The Fermi \ac{GBM} team conducts an analogous
effort when searching through \ac{GBM} data for gamma-ray transients
coincident with confirmed events and low-significance candidates
reported by LIGO--Virgo offline
analyses~\citep{Hamburg:2020dtg}. Similarly, the {\it Swift}/\ac{BAT}
team has developed their own autonomous pipeline to enable
subthreshold \ac{GRB} searches for externally triggered
events~\citep{2020ApJ...900...35T}.

This first part of the third observing run took place between
\OThreeAStart and \OThreeAEnd.  Setting the false-alarm-rate threshold
to two per year, \CBCsInOThreeA compact binary coalescence events were
identified in \ac{O3a}~\citep{gwtc2}.  The majority of these have been
classified as signals emitted by binary \ac{BH} mergers; however,
\PossibleNSBsInOThreeA events have the possibility of coming from a
binary with at least one \ac{NS}, that is, a potential short \ac{GRB}
progenitor.
\begin{enumerate}
\item \BNSOThreeA~\citep{Abbott:2020uma} was a compact binary
  coalescence with primary mass \BNSOThreeAMOne and secondary mass
  \BNSOThreeAMTwo (all measurements quoted at the 90\% credible level)
  and is therefore consistent with being the result of a \ac{BNS}
  merger~\citep{Abbott:2020uma, gwtc2}.
\item \NSBHAprOThreeA was the \ac{GW} candidate event with the highest
  false-alarm rate in~\citet{gwtc2}; assuming it is a real signal, its
  inferred component masses of \NSBHAprOThreeAMOne and
  \NSBHAprOThreeAMTwo indicate that it may have originated from an
  \ac{NSBH}, or a binary \ac{BH} merger.
\item \NSBHAugOThreeA~\citep{Abbott:2020khf} could have originated
  from an \ac{NSBH}, or a binary \ac{BH} merger, as it has a primary
  mass measurement of \NSBHAugOThreeAMOne and posterior support for a
  secondary mass \NSBHAugOThreeAMTwo.  This makes the secondary
  compact object either the lightest \ac{BH} or the heaviest \ac{NS}
  known to be in a compact binary system.
\end{enumerate}
While there is considerable uncertainty in source type for all three
of these events, \BNSOThreeA is the one for which the prospects of
observing an associated \ac{GRB} were most promising, as it is
consistent with a \ac{BNS} merger, rather than a binary \ac{BH} merger
or an \ac{NSBH} merger with high or moderately high mass ratio.
However, no confirmed electromagnetic or neutrino counterparts were
observed in association with this
event~(\citeauthor{Hosseinzadeh:2019ifm}~\citeyear{Hosseinzadeh:2019ifm},
\citeauthor{Lundquist:2019cty}~\citeyear{Lundquist:2019cty},
\citeauthor{Abbott:2020uma}~\citeyear{Abbott:2020uma},
\citeauthor{Coughlin:2019zqi}~\citeyear{Coughlin:2019zqi}; see
also~\citeauthor{Pozanenko:2019lwh}~\citeyear{Pozanenko:2019lwh})
despite extensive searches, which are logged in the \ac{GCN} Circular
archive.\footnote{All \ac{GCN} Circulars related to this event are
  archived at \burl{https://gcn.gsfc.nasa.gov/other/S190425z.gcn3}.}
There are a number of reasons for which an electromagnetic counterpart
associated with \BNSOThreeA may not have been detected.  First, the
large area covered by the localization region of \BNSOThreeA
determined from \ac{GW} data (\BNSOThreeAArea) posed a considerable
challenge for electromagnetic follow-up.  45.4\% of this localization
region was occulted by the Earth for the Fermi satellite, so, if
gamma-rays were emitted from the source, it is possible they were not
detectable.  Other gamma-ray observatories with lower sensitivities to
short \acp{GRB}, such as INTEGRAL and KONUS-Wind, were covering
relevant fractions of the localization region,
however~\citep{GCN:24169,GCN:24417}.  Second, \ac{GRB} jets are
expected to be aligned with the total angular momentum of the binary
system, and thus more easily detectable at small viewing angles.  The
binary inclination angle of \BNSOThreeA was poorly constrained, so it
is possible that a jet from this system was formed but was oriented
away from our line of sight.  Additionally, the luminosity distance
inferred for \BNSOThreeA (\BNSOThreeALumDist) was significantly larger
than that for GW170817 ($\sim 40$\,Mpc).  \ac{GRB} 170817A, which
followed GW170817, was such an exceptionally faint short
\ac{GRB}~\citep{Monitor:2017mdv} that its prompt emission photon flux
would have dipped below the detection threshold for
Fermi-\ac{GBM}, had the source been farther than $\sim
75$\,Mpc~\citep{Monitor:2017mdv, Goldstein:2017mmi}, and by $\sim
100$\,Mpc it would become undetectable by
Swift/BAT~\citep{2020ApJ...900...35T}. Thus, if emission from the
system that produced \BNSOThreeA was similarly faint, it would not
have been detectable by Swift/BAT or 
Fermi-\ac{GBM}. Therefore, we do not necessarily expect a \ac{GRB}
detection to be associated with \BNSOThreeA due to its almost
unconstrained inclination angle, large localization region, and
distance, even if gamma-rays were emitted from this system.  Scenarios
like this one further motivate the need for \ac{GW} follow-up analyses
of \ac{GRB} events which, by definition, constrain the sky
localization and inclination angle of the progenitor.

In Section~\ref{sec:grb_sample} we discuss the set of \acp{GRB}
analyzed in this paper. In Section~\ref{sec:search_methods}, we
summarize the two targeted search methods used to follow up \acp{GRB}.
Section~\ref{sec:results} presents the results obtained with these two
methods.
We also consider each of the two sets of results collectively and
quantify its consistency with the no-signal hypothesis.  Finally, in
Section~\ref{sec:conclusions} we provide our concluding remarks.
\section{GRB Sample}
\label{sec:grb_sample}

The sample of \acp{GRB} analyzed in this paper includes events
circulated by the \ac{GCN},\footnote{GCN Circulars Archive:
  \url{http://gcn.gsfc.nasa.gov/gcn3_archive.html}.} complemented
with information from the Swift/\ac{BAT}
catalog~\citep{Lien:2016zny},\footnote{Swift/\ac{BAT} Gamma-Ray
  Burst Catalog: \url{http://swift.gsfc.nasa.gov/results/batgrbcat/}.}
the online Swift \acp{GRB} Archive,\footnote{Swift
  \ac{GRB} Archive:
  \url{http://swift.gsfc.nasa.gov/archive/grb_table/}.} and the
  Fermi \ac{GBM} Catalog.\footnote{FERMIGBRST - Fermi GBM Burst
  Catalog:
  \url{https://heasarc.gsfc.nasa.gov/W3Browse/fermi/fermigbrst.html}.}~\citep{Gruber:2014iza,
  vonKienlin:2014nza, Bhat:2016odd} Once an alert detailing an event
has been received via the \ac{GCN}, the dedicated
\acl{VALID}~\citep{RCoyneThesis2015} is
applied to find the latest \ac{GRB} results by comparing the time and
localization parameters with those in tables relating to each
satellite, the published catalogs, and an automatic literature search.
The \ac{GCN} notices provide a set of \nAllGRB \acp{GRB} during the
\ac{O3a} data-taking period (\OThreeAStart~-- \OThreeAEnd).

As mentioned in the Introduction, we carry out two searches with
distinct assumptions about signal morphology (see
Sec.\,\ref{sec:search_methods} for details on both methods): a search
for generic \ac{GW} transients and a modeled search for \ac{GW}
signals from \ac{NS} binary inspirals, i.e., \ac{BNS} and \ac{NSBH}.
We do this because \acp{GRB} of different durations are expected to
have different origins and therefore different \ac{GW} signal
morphologies.  In particular, if a compact binary merger were to
produce a \ac{GRB} it would be expected to have a short duration.  In
order to specifically target such phenomena with the modeled search,
we classify each \ac{GRB} as {\it long}, {\it short}, or {\it
  ambiguous}. This classification relies on the measurement of the
time interval over which $90\%$ of the total background-subtracted
photon counts are observed ($T_{90}$, with error $|\delta T_{90}|$).
When $T_{90}+|\delta T_{90}| < 2$\,s the \acp{GRB} are labeled as
short, when $T_{90}-|\delta T_{90}| > 4$\,s the \acp{GRB} are
labeled as long, the rest are labeled as ambiguous. The
unmodeled search for generic transients is applied to \acp{GRB} of
all classifications. All of the short and ambiguous \acp{GRB} are
additionally analyzed with the modeled search in order to maximize
the chances of uncovering any potential binary coalescence candidate.

The classification process results in \nShortRun short \acp{GRB},
\nLongRun long \acp{GRB}, and \nAmbiguousRun ambiguous \acp{GRB}.  As
in~\citet{Authors:2019fue}, we require a minimum amount of coincident
data from at least two \ac{GW} detectors around the time of a \ac{GRB}
for the generic unmodeled \ac{GW} transient search to assess the
significance of a \ac{GW} candidate with sub-percent level accuracy
(see Sec.\,\ref{sec:burst-search} for technical details).  This
requirement is applied to \acp{GRB} of all classifications and results
in \nBurst \acp{GRB} being analyzed with this method, out of the
\nAllGRB \acp{GRB} recorded by Fermi and Swift during
\ac{O3a}.  This amounts to \FracBurst, a percentage of events that is
compatible with the fraction of observing time during which at least
two interferometers in the network were operating in observing
mode~\citep[\DoubleDF;][]{gwtc2}.  Similarly, requirements from the
modeled search (see Sec.\,\ref{sec:cbc-search} for technical details)
set the minimum amount of data needed from at least one detector
around the time of the \acp{GRB}.  It leads to \nCBC short and
ambiguous \acp{GRB} being analyzed with this method,\footnote{The
  single \ac{GRB} we were unable to follow up with the modeled search
  is \NoDataPyGRB.  The GRBs we were unable to analyze with either of
  the searches are: GRB 190401139,
GRB 190406745,
GRB 190411407,
GRB 190422A,
GRB 190424A,
GRB 190508808,
GRB 190515B,
GRB 190530430,
GRB 190531840,
GRB 190604B,
GRB 190605974,
GRB 190607071,
GRB 190609315,
GRB 190611A,
GRB 190611950,
GRB 190622368,
GRB 190626254,
GRB 190706B,
GRB 190714573,
GRB 190716917,
GRB 190719113,
GRB 190723309,
GRB 190731943,
GRB 190804792,
GRB 190806675,
GRB 190808498,
GRB 190814837,
GRB 190821A,
GRB 190821716,
GRB 190828614
.} that is, \FracCBC of
the \nShortAndAmbiguousRun possible ones.  This value matches the
fraction of observing time during which at least one interferometer in
the network was operating in observing mode during
\ac{O3a}~\citep[\SingleDF;][]{gwtc2}.

Of the \nAllGRB ~Fermi and Swift \acp{GRB} in our sample,
the vast majority do not have redshift measurements. Those that do are
the ambiguous \ac{GRB} 190627A at $z=1.942$~\citep{GCN:24916}, and the
two long \acp{GRB} 190719C and 190829A at $z=2.469$ and $z=0.0785$,
respectively~\citep{GCN:25252, GCN:25565}.  All three fall beyond the
detection range of our interferometers, and are not expected to
produce measurable \ac{GW} results. Regardless of availability of
redshift information, however, we followed up as many \acp{GRB} as we
could and we were indeed able to analyze these three cases.
\section{Search Methods} \label{sec:search_methods}

We now provide a description of the two targeted search methods used
in this paper.  These are the same methods applied to \ac{GW} data
coincident with \acp{GRB} that occurred during the
first~\citep{Abbott:2016cjt} and second~\citep{Monitor:2017mdv,
  Authors:2019fue} Advanced LIGO and Virgo observing runs.  In
Sec.\,\ref{sec:cbc-search} we summarize the modeled search method
that aims at uncovering sub-threshold \ac{GW} signals emitted by
\ac{BNS} and \ac{NSBH} binaries \citep[\PYGRB;][]{Harry:2010fr,
  Williamson:2014wma}.  In Sec.\,\ref{sec:burst-search} we discuss the
search for generic \ac{GW}
transients~\citep[\Xpipeline;][]{Sutton:2009gi, Was:2012zq}.  Results
from these two searches are presented in Sec.\,\ref{sec:results}.

\subsection{Modeled search for binary mergers}
\label{sec:cbc-search}

This analysis searches for a \ac{GW} signal compatible with the
inspiral of a \ac{BNS} or \ac{NSBH} binary --- collectively \ac{NS}
binaries --- within 6\,s of data associated with an observed short
\ac{GRB}.  This stretch of data is the \textit{on-source window} and
runs from $-5\,$s to $+1\,$s around the start of the \ac{GRB} emission
(i.e., the \ac{GRB} trigger time).  The surrounding $\sim$30--90
minutes of data are split into 6\,s \textit{off-source trials} which
are also analyzed in order to build a background.  Around 30 minutes
allows the modeled search to accurately estimate the power spectral
density of the available instruments and ensures that it can assess at
sub-percent level accuracy the significance of any candidate events
found in the on-source window.
All the data are processed using \PYGRB~\citep{Harry:2010fr,
  Williamson:2014wma}, a coherent matched filtering pipeline that is
part of the general open-source software
\PYCBC~\citep{alex_nitz_2020_3961510} and has core elements in the
\LALSuite software library~\citep{LALSuite}.  We scan each trial of
data and the on-source window in the $30$--$1000$\,Hz frequency band
using a predefined bank of waveform templates~\citep{Owen:1998dk}
created with a hybrid geometric-stochastic
method~\citep{Capano:2016dsf, DalCanton:2017ala} and using a
phenomenological inspiral-merger-ringdown waveform model for
non-precessing point-particle
binaries~\citep[\texttt{IMRPhenomD};][]{Husa:2015iqa,
  Khan:2015jqa}.\footnote{All waveforms mentioned in this section are
  generated with the \texttt{LALSimulation} package that is part of
  the \texttt{LALSuite} software library~\citep{LALSuite}.}
The waveform template bank includes waveforms corresponding to a range
of masses ($[1.0, 2.8]\Msun$ for \acp{NS}, $[1.0, 25.0]\Msun$ for
\acp{BH}) and dimensionless spin magnitudes ($[0, 0.05]$ for \acp{NS},
$[0, 0.998]$ for \acp{BH}) for
aligned-spin, zero-eccentricity \ac{BNS} or \ac{NSBH}
  systems
that may produce an electromagnetic counterpart via the tidal
disruption of the \ac{NS}~\citep{Pannarale:2014rea}.  Aside from the
updated sensitivity of our detectors, the only difference with respect
to the second LIGO--Virgo observing run~\citep{Authors:2019fue} is that
the generation of the bank has been updated to apply more accurate
physics to determine whether an \ac{NSBH} system could produce an
accretion disk from this disruption~\citep{Foucart:2018rjc}.  We only
search for circularly polarized \acp{GW}, which may be emitted by
binaries with inclinations of $0^\circ$ or $180^\circ$: such systems
have \ac{GW} amplitudes that are consistent~\citep{Williamson:2014wma}
with those of binary progenitors with inclination angles over the full
range of viewing angles that we expect for typical brightness
\acp{GRB}~\citep[$\lesssim30^\circ$;][]{Fong:2015oha}, such as those in
our sample.

The strength of any potential signal is ranked via a coherent matched
filter \acl{SNR}~\citep[\acs{SNR}\acused{SNR};][]{Harry:2010fr,
  Williamson:2014wma} which is re-weighted according to a $\chi^2$
goodness-of-fit between the template that identified it and the signal
itself.  The significance of the latter is quantified as the
probability of background alone producing such an event.  This is
evaluated by comparing the re-weighted \ac{SNR} of the loudest trigger
within the $6\,$s on-source to the distribution of the re-weighted
\acp{SNR} of the loudest triggers in the $6\,$s off-source trials.
When data from more than one detector are available, this background
\ac{SNR} distribution is extended by generating additional off-source
trials via \textit{time slides}, that is, by combining data from
detectors after introducing time shifts longer than the light-travel
time across the network.  Specifically, our time shifts are $6$\,s
long, in order to match the width of the on-source window and the
off-source trials.

In order to derive the sensitivity of this search to potential
\ac{GRB} sources, simulated signals are injected in software into the
off-source data.  The 90\% exclusion distances, $D_{90}$, are defined
as the distances within which 90\% of the injected simulated signals
are recovered with a greater ranking statistic than the loudest
on-source event.  Three different astrophysical populations are
considered: \ac{BNS} binaries with generically oriented --- i.e.,
precessing --- spins, aligned spin \ac{NSBH} binaries, and \ac{NSBH}
binaries with generically oriented spins.  These simulated signals
cover a portion of parameter space that extends beyond that covered
by the template bank, as they include \ac{NS} dimensionless spin
values up to $0.4$ and, for two families of injected signals, admit
precession.  As stated previously, the templates used to filter the
data are produced using \texttt{IMRPhenomD}.  In order to factor into
the sensitivity assessment any potential loss due to uncertainties in
\ac{GW} signal modeling, the injected signals are not produced with
the same model used for the templates.  Precessing \ac{BNS} signals
are simulated using the TaylorT2 time-domain, post-Newtonian inspiral
approximant~\citep[\texttt{SpinTaylorT2};][]{Sathyaprakash:1991mt,
  Blanchet:1996pi, Mikoczi:2005dn, Arun:2008kb, Bohe:2013cla,
  Bohe:2015ana, Mishra:2016whh}, while \ac{NSBH}-injected waveforms
are generated assuming a point-particle effective-one-body model tuned
to numerical simulations which can allow for precession effects from
misaligned spins~\citep[\texttt{SEOBNRv3};][]{Pan:2013rra,
  Taracchini:2013rva, Babak:2016tgq}.  The three populations used to
build the injected signals are defined as in the first two LIGO--Virgo
observing runs, to allow for direct comparisons~\citep{Abbott:2016cjt,
  Authors:2019fue}.  \ac{NS} masses for the injections are taken
between 1~\Msun and 3~\Msun from a normal distribution centered at
1.4~\Msun with a standard deviation of
0.2~\Msun~\citep{Kiziltan:2013oja} and 0.4~\Msun for \ac{BNS} and
\ac{NSBH} systems, respectively.  \ac{BH} masses are taken to be
between 3~\Msun and 15~\Msun from a normal distribution centered at
10~\Msun with a standard deviation of 6~\Msun.  Spins are drawn
uniformly in magnitude and, when applicable, with random orientation;
the maximum allowed \ac{NS} spin magnitude is $0.4$, from the fastest
observed pulsar spin~\citep{Hessels:2006ze}, while the maximum \ac{BH}
spin magnitude is set to 0.98, motivated by X-ray binary
observations~\citep[e.g.,][]{Ozel:2010su, Kreidberg:2012ud,
  Miller:2014aaa}.  Injected signals have a range of total
inclinations from $0^\circ$--$30^\circ$ and $150^\circ$--$180^\circ$
whilst removing any systems which could not feasibly produce a short
\ac{GRB}~\citep{Pannarale:2014rea}.

\subsection{Unmodelled search for generic transients}
\label{sec:burst-search}

\Xpipeline looks for excess power that is coherent across the network
of \ac{GW} detectors and consistent with the sky localization and time
window for each \ac{GRB}.  As in the first two observing runs, we use
a search time window that begins 600\,s before the \ac{GRB} trigger
time and ends 60\,s after it, or at the $T_{90}$ time itself
(whichever is larger).  This window is long enough to encapsulate the
time delay between \ac{GW} emission from a progenitor and the \ac{GRB}
prompt emission~\citep{Koshut1995, Aloy:1999ai, MacFadyen:1999mk,
  Zhang:2002yk, Lazzati:2004af, Wang:2007nta, Burlon:2008yu,
  Burlon:2009tu, Lazzati:2009xx, Vedreenne2009book}. Our frequency
range is restricted to the most sensitive band of the \ac{GW}
detectors, namely 20--500\,Hz. While gravitational radiation from
core-collapse supernovae is expected to contain frequency content
above this band~\citep{Radice:2018usf}, detection of bursts above a
few hundred hertz is not energetically favorable \citep[see, e.g.,
Fig.\,4 in][]{Abbott:2019prv} and increasing the frequency upper limit
also increases the computational cost.

The generic transient search pipeline coherently combines data from
all detectors and produces time-frequency maps of this \ac{GW} data
stream.  The maps are scanned for clusters of pixels with excess
energy, referred to as \textit{events}.  The events obtained this way
are first ranked according to a detection statistic based on energy
and then subject to coherent consistency tests.  These are based on
correlations between data in different detectors and reject events
associated with noise transients.  The surviving event with the
largest ranking statistic is taken to be the best candidate for a
\ac{GW} detection.  Its significance is evaluated in the same way as
in the modeled analysis, but with 660\,s long off-source trials.  In
order to ensure that the significance is assessed at a sub-percent
level, we require at least $\sim 1.5$ hr of coincident data from at
least two detectors around the time of a \ac{GRB}.  Non-Gaussian noise
transients, or \textit{glitches}, are handled as described
in~\citet{Authors:2019fue}.

Similarly to the modeled search, we quantify the sensitivity of the
generic transient search by injecting simulated signals into
off-source data in software and recovering them.  Calibration errors
are accounted for by jittering the amplitude and arrival time of the
injections according to a Gaussian distribution representative of the
calibration uncertainties in \ac{O3a}~\citep{Abbott:2016cjt}.  We
report results obtained for four distinct sets of \ac{CSG} waveforms,
with fixed quality factor $Q=9$ and with central frequencies of
70, 100,
150, and 300\,Hz~\citep[see Equation~1 and
Section~3.2 of][]{Abbott:2016cjt}.  These models are intended to
represent the \acp{GW} from stellar collapses.  In all four cases, we
set the total radiated energy to $E_{\mathrm{GW}} = 10^{-2} \Msun
c^2$, a choice that is about an order of magnitude higher than
the results presented in~\citet{Abbott:2019pxc} for the detectability
of core-collapse supernovae.  As optimistic
representatives~\citep{ADIimplemtation} of longer-duration \ac{GW}
signals detectable by the unmodeled search, we use \ac{ADI}
waveforms~\citep{vanPutten:2001gi, vanPutten:2014kja}.  In these
\ac{ADI} models, instabilities form in a magnetically suspended torus
around a rapidly spinning \ac{BH}, causing \acp{GW} to be emitted.
The model specifics and parameters used to generate the five families
of \ac{ADI} signals that we consider are the same as in Table~1 and
Section~3.2 of~\citet{Abbott:2016cjt}.
\section{Results}
\label{sec:results}

\begin{figure}[!t]
  \begin{center}
    \includegraphics[width=\linewidth]{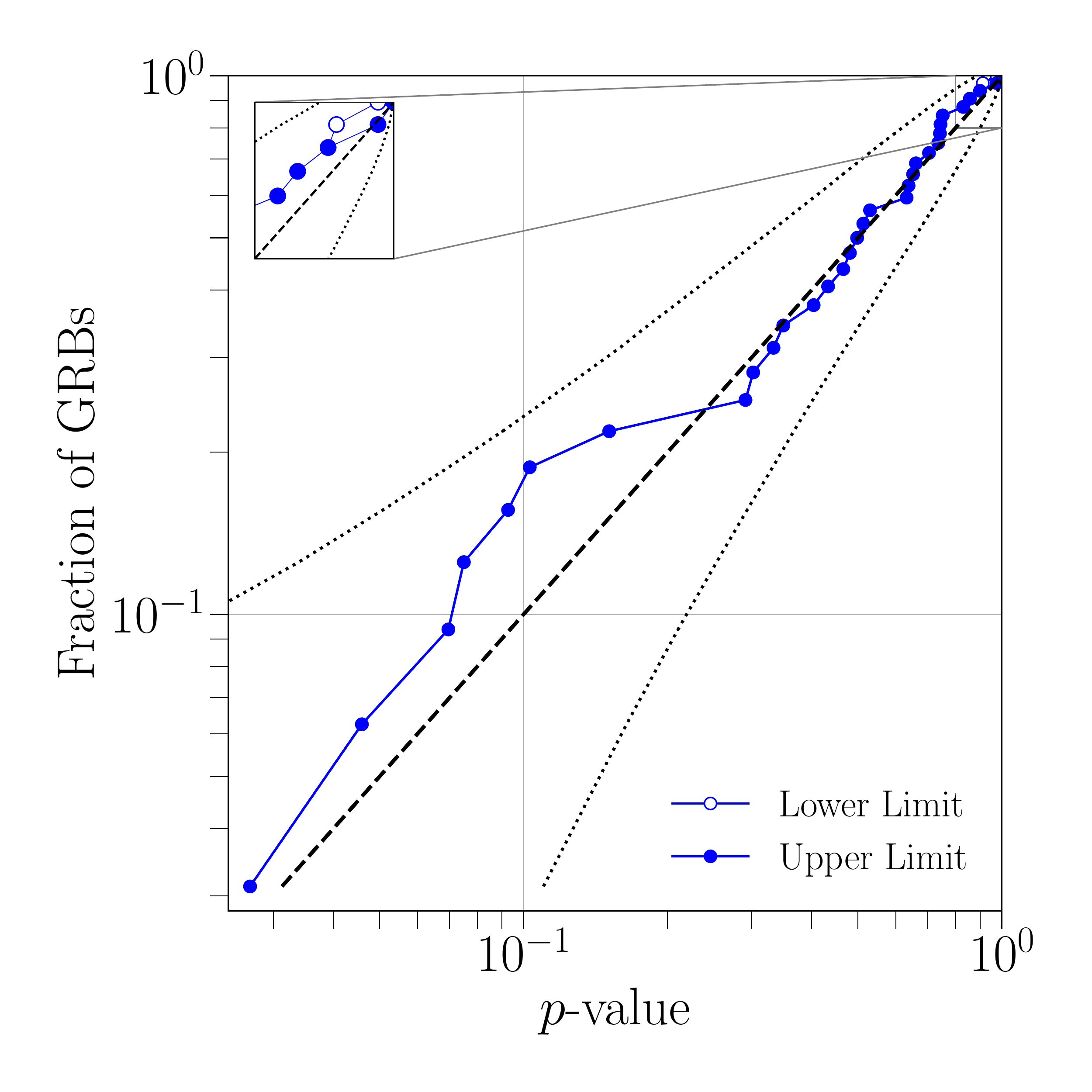}
  \end{center}
  \caption{Cumulative distribution of loudest on-source event
    $p$-values for the neutron star binary modeled search in \ac{O3a}.  If
    the search reports no trigger in the on-source, we plot an upper
    limit on the $p$-value of 1 (open circles), and a lower limit
    equal to the fraction of off-source trials that contained no
    trigger (full circles).  The dashed line indicates the expected
    distribution of $p$-values under the no-signal hypothesis, with
    the corresponding $2\sigma$ envelope marked by dotted lines.}
  \label{fig:pygrb-pvalue}
\end{figure}

\begin{figure}[!t]
  \centering
  \includegraphics[width=\linewidth]{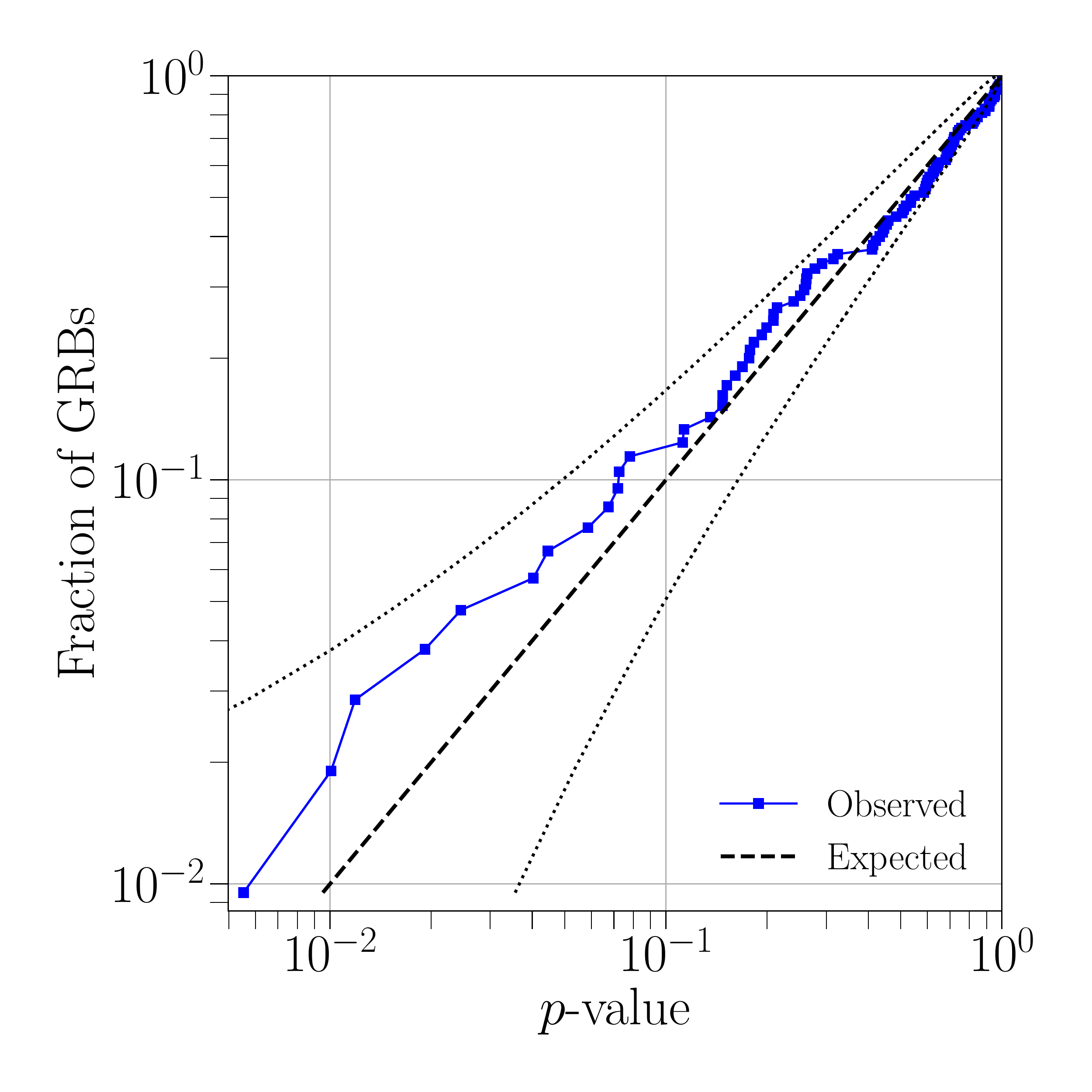}
    \caption{
      \label{fig:burst_pvals} Cumulative distribution of $p$-values
      from the unmodeled search for transient gravitational waves associated with
      \nBurst gamma-ray bursts.  The dashed line represents the expected
      distribution under the no-signal hypothesis, with dotted lines
      indicating a $2\sigma$ deviation from this distribution.}
\end{figure}

During \ac{O3a} we used the generic transient method to follow up a
total of \nBurst \acp{GRB}, whereas the modeled search was applied to
the \nCBC \ac{GRB} triggers classified as short or ambiguous.  For all
of the most \ac{GW}-signal-like triggers associated with the examined
\acp{GRB}, the searches returned no significant probability of
incompatibility with background alone ($p$-value).  This indicates
that no \ac{GW} signal was uncovered in association with any of these
\acp{GRB}.  This is consistent with the estimated GW--GRB joint
detection rate with Fermi-\ac{GBM} of $0.07$--$1.80$ per year
reported in~\citet{Authors:2019fue} for the 2019--2020 LIGO--Virgo
observing run.  The most significant events found by the generic
transient method and by the modeled search had $p$-values of
\pvalBurstLowest (\nameBurstLowest) and \pvalCBCLowestFAP
(\nameCBCLowestFAP), respectively.

Figures~\ref{fig:pygrb-pvalue} and \ref{fig:burst_pvals} show the
cumulative distributions of $p$-values returned by the modeled search
and the generic transient search, respectively.  For cases in which no
associated on-source trigger survived the analysis cuts of the
modeled search, the associated $p$-value ranges between 1 --- i.e.
an upper bound on a probability --- and the fraction of background
trials for the \ac{GRB} that also yielded no associated \ac{GW}
trigger.  In both figures, the expected background distribution under
the no-signal hypothesis is shown by the dashed line, and its
2$\sigma$ limits are indicated by the two dotted lines.  Both
cumulative distributions are within the 2$\sigma$ lines and therefore
compatible with the no-signal hypothesis.  These figures indicate that
the lowest $p$-value found by each search is compatible with the
no-signal hypothesis.

\begin{figure}[!t]
  \begin{center}
    \includegraphics[width=\linewidth]{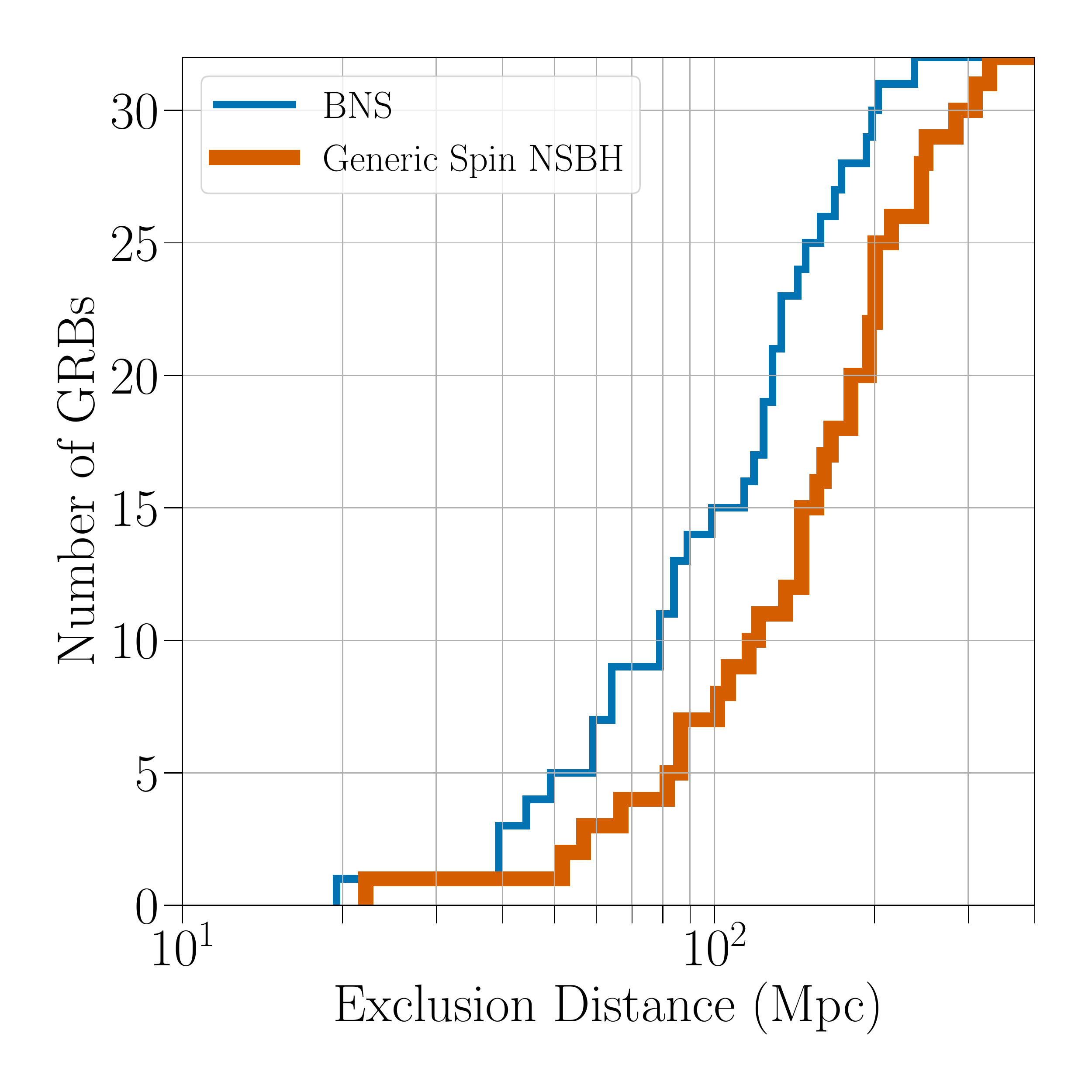}
  \end{center}
  \caption{Cumulative histograms of the 90\% confidence exclusion
    distances, $D_{90}$, for the binary neutron star (blue, thin line) and
    generically spinning neutron star--black hole (orange, thick line) signal models,
    shown for the sample of \nCBC short and ambiguous gamma-ray bursts that
    were followed up by the \ac{NS} binary modeled search during
    \ac{O3a}, none of which had an identified gravitational wave
    counterpart.  For a given \ac{GRB} event and signal model,
    $D_{90}$ is the distance within which 90\% of simulated signals
    inserted into off-source data are recovered with greater
    significance than the most significant on-source trigger.  These
    simulated signals have inclinations $\theta_{JN}$ -- the angle
    between the total angular momentum and the line of sight -- drawn
    uniformly in $\sin \theta_{JN}$ with $\theta_{JN}$ restricted to
    $[0\degr,30\degr] \cup [150\degr,180\degr]$.}
  \label{pygrb-90exclusion}
\end{figure}

Having found no \ac{GW} signal associated with the \acp{GRB} followed
up by our searches, we consider the set of modeled search results and
the set of generic transient search results, collectively.  We apply a
weighted binomial test described in~\citet{2012ApJ...760...12A} to
evaluate how consistent each set of results is collectively with the
no-signal hypothesis.  This test is conducted using the most
significant $5\%$ of $p$-values in the sample weighted by a prior
probability of detection estimated using the network detector
sensitivity at the time and location of each \ac{GRB}.  This final
probability of observing this distribution of $p$-values given
background alone, i.e. under the no-signal hypothesis, was \pvalCBC
(\pvalBurst) for the modeled (generic transient) search method.
Therefore, both searches gave no significant evidence for a population
of unidentified subthreshold \ac{GW} signals.  For the analyses
carried out in \acl{O1}, the combined $p$-values were
\pvalCBCOOne and \pvalBurstOOne for the modeled and generic transient
search, respectively~\citep{Abbott:2016cjt}; in
\acl{O2}, they were \pvalCBCOTwo and
\pvalBurstOTwo~\citep{Authors:2019fue}.

\begin{table}
  \hspace{0.5cm}
  \caption{\label{tab:exclDist} Median 90\% Confidence Level Exclusion Distances, $D_{90}$, for the Searches during \ac{O3a}.}
    \begin{tabularx}{\columnwidth}{ c c c c }
        \hline
        \hline
        \rule{0pt}{4ex}
        Modeled search &     & NSBH          & NSBH          \\
        (Short GRBs)   & BNS & Generic Spins & Aligned Spins \\
        \hline
        \rule[-2ex]{0pt}{4ex}
        $D_{90}$ [Mpc] & \DBNS & \DNSBHGen &  \DNSBHAli
    \end{tabularx}
    \begin{tabularx}{\columnwidth}{ c c c c c}
        \hline
        \hline
        \rule{0pt}{4ex}
        Unmodeled search & CSG    & CSG     & CSG     & CSG     \\ 
        (All GRBs)       & 70\,Hz & 100\,Hz & 150\,Hz & 300\,Hz \\
        \hline
        \rule[-2ex]{0pt}{4ex}
        $D_{90}$ [Mpc] & \DCSGSeventy & \DCSGHundred & \DCSGOneFifty & \DCSGThreeHundred
    \end{tabularx}
    \begin{tabularx}{\columnwidth}{ c c c c c c }
        \hline
        \hline
        \rule{0pt}{4ex}
        Unmodeled search & ADI & ADI & ADI & ADI & ADI \\
        (All GRBs)       & A   & B   & C   & D   & E   \\
        \hline
        \rule[-2ex]{0pt}{4ex}
        $D_{90}$ [Mpc] & \DADIA & \DADIB & \DADIC & \DADID & \DADIE \\
        \hline
    \end{tabularx}
    \vspace{1ex}

     {\raggedright {\bf Note.} Modeled search results are shown for three classes of \ac{NS} binary progenitor model, and unmodeled search results are shown for \acl{CSG}~\citep[\ac{CSG};][]{Abbott:2016cjt} and \acl{ADI}~\citep[\ac{ADI};][]{vanPutten:2001gi,vanPutten:2014kja} models.}
\end{table}

\begin{figure}[!t]
  \centering
  \includegraphics[width=\linewidth]{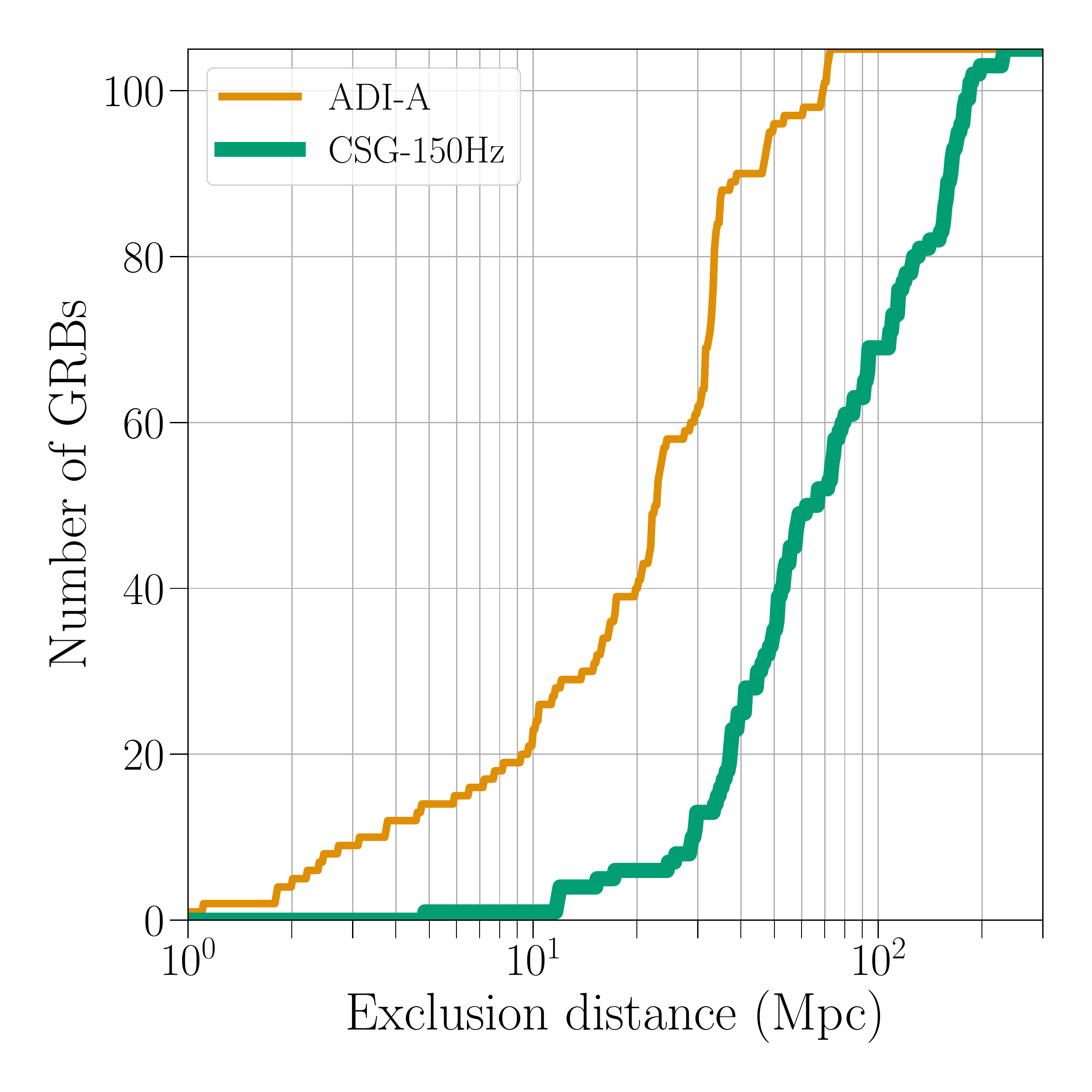}
    \caption{
      \label{fig:burst_dists} Cumulative histograms of the 90\%
      confidence exclusion distances, $D_{90}$, for
      \acl{ADI}~\citep[\ac{ADI};][]{vanPutten:2001gi,
        vanPutten:2014kja} signal model A (orange, thin line) and
      \acf{CSG} 150\,Hz~\citep{Abbott:2016cjt} model (green, thick
      line).  For a given \ac{GRB} and signal model this is the
      distance within which 90\% of simulated signals inserted into
      off-source data are successfully recovered with a significance
      greater than the loudest on-source trigger.  The median values
      for \ac{ADI}-A and \ac{CSG}-150 waveforms are \DADIA\,Mpc and
      \DCSGOneFifty\,Mpc respectively.}
\end{figure}

In Fig.\,\ref{pygrb-90exclusion}, we show the cumulative $90\%$
exclusion distances for the \nCBC short and ambiguous \acp{GRB}
followed up with the modeled search.  The lowest exclusion distance
values ($\sim 20\,$Mpc) were obtained for ambiguous GRB 190409901.
This is due to the fact that only Virgo data were available for this
\ac{GRB} and that the sky location of this event was in a direction in
which Virgo had $\sim 30$\% sensitivity with respect to an optimal sky
location.  For each of the three simulated signal classes, we quote
the median of the \nCBC $D_{90}$ results in the top part of Table
\ref{tab:exclDist}.  All three values are $40$\,\%--$60$\,\% times higher
than those reported in~\citet{Authors:2019fue} for the previous
LIGO--Virgo observing run.  The individual $D_{90}$ values for each
class of simulated signals are reported in Table \ref{tab:combined},
at the end of this paper.  As a term of comparison, during the six
month duration of \ac{O3a}, the Hanford and Livingston Advanced LIGO
instruments, and the Virgo interferometer had \ac{BNS} ranges of
\HBNSRange, \LBNSRange, and \VBNSRange, respectively.\footnote{The BNS
  inspiral range is defined as the distance at which the coalescence
  of two $1.4\,M_\odot$ \acp{NS} can be detected with an \ac{SNR} of
  8, averaged over all directions in the sky, source orientation, and
  polarization~\citep{Finn:1992xs, Allen:2005fk, Chen:2017wpg}.} 
We also place a $90\%$ confidence level lower limit on the distance
for each of the \nBurst \acp{GRB} analyzed by the generic transient
search, assuming the various emission models discussed in
Sec.\,\ref{sec:burst-search}~\citep[see also][]{Abbott:2016cjt}.
Figure \ref{fig:burst_dists} shows the distribution of $D_{90}$ values
for the \ac{ADI} model A~\citep{vanPutten:2001gi, vanPutten:2014kja}
and for a \ac{CSG} with central frequency of
$150$~Hz~\citep{Abbott:2016cjt}.  These limits depend on the
sensitivity of the detector network, which, in turn, varies over time
and with sky location, and have been marginalized over errors
introduced by detector calibration.  For the \ac{ADI} and the \ac{CSG}
models mentioned above, as well as for the other seven models used in
the generic transient method search (see
Sec.\,\ref{sec:burst-search}), we provide population median exclusion
limits, $D_{90}$, in Table \ref{tab:exclDist}.  These vary roughly
over one order of magnitude, which reflects the wide range of models
used in the analysis.  We report the $D_{90}$ values found for each
\ac{GRB} in the case of \ac{ADI} model A simulated signals and
\ac{CSG} simulated signals with central frequency of $150$~Hz in Table
\ref{tab:combined}, at the end of this paper.

\subsection{GRB 190610A}
\label{sec:GRB190610A}

\begin{figure}[!t]
  \centering
  \includegraphics[width=\linewidth]{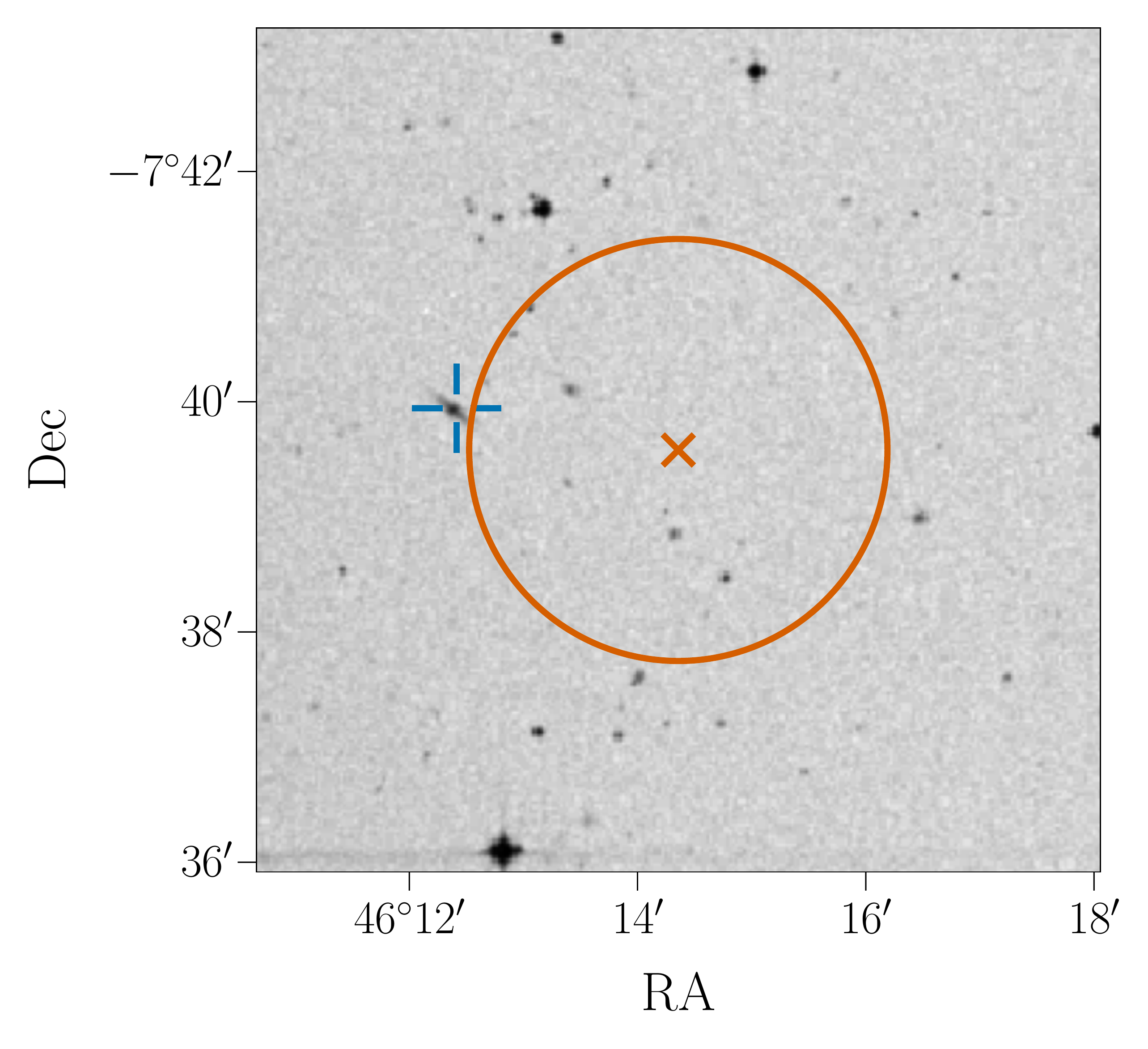}
  \caption{\label{fig:GRB190610A_loc} Overlay of the estimated 90\%
    Swift/\ac{BAT} error radius for GRB 190610A (orange circle) on
    the sky. A galaxy at around
    \NearbyGRBGalaxyLumDist~\citep{Dalya:2018cnd} compatible with this
    localization is indicated by the blue crosshair.}
\end{figure}

For each event in the \ac{O3a} sample that was localized with an error
radius smaller than $0^\circ\hspace{-0.8ex}.\hspace{0.1ex}5$, we searched
GLADE~\citep{Dalya:2018cnd} for galaxies within 200\,Mpc.  We then
compared the angular separation between each \ac{GRB} and galaxy, and
recorded all separations less than or equal to twice the error radius
for each GRB.  Of the \nAllGRB events in our sample, \nNearbyGalaxy
had nearby galaxies according to the definition above: GRB 190530430,
GRB 190531840, GRB 190610A, and GRB 190731943.  Data for our \ac{GW}
follow-up analysis were available only in the case of the short GRB
190610A, first observed by Swift/\ac{BAT}~\citep{GCN:24775} and
localized to within a 90\% error radius of
\NearbyGRBPosErr~\citep{Lien:2016zny, 2019GCN.24783....1P}.  On the
edge of its localization region, there is a nearby galaxy at a
luminosity distance of approximately \NearbyGRBGalaxyLumDist
(\NearbyGRBGalaxyRedshift), as reported in GLADE (see
Fig.\,\ref{fig:GRB190610A_loc}).\footnote{This galaxy can be found in
  the HyperLeda database (\url{http://leda.univ-lyon1.fr/}) under the
  identifier PGC 1015066~\citep{2014A&A...570A..13M}, as well as the
  Sloan Digital Sky Survey under the identifier
  J030449.65-073956.6~\citep{2015ApJS..219...12A}.}  The angular
separation between the center of the localization region and the
nearby galaxy is at the \NearbyGRBSigma level relative to the formal
fit error, which is slightly less conservative than the quoted 90\%
localization derived from the \ac{SNR}, and is consistent with
expectations of angular offsets from a host galaxy at that
distance~\citep{2013ApJ...776...18F}.

We did not find any \ac{GW} signal associated with GRB 190610A in the
data available from the two LIGO detectors (Virgo data were not in
observing mode at that particular time).  Our modeled search
described in Sec.\,\ref{sec:cbc-search}, which uses an on-source
window from $-5\,$s to $+1\,$s around the \ac{GRB} trigger time,
placed 90\% confidence exclusion distances of \BNSExclJuneTen,
\NSBHExclJuneTen, and \NSBHAlignExclJuneTen for \ac{BNS} binaries with
generically oriented spins, \ac{NSBH} binaries with generically
oriented spins, and aligned spin \ac{NSBH} binaries (see
Sec.\,\ref{sec:cbc-search} for more details on these three
populations).  In general, a distance of \NearbyGRBGalaxyLumDist can
be within the reach of our modeled search, but GRB 190610A was in a
sky location such that the sensitivity of both detectors was less than
30\% of what it would have been in an optimal sky location.

\section{Conclusions}
\label{sec:conclusions}

We carried out targeted analyses for \acp{GW} associated with 
Fermi and Swift \acp{GRB} reported during the \ac{O3a}
LIGO--Virgo observing run.  In the case of short and ambiguous
\acp{GRB} events (see Sec.\,\ref{sec:grb_sample}), we ran a modeled
search for \ac{NS} binary merger signals~\citep{Harry:2010fr,
  Williamson:2014wma}, while an unmodeled search for \ac{GW}
transient signals was performed for all
\acp{GRB}~\citep{Sutton:2009gi, Was:2012zq}.
As a result of our analyses, we found no \ac{GW} signal in association
with the \acp{GRB} that we followed up. This is consistent with the
previously predicted rate of coincident detections of $0.1$--$1.4$ per
year for the third observing run of Advanced LIGO and Advanced
Virgo~\citep{Monitor:2017mdv}.  Additionally, by carrying out a
weighted binomial test, we found no strong evidence for a population
of unidentified subthreshold \ac{GW} signals in our results. We set
lower bounds on the distances to the progenitors of all \acp{GRB} we
analyzed for a number of emission models.  These $D_{90}$ values are
reported in Table \ref{tab:combined}, along with other information about each GRB that
we considered; this includes timing, sky location, observing
instrument, and \ac{GW} detectors with available data.  The 90\%
confidence level exclusion distances achieved in this run include the
largest values published so far for some individual
\acp{GRB}~\citep[see][]{Abbott:2016cjt, Authors:2019fue}.  Among the
\acp{GRB} we analyzed is GRB 190610A, the sky localization of which
included a nearby galaxy at a luminosity distance of
\NearbyGRBGalaxyLumDist.  We placed 90\% confidence level exclusion
distances lower than this value for \ac{NS} binary merger \ac{GW}
signals and are therefore unable to rule out the possibility that GRB
190610A occurred in such galaxy.

\acknowledgments
The authors gratefully acknowledge the support of the United States
National Science Foundation (NSF) for the construction and operation of the
LIGO Laboratory and Advanced LIGO as well as the Science and Technology Facilities Council (STFC) of the
United Kingdom, the Max Planck Society (MPS), and the State of
Niedersachsen/Germany for support of the construction of Advanced LIGO
and construction and operation of the GEO600 detector.
Additional support for Advanced LIGO was provided by the Australian Research Council.
The authors gratefully acknowledge the Italian Istituto Nazionale di Fisica Nucleare (INFN),
the French Centre National de la Recherche Scientifique (CNRS) and
the Netherlands Organization for Scientific Research,
for the construction and operation of the Virgo detector
and the creation and support  of the EGO consortium.
The authors also gratefully acknowledge research support from these agencies as well as by
the Council of Scientific and Industrial Research of India,
the Department of Science and Technology, India,
the Science \& Engineering Research Board (SERB), India,
the Ministry of Human Resource Development, India,
the Spanish Agencia Estatal de Investigaci\'on,
the Vicepresid\`encia i Conselleria d'Innovaci\'o, Recerca i Turisme and the Conselleria d'Educaci\'o i Universitat del Govern de les Illes Balears,
the Conselleria d'Innovaci\'o, Universitats, Ci\`encia i Societat Digital de la Generalitat Valenciana and
the CERCA Programme Generalitat de Catalunya, Spain,
the National Science Centre of Poland and the Foundation for Polish Science (FNP),
the Swiss National Science Foundation (SNSF),
the Russian Foundation for Basic Research,
the Russian Science Foundation,
the European Commission,
the European Regional Development Funds (ERDF),
the Royal Society,
the Scottish Funding Council,
the Scottish Universities Physics Alliance,
the Hungarian Scientific Research Fund (OTKA),
the French Lyon Institute of Origins (LIO),
the Belgian Fonds de la Recherche Scientifique (FRS-FNRS),
Actions de Recherche Concertées (ARC) and
Fonds Wetenschappelijk Onderzoek–Vlaanderen (FWO), Belgium,
the Paris \^{I}le-de-France Region,
the National Research, Development and Innovation Office Hungary (NKFIH),
the National Research Foundation of Korea,
the Natural Science and Engineering Research Council Canada,
Canadian Foundation for Innovation (CFI),
the Brazilian Ministry of Science, Technology, Innovations, and Communications,
the International Center for Theoretical Physics South American Institute for Fundamental Research (ICTP-SAIFR),
the Research Grants Council of Hong Kong,
the National Natural Science Foundation of China (NSFC),
the Leverhulme Trust,
the Research Corporation,
the Ministry of Science and Technology (MOST), Taiwan
and
the Kavli Foundation.
The authors gratefully acknowledge the support of the NSF, STFC, INFN and CNRS for provision of computational resources.

We would like to thank all of the essential workers who put their
health at risk during the COVID-19 pandemic, without whom we would not
have been able to complete this work.

We would also like to thank Christian Malacaria and Aaron Tohuvavohu for providing useful comments that helped improve this paper.

\facilities{
    LIGO,
    EGO:Virgo,
    Fermi(GBM),
    Swift(BAT).
}

\software{
    \LALSuite~software library~\citep{LALSuite},
    \texttt{Matplotlib}~\citep{Hunter:2007, thomas_a_caswell_2018_1343133},
    \PYCBC~\citep{alex_nitz_2020_3961510},
    \Xpipeline~\citep{Sutton:2009gi, Was:2012zq}.
}

\begin{longrotatetable}
\begin{deluxetable*}{lcRRlllRRRRR}
  \tablecaption{\label{tab:combined}
    Information and Limits on Associated \ac{GW} Emission for Each of
    the Fermi and Swift \acp{GRB} Followed Up during the
    LIGO--Virgo Run O3a.}
  \tablewidth{700pt}
  \tabletypesize{\scriptsize}
  \tablehead{
    \colhead{} & \colhead{} & \colhead{} & \colhead{} & \colhead{} &
    \colhead{} & \colhead{} & \multicolumn{5}{c}{$D_{90}$ (Mpc)} \\
    \cmidrule{8-12}
    \colhead{GRB Name} & \colhead{UTC Time} & 
    \colhead{R.A.} & \colhead{Decl.} & 
    \colhead{Satellite} & \colhead{Type} & 
    \colhead{Network} & \colhead{\ac{BNS}} & 
    \colhead{Generic \ac{NSBH}} & \colhead{Aligned \ac{NSBH}} &
    \colhead{ADI-A} & \colhead{CSG 150 Hz}
  } 
  \startdata
  190404293 & 07:01:21 & $  8^{\mathrm{h}} 05^{\mathrm{m}} 33^{\mathrm{s}}$ & $ 55^{\circ} 25' $ & Fermi & Long & H1L1  & - & - & - & 35 & 152\\
190406450 & 10:47:20 & $ 23^{\mathrm{h}} 46^{\mathrm{m}} 21^{\mathrm{s}}$ & $ 20^{\circ} 23' $ & Fermi & Long & H1L1V1  & - & - & - & 2 & 115\\
190406465 & 11:09:47 & $ 19^{\mathrm{h}} 05^{\mathrm{m}} 21^{\mathrm{s}}$ & $ 61^{\circ} 30' $ & Fermi & Long & H1L1V1  & - & - & - & 69 & 186\\
190407575 & 13:48:36 & $  6^{\mathrm{h}} 02^{\mathrm{m}} 07^{\mathrm{s}}$ & $ -64^{\circ} 08' $ & Fermi & Long & H1L1V1$^\dagger$  & - & - & - & 61 & 171\\
190407672 & 16:07:26 & $ 12^{\mathrm{h}} 07^{\mathrm{m}} 16^{\mathrm{s}}$ & $ 40^{\circ} 37' $ & Fermi & Long & L1V1  & - & - & - & 32 & 49\\
190407788 & 18:54:41 & $ 13^{\mathrm{h}} 30^{\mathrm{m}} 57^{\mathrm{s}}$ & $ -7^{\circ} 57' $ & Fermi & Ambiguous & L1V1  & 169 & 311 & 395 & 34 & 54\\
190409901 & 21:38:05 & $15^{\mathrm{h}}19^{\mathrm{m}}53^{\mathrm{s}}$ & $-33^{\circ}52'$ & Fermi & Ambiguous & V1 & 19 & 22 & 34 & - & -\\
190411579 & 13:53:58 & $  3^{\mathrm{h}} 02^{\mathrm{m}} 31^{\mathrm{s}}$ & $ 48^{\circ} 38' $ & Fermi & Long & H1L1V1$^\dagger$  & - & - & - & 10 & 108\\
190415173 & 04:09:49 & $  1^{\mathrm{h}} 50^{\mathrm{m}} 50^{\mathrm{s}}$ & $ 17^{\circ} 26' $ & Fermi & Long & H1V1  & - & - & - & 2 & 5\\
190419414 & 09:55:37 & $  7^{\mathrm{h}} 05^{\mathrm{m}} 48^{\mathrm{s}}$ & $ -40^{\circ} 08' $ & Fermi & Long & H1V1$^\dagger$  & - & - & - & 34 & 54\\
190420981 & 23:32:24 & $ 21^{\mathrm{h}} 17^{\mathrm{m}} 09^{\mathrm{s}}$ & $ -66^{\circ} 25' $ & Fermi & Ambiguous & L1V1  & 175 & 215 & 315 & 35 & 52\\
190422284 & 06:48:17 & $ 20^{\mathrm{h}} 26^{\mathrm{m}} 38^{\mathrm{s}}$ & $ -73^{\circ} 01' $ & Fermi & Long & H1L1$^\dagger$  & - & - & - & 14 & 127\\
190422670 & 16:05:04 & $ 12^{\mathrm{h}} 36^{\mathrm{m}} 55^{\mathrm{s}}$ & $ -54^{\circ} 57' $ & Fermi & Long & H1L1V1  & - & - & - & 2 & 167\\
190425089 & 02:07:43 & $ 21^{\mathrm{h}} 01^{\mathrm{m}} 43^{\mathrm{s}}$ & $ -15^{\circ} 13' $ & Fermi & Ambiguous & L1V1(H1L1)  & 204 & 247 & 440 & 23 & 38\\
190427A & 04:34:15 & $ 18^{\mathrm{h}} 40^{\mathrm{m}} 52^{\mathrm{s}}$ & $ 40^{\circ} 19' $ & Swift & Short & L1V1  & 138 & 199 & 253 & 33 & 92\\
190428783 & 18:48:12 & $  1^{\mathrm{h}} 55^{\mathrm{m}} 45^{\mathrm{s}}$ & $ 15^{\circ} 51' $ & Fermi & Long & L1V1  & - & - & - & 29 & 38\\
190429743 & 17:49:50 & $ 13^{\mathrm{h}} 20^{\mathrm{m}} 12^{\mathrm{s}}$ & $ -7^{\circ} 60' $ & Fermi & Long & H1L1V1  & - & - & - & 70 & 126\\
190501794 & 19:03:42 & $ 10^{\mathrm{h}} 25^{\mathrm{m}} 09^{\mathrm{s}}$ & $ -22^{\circ} 00' $ & Fermi & Long & L1V1$^\dagger$  & - & - & - & 18 & 35\\
190502168 & 04:01:30 & $  6^{\mathrm{h}} 16^{\mathrm{m}} 43^{\mathrm{s}}$ & $ 3^{\circ} 17' $ & Fermi & Long & H1L1V1  & - & - & - & 34 & 92\\
190504415 & 09:57:34 & $  4^{\mathrm{h}} 41^{\mathrm{m}} 57^{\mathrm{s}}$ & $ 39^{\circ} 34' $ & Fermi & Long & H1L1V1$^\dagger$  & - & - & - & 16 & 50\\
190504678 & 16:16:28 & $  9^{\mathrm{h}} 09^{\mathrm{m}} 43^{\mathrm{s}}$ & $ 33^{\circ} 01' $ & Fermi & Short & L1V1  & 93 & 124 & 189 & 17 & 68\\
190505051 & 01:14:09 & $ 22^{\mathrm{h}} 21^{\mathrm{m}} 33^{\mathrm{s}}$ & $ 42^{\circ} 11' $ & Fermi & Short & L1V1  & 100 & 149 & 206 & 24 & 36\\
190507270 & 06:28:23 & $ 10^{\mathrm{h}} 23^{\mathrm{m}} 50^{\mathrm{s}}$ & $ -12^{\circ} 48' $ & Fermi & Long & H1L1$^\dagger$  & - & - & - & 39 & 111\\
190507712 & 17:05:16 & $05^{\mathrm{h}}44^{\mathrm{m}}53^{\mathrm{s}}$ & $-61^{\circ}7'$ & Fermi & Short & V1 & 42 & 58 & 70 & - & -\\
190507970 & 23:16:29 & $ 19^{\mathrm{h}} 11^{\mathrm{m}} 16^{\mathrm{s}}$ & $ -22^{\circ} 49' $ & Fermi & Long & H1L1V1  & - & - & - & 32 & 231\\
190508987 & 23:41:24 & $  6^{\mathrm{h}} 54^{\mathrm{m}} 02^{\mathrm{s}}$ & $ 27^{\circ} 02' $ & Fermi & Long & H1L1V1$^\dagger$  & - & - & - & 30 & 178\\
190510120 & 02:52:13 & $  8^{\mathrm{h}} 18^{\mathrm{m}} 09^{\mathrm{s}}$ & $ -53^{\circ} 04' $ & Fermi & Long & H1V1$^\dagger$  & - & - & - & 8 & 53\\
190510430 & 10:19:16 & $  8^{\mathrm{h}} 32^{\mathrm{m}} 31^{\mathrm{s}}$ & $ 33^{\circ} 33' $ & Fermi & Short & H1L1  & 128 & 196 & 253 & 48 & 116\\
190511A & 07:14:48 & $  8^{\mathrm{h}} 25^{\mathrm{m}} 46^{\mathrm{s}}$ & $ -20^{\circ} 15' $ & Swift & Long & H1L1  & - & - & - & 50 & 142\\
190512A & 14:40:09 & $  5^{\mathrm{h}} 29^{\mathrm{m}} 35^{\mathrm{s}}$ & $ -7^{\circ} 35' $ & Swift & Long & L1V1  & - & - & - & 20 & 56\\
190515190 & 04:33:03 & $  9^{\mathrm{h}} 10^{\mathrm{m}} 45^{\mathrm{s}}$ & $ 29^{\circ} 17' $ & Fermi & Short & L1V1  & 122 & 148 & 194 & 22 & 42\\
190517813 & 19:30:10 & $ 18^{\mathrm{h}} 00^{\mathrm{m}} 04^{\mathrm{s}}$ & $ 25^{\circ} 46' $ & Fermi & Long & H1L1  & - & - & - & 30 & 74\\
190519A & 07:25:39 & $  7^{\mathrm{h}} 39^{\mathrm{m}} 01^{\mathrm{s}}$ & $ -38^{\circ} 49' $ & Swift & Long & H1L1V1  & - & - & - & 10 & 190\\
190525032 & 00:45:47 & $ 22^{\mathrm{h}} 32^{\mathrm{m}} 04^{\mathrm{s}}$ & $ 5^{\circ} 27' $ & Fermi & Short & H1L1V1  & 128 & 248 & 385 & 22 & 165\\
190531312 & 07:29:11 & $  1^{\mathrm{h}} 24^{\mathrm{m}} 28^{\mathrm{s}}$ & $ 16^{\circ} 21' $ & Fermi & Long & L1V1  & - & - & - & 21 & 73\\
190531568 & 13:38:03 & $ 18^{\mathrm{h}} 16^{\mathrm{m}} 40^{\mathrm{s}}$ & $ 38^{\circ} 52' $ & Fermi & Short & H1V1  & 86 & 150 & 187 & 3 & 29\\
190601325 & 07:47:24 & $ 10^{\mathrm{h}} 51^{\mathrm{m}} 55^{\mathrm{s}}$ & $ 54^{\circ} 35' $ & Fermi & Short & H1V1(H1L1V1)  & 136 & 169 & 248 & 17 & 34\\
190603795 & 19:04:25 & $  1^{\mathrm{h}} 20^{\mathrm{m}} 19^{\mathrm{s}}$ & $ 40^{\circ} 55' $ & Fermi & Long & H1L1  & - & - & - & 3 & 156\\
190604446 & 10:42:37 & $ 22^{\mathrm{h}} 50^{\mathrm{m}} 12^{\mathrm{s}}$ & $ 46^{\circ} 22' $ & Fermi & Long & H1L1  & - & - & - & 72 & 174\\
190606080 & 01:55:07 & $  5^{\mathrm{h}} 06^{\mathrm{m}} 09^{\mathrm{s}}$ & $ -0^{\circ} 41' $ & Fermi & Short & H1V1  & 52 & 68 & 81 & 8 & 37\\
190608009 & 00:12:18 & $ 15^{\mathrm{h}} 02^{\mathrm{m}} 57^{\mathrm{s}}$ & $ -31^{\circ} 25' $ & Fermi & Long & L1V1$^\dagger$  & - & - & - & 15 & 30\\
190610750 & 17:59:49 & $ 21^{\mathrm{h}} 49^{\mathrm{m}} 31^{\mathrm{s}}$ & $ 42^{\circ} 25' $ & Fermi & Long & L1V1  & - & - & - & 1 & 40\\
190610834 & 20:00:23 & $ 20^{\mathrm{h}} 59^{\mathrm{m}} 19^{\mathrm{s}}$ & $ -15^{\circ} 56' $ & Fermi & Ambiguous & L1V1  & 149 & 202 & 306 & 34 & 58\\
190610A & 11:27:45 & $  3^{\mathrm{h}} 04^{\mathrm{m}} 57^{\mathrm{s}}$ & $ -7^{\circ} 40' $ & Swift & Short & H1L1  & 63 & 82 & 114 & 23 & 58\\
190612165 & 03:57:24 & $ 14^{\mathrm{h}} 55^{\mathrm{m}} 48^{\mathrm{s}}$ & $ 62^{\circ} 06' $ & Fermi & Long & H1L1V1$^\dagger$  & - & - & - & 48 & 178\\
190613A & 04:07:18 & $ 12^{\mathrm{h}} 10^{\mathrm{m}} 12^{\mathrm{s}}$ & $ 67^{\circ} 15' $ & Swift & Long & H1L1V1  & - & - & - & 70 & 200\\
190613B & 10:47:02 & $ 20^{\mathrm{h}} 21^{\mathrm{m}} 45^{\mathrm{s}}$ & $ -4^{\circ} 39' $ & Swift & Long & H1L1$^\dagger$  & - & - & - & 54 & 160\\
190615636 & 15:16:27 & $ 12^{\mathrm{h}} 45^{\mathrm{m}} 36^{\mathrm{s}}$ & $ 49^{\circ} 23' $ & Fermi & Long & H1L1V1  & - & - & - & 4 & 45\\
190619018 & 00:26:01 & $ 23^{\mathrm{h}} 17^{\mathrm{m}} 14^{\mathrm{s}}$ & $ 12^{\circ} 52' $ & Fermi & Long & H1L1V1$^\dagger$  & - & - & - & 6 & 132\\
190619595 & 14:16:25 & $ 19^{\mathrm{h}} 24^{\mathrm{m}} 16^{\mathrm{s}}$ & $ 20^{\circ} 10' $ & Fermi & Long & H1L1V1$^\dagger$  & - & - & - & 2 & 47\\
190620507 & 12:10:10 & $ 10^{\mathrm{h}} 48^{\mathrm{m}} 19^{\mathrm{s}}$ & $ 30^{\circ} 29' $ & Fermi & Long & H1L1  & - & - & - & 5 & 94\\
190623461 & 11:03:27 & $ 22^{\mathrm{h}} 21^{\mathrm{m}} 57^{\mathrm{s}}$ & $ -23^{\circ} 20' $ & Fermi & Long & H1L1V1  & - & - & - & 10 & 95\\
190627481 & 11:31:59 & $ 23^{\mathrm{h}} 29^{\mathrm{m}} 02^{\mathrm{s}}$ & $ -8^{\circ} 53' $ & Fermi & Long & H1L1  & - & - & - & 16 & 116\\
190627A & 11:18:31 & $ 16^{\mathrm{h}} 19^{\mathrm{m}} 29^{\mathrm{s}}$ & $ -5^{\circ} 18' $ & Swift & Ambiguous & H1L1  & 115 & 139 & 211 & 21 & 77\\
190628521 & 12:30:55 & $  9^{\mathrm{h}} 36^{\mathrm{m}} 19^{\mathrm{s}}$ & $ -77^{\circ} 04' $ & Fermi & Long & H1L1  & - & - & - & 47 & 164\\
190630257 & 06:09:58 & $ 20^{\mathrm{h}} 27^{\mathrm{m}} 55^{\mathrm{s}}$ & $ -1^{\circ} 20' $ & Fermi & Short & H1V1  & 47 & 91 & 121 & 16 & 25\\
190630B & 06:02:08 & $ 14^{\mathrm{h}} 54^{\mathrm{m}} 55^{\mathrm{s}}$ & $ 41^{\circ} 32' $ & Swift & Long & H1V1  & - & - & - & 10 & 12\\
190630C & 23:52:59 & $ 19^{\mathrm{h}} 35^{\mathrm{m}} 33^{\mathrm{s}}$ & $ -32^{\circ} 46' $ & Swift & Long & H1L1V1  & - & - & - & 47 & 118\\
190701A & 09:45:20 & $  1^{\mathrm{h}} 52^{\mathrm{m}} 31^{\mathrm{s}}$ & $ 58^{\circ} 54' $ & Swift & Long & H1L1V1  & - & - & - & 12 & 157\\
190707285 & 06:50:05 & $ 10^{\mathrm{h}} 11^{\mathrm{m}} 28^{\mathrm{s}}$ & $ -30^{\circ} 59' $ & Fermi & Long & H1L1V1  & - & - & - & 49 & 163\\
190707308 & 07:23:01 & $ 12^{\mathrm{h}} 17^{\mathrm{m}} 19^{\mathrm{s}}$ & $ -9^{\circ} 31' $ & Fermi & Long & H1L1  & - & - & - & 34 & 75\\
190708365 & 08:45:11 & $ 13^{\mathrm{h}} 59^{\mathrm{m}} 24^{\mathrm{s}}$ & $ -1^{\circ} 18' $ & Fermi & Long & H1L1V1  & - & - & - & 18 & 58\\
190712018 & 00:25:20 & $22^{\mathrm{h}}44^{\mathrm{m}}14^{\mathrm{s}}$ & $-38^{\circ}35'$ & Fermi & Ambiguous & H1L1 & 68 & 204 & 357 & - & -\\
190712095 & 02:16:41 & $ 19^{\mathrm{h}} 13^{\mathrm{m}} 33^{\mathrm{s}}$ & $ 56^{\circ} 09' $ & Fermi & Long & H1L1V1  & - & - & - & 38 & 159\\
190716019 & 00:27:59 & $  4^{\mathrm{h}} 41^{\mathrm{m}} 40^{\mathrm{s}}$ & $ 16^{\circ} 28' $ & Fermi & Long & H1L1$^\dagger$  & - & - & - & 5 & 12\\
190718A & 04:41:15 & $ 22^{\mathrm{h}} 26^{\mathrm{m}} 25^{\mathrm{s}}$ & $ -41^{\circ} 11' $ & Swift & Long & H1L1$^\dagger$  & - & - & - & 12 & 93\\
190719499 & 11:57:51 & $  6^{\mathrm{h}} 34^{\mathrm{m}} 26^{\mathrm{s}}$ & $ 6^{\circ} 42' $ & Fermi & Long & H1L1V1  & - & - & - & 33 & 94\\
190719C & 14:58:34 & $ 16^{\mathrm{h}} 00^{\mathrm{m}} 49^{\mathrm{s}}$ & $ 13^{\circ} 00' $ & Swift & Long & H1L1V1$^\dagger$  & - & - & - & 4 & 157\\
190720613 & 14:42:09 & $ 13^{\mathrm{h}} 30^{\mathrm{m}} 52^{\mathrm{s}}$ & $ 41^{\circ} 47' $ & Fermi & Long & H1L1V1  & - & - & - & 10 & 38\\
190720964 & 23:08:38 & $  9^{\mathrm{h}} 15^{\mathrm{m}} 28^{\mathrm{s}}$ & $ -55^{\circ} 35' $ & Fermi & Long & H1L1  & - & - & - & 10 & 34\\
190724031 & 00:43:56 & $11^{\mathrm{h}}21^{\mathrm{m}}24^{\mathrm{s}}$ & $15^{\circ}9'$ & Fermi & Short & H1L1 & 197 & 286 & 329 & - & -\\
190726642 & 15:24:53 & $ 20^{\mathrm{h}} 41^{\mathrm{m}} 02^{\mathrm{s}}$ & $ 34^{\circ} 17' $ & Fermi & Long & H1L1V1$^\dagger$  & - & - & - & 34 & 85\\
190726843 & 20:14:30 & $ 22^{\mathrm{h}} 50^{\mathrm{m}} 43^{\mathrm{s}}$ & $ -55^{\circ} 59' $ & Fermi & Long & H1L1V1$^\dagger$  & - & - & - & 72 & 180\\
190727668 & 16:01:52 & $ 14^{\mathrm{h}} 57^{\mathrm{m}} 57^{\mathrm{s}}$ & $ 19^{\circ} 26' $ & Fermi & Long & H1L1  & - & - & - & 24 & 109\\
190727B & 20:18:17 & $  8^{\mathrm{h}} 25^{\mathrm{m}} 59^{\mathrm{s}}$ & $ -13^{\circ} 16' $ & Swift & Long & L1V1  & - & - & - & 34 & 68\\
190728271 & 06:30:36 & $ 23^{\mathrm{h}} 46^{\mathrm{m}} 45^{\mathrm{s}}$ & $ 5^{\circ} 26' $ & Fermi & Short & H1L1V1  & 160 & 204 & 272 & 32 & 79\\
190804058 & 01:23:27 & $  7^{\mathrm{h}} 12^{\mathrm{m}} 04^{\mathrm{s}}$ & $ -64^{\circ} 52' $ & Fermi & Ambiguous & H1V1  & 132 & 184 & 240 & 34 & 74\\
190805106 & 02:32:30 & $ 11^{\mathrm{h}} 10^{\mathrm{m}} 36^{\mathrm{s}}$ & $ -23^{\circ} 46' $ & Fermi & Long & H1L1V1  & - & - & - & 35 & 121\\
190805199 & 04:46:00 & $ 13^{\mathrm{h}} 59^{\mathrm{m}} 00^{\mathrm{s}}$ & $ 19^{\circ} 28' $ & Fermi & Long & H1V1  & - & - & - & 33 & 75\\
190806535 & 12:50:02 & $ 20^{\mathrm{h}} 22^{\mathrm{m}} 14^{\mathrm{s}}$ & $ 0^{\circ} 33' $ & Fermi & Long & H1L1V1  & - & - & - & 20 & 52\\
190808752 & 18:03:17 & $ 11^{\mathrm{h}} 12^{\mathrm{m}} 12^{\mathrm{s}}$ & $ 39^{\circ} 43' $ & Fermi & Long & L1V1  & - & - & - & 32 & 62\\
190810675 & 16:12:01 & $ 12^{\mathrm{h}} 55^{\mathrm{m}} 07^{\mathrm{s}}$ & $ -37^{\circ} 34' $ & Fermi & Short & H1L1V1  & 85 & 159 & 222 & 2 & 50\\
190813520 & 12:29:09 & $  7^{\mathrm{h}} 05^{\mathrm{m}} 31^{\mathrm{s}}$ & $ -23^{\circ} 16' $ & Fermi & Short & H1L1  & 84 & 121 & 161 & 23 & 56\\
190816A & 14:42:24 & $ 22^{\mathrm{h}} 44^{\mathrm{m}} 43^{\mathrm{s}}$ & $ -29^{\circ} 45' $ & Swift & Long & L1V1$^\dagger$  & - & - & - & 34 & 75\\
190817953 & 22:52:25 & $ 18^{\mathrm{h}} 20^{\mathrm{m}} 40^{\mathrm{s}}$ & $ -31^{\circ} 08' $ & Fermi & Ambiguous & H1L1  & 61 & 102 & 109 & 1 & 30\\
190822705 & 16:55:29 & $  8^{\mathrm{h}} 49^{\mathrm{m}} 04^{\mathrm{s}}$ & $ -8^{\circ} 05' $ & Fermi & Short & L1V1  & 148 & 181 & 278 & 2 & 17\\
190824A & 14:46:39 & $ 14^{\mathrm{h}} 21^{\mathrm{m}} 17^{\mathrm{s}}$ & $ -41^{\circ} 54' $ & Swift & Long & H1L1V1$^\dagger$  & - & - & - & 24 & 160\\
190825171 & 04:06:56 & $ 14^{\mathrm{h}} 03^{\mathrm{m}} 26^{\mathrm{s}}$ & $ -74^{\circ} 08' $ & Fermi & Long & L1V1  & - & - & - & 17 & 38\\
190827467 & 11:12:48 & $ 11^{\mathrm{h}} 43^{\mathrm{m}} 14^{\mathrm{s}}$ & $ 46^{\circ} 27' $ & Fermi & Long & H1L1  & - & - & - & 7 & 26\\
190828B & 12:59:59 & $ 16^{\mathrm{h}} 47^{\mathrm{m}} 21^{\mathrm{s}}$ & $ 27^{\circ} 17' $ & Swift & Long & H1V1$^\dagger$  & - & - & - & 22 & 52\\
190829A & 19:56:44 & $  2^{\mathrm{h}} 58^{\mathrm{m}} 10^{\mathrm{s}}$ & $ -8^{\circ} 57' $ & Swift & Long & L1V1$^\dagger$  & - & - & - & 33 & 51\\
190830023 & 00:32:48 & $  7^{\mathrm{h}} 27^{\mathrm{m}} 36^{\mathrm{s}}$ & $ -23^{\circ} 46' $ & Fermi & Long & L1V1  & - & - & - & 33 & 59\\
190830264 & 06:20:46 & $ 10^{\mathrm{h}} 36^{\mathrm{m}} 48^{\mathrm{s}}$ & $ -54^{\circ} 43' $ & Fermi & Ambiguous & H1V1(H1L1V1)  & 242 & 331 & 478 & 31 & 45\\
190831332 & 07:57:31 & $  4^{\mathrm{h}} 22^{\mathrm{m}} 31^{\mathrm{s}}$ & $ 14^{\circ} 53' $ & Fermi & Long & L1V1  & - & - & - & 22 & 40\\
190831693 & 16:38:37 & $ 11^{\mathrm{h}} 19^{\mathrm{m}} 31^{\mathrm{s}}$ & $ -22^{\circ} 21' $ & Fermi & Long & H1L1V1  & - & - & - & 28 & 86\\
190901890 & 21:21:49 & $ 14^{\mathrm{h}} 41^{\mathrm{m}} 12^{\mathrm{s}}$ & $ 0^{\circ} 56' $ & Fermi & Long & L1V1$^\dagger$  & - & - & - & 22 & 29\\
190903722 & 17:19:36 & $04^{\mathrm{h}}09^{\mathrm{m}}43^{\mathrm{s}}$ & $-64^{\circ}8'$ & Fermi & Short & V1 & 66 & 87 & 133 & - & -\\
190904174 & 04:11:00 & $  2^{\mathrm{h}} 23^{\mathrm{m}} 40^{\mathrm{s}}$ & $ -25^{\circ} 02' $ & Fermi & Ambiguous & L1V1(H1V1)  & 84 & 109 & 154 & 7 & 12\\
190905985 & 23:38:28 & $15^{\mathrm{h}}37^{\mathrm{m}}55^{\mathrm{s}}$ & $3^{\circ}7'$ & Fermi & Short & V1 & 42 & 54 & 77 & - & -\\
190906767 & 18:25:09 & $ 11^{\mathrm{h}} 27^{\mathrm{m}} 21^{\mathrm{s}}$ & $ -71^{\circ} 34' $ & Fermi & Long & H1L1V1  & - & - & - & 36 & 111\\
190910028 & 00:39:37 & $ 15^{\mathrm{h}} 18^{\mathrm{m}} 00^{\mathrm{s}}$ & $ 9^{\circ} 04' $ & Fermi & Long & H1V1  & - & - & - & 33 & 54\\
190913155 & 03:43:09 & $ 16^{\mathrm{h}} 53^{\mathrm{m}} 21^{\mathrm{s}}$ & $ 44^{\circ} 58' $ & Fermi & Short & H1V1(H1L1V1)  & 201 & 250 & 382 & 9 & 15\\
190914345 & 08:16:34 & $  1^{\mathrm{h}} 13^{\mathrm{m}} 45^{\mathrm{s}}$ & $ 21^{\circ} 27' $ & Fermi & Long & L1V1$^\dagger$  & - & - & - & 23 & 36\\
190915240 & 05:44:57 & $  3^{\mathrm{h}} 13^{\mathrm{m}} 19^{\mathrm{s}}$ & $ 3^{\circ} 59' $ & Fermi & Long & L1V1  & - & - & - & 25 & 42\\
190916590 & 14:10:14 & $ 21^{\mathrm{h}} 25^{\mathrm{m}} 04^{\mathrm{s}}$ & $ -48^{\circ} 54' $ & Fermi & Long & H1L1V1  & - & - & - & 23 & 170\\
190919764 & 18:20:02 & $ 23^{\mathrm{h}} 49^{\mathrm{m}} 26^{\mathrm{s}}$ & $ -21^{\circ} 49' $ & Fermi & Long & H1L1V1  & - & - & - & 73 & 234\\
190921699 & 16:45:55 & $ 22^{\mathrm{h}} 33^{\mathrm{m}} 31^{\mathrm{s}}$ & $ -63^{\circ} 25' $ & Fermi & Long & H1V1  & - & - & - & 32 & 46\\
190923617 & 14:48:02 & $  0^{\mathrm{h}} 32^{\mathrm{m}} 48^{\mathrm{s}}$ & $ -11^{\circ} 01' $ & Fermi & Ambiguous & H1L1  & 133 & 162 & 239 & 32 & 80\\
190926A & 09:52:16 & $  6^{\mathrm{h}} 42^{\mathrm{m}} 27^{\mathrm{s}}$ & $ 59^{\circ} 32' $ & Swift & Long & H1L1$^\dagger$  & - & - & - & 72 & 186\\
190930400 & 09:36:06 & $ 15^{\mathrm{h}} 52^{\mathrm{m}} 52^{\mathrm{s}}$ & $ -6^{\circ} 05' $ & Fermi & Long & L1V1$^\dagger$  & - & - & - & 22 & 30\\
191001279 & 06:41:50 & $ 20^{\mathrm{h}} 20^{\mathrm{m}} 47^{\mathrm{s}}$ & $ 15^{\circ} 05' $ & Fermi & Long & H1V1  & - & - & - & 12 & 41\\
  \enddata
\vspace{1ex}                                                                                                                                                

{\raggedright {\bf Note.} 
   The Satellite column lists the instrument the
    sky localization of which was used for \ac{GW} analysis purposes.
    The Network column lists the \ac{GW} detector network used in the
    analysis of each \ac{GRB}: H1 = LIGO Hanford, L1 = LIGO
    Livingston, V1 = Virgo.  A $^\dagger$ denotes cases in which
    $T_{90} > 60\,\mathrm{s}$, so the on-source window of the generic
    transient search was extended to cover the \ac{GRB} duration.  For
    cases in which the generic transient search
    (Sec.\,\ref{sec:burst-search}) and the neutron star binary search
    (Sec.\,\ref{sec:cbc-search}) used a different network, we report
    the network used by the latter in parentheses.  Columns 8--12
    display the 90\% confidence exclusion distances to the \ac{GRB}
    ($D_{90}$) for several emission scenarios: \ac{BNS}, generic and
    aligned spin \ac{NSBH}, \ac{ADI}-A, and \ac{CSG} \ac{GW} burst at
    150\,Hz with total radiated energy $E_{\text{GW}} =
    10^{-2}\,\mathrm{\Msun c^2}$.  The first three are determined with
    the neutron star binary search, while the last two are calculated
    with the generic transient search.}
\end{deluxetable*}
\end{longrotatetable}

\end{document}